\documentclass[aip]{revtex4-1}

\usepackage{graphicx}
\usepackage{appendix}
\usepackage{mciteplus}
\usepackage{multirow}
\usepackage{amsfonts}
\usepackage{amsmath}
\usepackage{verbatim}
\usepackage{enumerate}
\usepackage[T1]{fontenc}
\usepackage{bm}
\usepackage{siunitx}

\usepackage[utf8]{inputenc}
\usepackage{xr-hyper}
\usepackage{hyperref}
\hypersetup{breaklinks=true,colorlinks=true,linkcolor=blue,citecolor=blue,filecolor=magenta,urlcolor=blue}
\usepackage{siunitx}
\usepackage{orcidlink}

\begin{document}


\title{Lattice Instabilities Along the Transformation from Hexagonal to Cuboidal Structures in Hard- and Soft-Sphere Models}

\author{Andres Robles-Navarro\,\orcidlink{0000-0001-9586-1003}}\email{andres.robles.n@gmail.com}
\affiliation{Centre for Theoretical Chemistry and Physics, The New Zealand Institute for Advanced Study (NZIAS), Massey University Albany, Private Bag 102904, Auckland 0745, New Zealand}
\author{Shaun Cooper\,\orcidlink{0000-0001-5103-0400}}\email{S.Cooper@massey.ac.nz}
\affiliation{School of Natural and Computational Sciences, Massey University Albany, Private Bag 102904,
Auckland 0745, New Zealand.}
\author{Andreas W. Hauser\,\orcidlink{0000-0001-6918-3106}}\email{andreas.w.hauser@gmail.com}
\affiliation{Institute of Experimental Physics, Graz University of Technology, Petersgasse 16, 8010 Graz, Austria}
\author{Fabian Zehetmair\,\orcidlink{0009-0009-8842-0512}}\email{fabian.zehetmair@student.tugraz.at}
\affiliation{Institute of Experimental Physics, Graz University of Technology, Petersgasse 16, 8010 Graz, Austria}
\author{Odile Smits\,\orcidlink{0000-0003-1259-147X}}\email{Smits.Odile.Rosette@gmail.com}
\affiliation{School of Mathematics and Physics, University of Queensland, Brisbane QLD 4072, Australia}
\author{Peter Schwerdtfeger\,\orcidlink{0000-0003-4845-686X}}\email{peter.schwerdtfeger@gmail.com}
\affiliation{Centre for Theoretical Chemistry and Physics, The New Zealand Institute for Advanced Study (NZIAS), Massey University Albany, Private Bag 102904, Auckland 0745, New Zealand}


\date{\today}


\begin{abstract}
The diffusionless Burgers-Bain phase transition from a hexagonal close-packed (hcp) arrangement to a cuboidal lattice (face-centered cubic, fcc, and body-centered cubic, bcc) is analysed in great detail for Lennard-Jones (LJ)solids. From the lattice vectors of an underlying bi-lattice smoothly connecting these phases, we are able to express the corresponding lattice sums for inverse power potentials in terms of fast converging Bessel function expansions resulting in an efficient evaluation to computer accuracy for cohesive energies. From the kissing hard-sphere limit we derive exact analytical expressions for the lattice parameters varying along the minimum energy path of the phase transition. This simple model suggests that for the Burgers-Bain transformation of a LJ solid requires a minimum of four lattice parameters, $(a,\alpha,\beta,\gamma=c/a)$, describing the change in the base lattice lengths $a$ and $c$, the shear force acting on the hexagonal base plane through a parameter $\alpha$, the sliding force of the middle layer in the original hexagonal packing arrangement through a single parameter $\beta$, and the cuboidal transformation through a parameter $\gamma=c/a$. This choice results in a smooth transformation through a two-step process: hcp$\to$fcc$\to$bcc. However, a further extension of the parameter space including an additional slide parameter for the middle layer, one suddenly observes a distinct symmetry-breaking effect along the hcp$\rightarrow$fcc transition path with a bifurcation point appearing joining the original Burgers with the Bain path of the bcc$\rightarrow$fcc cuboidal transition. Furthermore, for soft LJ potentials the bcc phase appears as a local minimum along the Burgers hcp$\rightarrow$fcc path with two transition states to either the hcp or fcc phase. The underlying topology of the Burgers-Bain phase transition also incorporates the rhombohedral distortion of the bcc phase, which is analyzed in detail. As a first application of our formalism, we discuss solid argon and compare the LJ results with variable-cell nudge elastic band optimizations using density functional theory. We find that the activation energy for the hcp$\rightarrow$fcc transformation is highly sensitive to the density functional applied, and that dispersion corrections are important as expected for weakly interacting systems.

\end{abstract}


\maketitle 


\section{Introduction}
\label{sec:intro}

The relative stability between the face-centred cubic (fcc) and hexagonal close-packed (hcp) structures, and their formation from the gas or the liquid phase known as the nucleation problem, has been a matter of intense discussion and debate over the past 50 years.\cite{Grimvall2012,Krainyukova-2012,Sanchez-Burgos2021} Both phases have the same hard-sphere packing density of $\rho=\pi/3\sqrt{2}=0.74048048969\dots$, as have all possible associated Barlow structures (mixtures between (AB)$_\infty$ and (ABC)$_\infty$ stackings)\cite{Barlow1883,Hopkins2011,Osang2021,Schwerdtfeger-2024b} for which there are infinitely many. It was only in recent times that Kepler's original conjecture, stating that the fcc packing density cannot be surpassed for hard (unit) sphere packings and therefore is optimal, was proven by Hales.\cite{Hales2011,hales-2005}

Optimal sphere packings predict only a very small difference in the free energy up to the melting point.\cite{FrenkelLadd1984,Woodcock1997,Sanchez-Burgos2021} For real atomic or molecular crystals the two close-packed structures are also found to be very close in energy.\cite{Krainyukova2011}  A prominent example is solid argon, where at low temperatures and pressures, vibrational effects need to be included to stabilize the experimentally observed fcc over the hcp phase.\cite{Moyano2007,Schwerdtfeger-2016} However, the cohesive energy difference between the two phases is predicted to be a mere $8.1\pm0.7$ J/mol\cite{Stoll-2000,Schwerdtfeger-2016} from relativistic coupled cluster calculations. Such small energy differences between the fcc and hcp phase for solids are also detected in nucleation processes \cite{Senger_JChemPhys110,Lovett_RepProgPhys2007,LuLi1998,Schwerdtfeger-2006}. On the other hand, for the heavier noble gas elements the hcp phase becomes the dominant phase at high pressures,\cite{Cynn2001,Rosa2018,Dewaele2021} which is most likely due to the fact that with increasing pressure vibrational effects become less important compared to the competing many-body electronic contributions. 

Why one phase dominates over the other at certain temperatures and pressures delicately depends on the different static and dynamic contributions to the free energy. It is however notoriously difficult to get a detailed mechanistic insight into solid-state phase transitions by both experimental and theoretical methods,\cite{Caspersen2005} and therefore such transitions are in general not so well understood.\cite{Stillinger2001,Ackland2002} One could naively slide some of the hexagonal layers to turn hcp into fcc, i.e. for the hexagonal layer sequences we have the transformation (ABA$\downarrow$B$\downarrow$A$\uparrow$B$\uparrow$)$_\infty \rightarrow$(ABCABC)$_\infty$,\cite{Ackland2002} which involves a supercell treatment. However, it is not known if this is the lowest minimum energy  path, for example, in such sliding of the hexagonal layers one may well access local minima such as dense Barlow packings, e.g. (ABABA$\downarrow$B)$_\infty \rightarrow$(ABABCB)$_\infty$. 

In 1934 Burgers suggested a very simple diffusionless hcp$\leftrightarrow$bcc (body centred cubic) phase transition path\cite{Burgers1934}, while Bain earlier in 1924 proposed a simple cuboidal fcc$\leftrightarrow$bcc transition path \cite{Bain1924}. Since then, there has been some controversy whether or not the fcc phase is actually required in a minimum energy path for the hcp$\rightarrow$bcc phase transition and if these transitions follow at all the diffusionless mechanism as suggested by Bain and Burgers.\cite{Cayron2013,lu2014,Cayron2015,Cayron2016} Moreover, it is not clear if the Burgers transformation belonging to the class of martensitic transitions is a first-order phase transition or not.\cite{Cook1975} More recently, Feng and Widom showed from DFT calculations of several transition metals that the hcp$\rightarrow$bcc transition exhibits an alternating slide instability corresponding to a tetragonal symmetry breaking\cite{Feng2018}. While this instability manifests in a distinct phonon instability, a detailed explanation is still missing.

Modeling solid-state phase transitions at the microscopic level poses significant challenges.\cite{raghavan1975solid,Gooding1988,Caspersen2005,Carter2008a,Torrents2017,Wenbin2021} This is mainly due to the fact that often stacking faults or defects are involved in such phase changes, which requires a computationally expensive super-cell treatment in a molecular dynamics simulation.\cite{Zangwill1985,Dovesi1994,Bingxi2017} Diffusionless martensitic transformations do not have these problems, but mapping out the correct minimum energy path for such a transition can still be a formidable task.\cite{rifkin1984,Sandoval_2009,Cayron2013,Cayron2015,Bingxi2017} Moreover, various theoretical approximations (such as the density functional approximation) may not result in the correct energy sequence between the different (often quasi-degenerate) polymorphs involved.\cite{Dronskowski2003} Moreover, when empirical model potentials are used, the stability of the bcc phase heavily depends on its analytical form and parameters chosen\cite{Beauchamp1983}. 

Caspersen and Carter \cite{Caspersen2005} generalized the climbing image-nudged elastic band algorithm to find transition paths in solid-state phase transitions, and subsequently applied this to the diffusionless martensitic transformation \cite{Bain1924,Burgers1934} from hcp to the bcc structure and further to fcc for metallic lithium. They gave detailed information for the transformation matrices acting on the corresponding lattice vectors. In a different approach, Raju Natarajan and Van der Ven used a set of two parameters derived from the Hencky logarithmic strain to map out the volume preserving potential energy surfaces for the Burgers-Bain transformation for metallic lithium, sodium and magnesium.\cite{Natarajan2019} Cayron introduced transformation matrices (which he termed angular distortive matrices) for continuous atomic displacements between the three different phases.\cite{Cayron2016} Bingxi Li et al. performed molecular dynamics simulations using a (12,6)-Lennard-Jones potential for argon calibrated at $T=40$ K and $P= 1$ bar.\cite{Bingxi2017} They discussed different paths for the hcp$\leftrightarrow$fcc phase transition. However, the topology of the cohesive energy surface as a function of the lattice parameters to map minimum energy paths in such phase transitions may critically depend on the chosen model applied describing the interactions between the atoms in the lattice. Moreover, molecular dynamics (MD) or Monte-Carlo simulations for the phase transitions in the solid state are computer time consuming and depend on the size of the supercell chosen. From the dynamics at elevated temperatures it is often not easy to map out different minimum energy paths. Hence, it would be advantageous to develop a simpler model capable of estimating the activation energy (or at least an upper bound) involved in such solid-state phase transitions.

In this paper we analyze in detail the Burgers-Bain hexagonal-to-cuboidal transformation path from hcp$\rightarrow$fcc/bcc for the case of a general $(n,m)$ Lennard-Jones (LJ) potential ($n>m, m>3$),\cite{Jones-1924b,Jones-1925,Wales2024}
\begin{equation} \label{eq:VLJ}
\phi_{\textrm{LJ}}(r)=\frac{\epsilon nm}{n-m} \;  \left[ \frac{1}{n}\left( \frac{r_e}{r} \right)^{n} - \frac{1}{m} \left( \frac{r_e}{r} \right)^{m} \right] 
\end{equation}
describing the interactions between the atoms in a solid using exact lattice summations to obtain the cohesive energy.\cite{borwein-2013,burrows-2020,Smits2021,Cooper2024} Here, $\epsilon$ and $r_e$ are the binding energy and the equilibrium distance of the diatomic molecule, respectively. Both of these parameters can be arbitrarily set to 1 in order to express them in dimensionless units. For very large exponents $(n,m)$, the LJ potential approaches the kissing hard-sphere (KHS) model with 
\begin{align}
\label{eq:KHS}
\phi_\text{KHS}(r)
&= \begin{cases}
\infty \quad &\text{for} \quad r<r_e, \\
-\epsilon \quad &\text{for} \quad r=r_e, \\
0 \quad &\text{for} \quad r>r_e.
\end{cases}
\end{align}
as originally introduced by Baxter \cite{Baxter1968,Stell1991}. A preliminary account for the (12,6)-LJ potential is given in Ref.\onlinecite{RoblesNavarro2025}. Here we focus more on the theoretical details of the combined Burgers-Bain path and derivations of lattice sums. Many insights gained through this simple LJ model become invaluable in future applications for real systems using, for example, density functional theory. We note that the LJ potential has in principle uncountable infinitely many solid state structures representing minima (proof by number of Barlow packings\cite{Cooper2024,Schwerdtfeger-2024b}), and many for unit cells which are not too large in size. For example, not too long ago a new metastable LJ phase has been discovered by Parinello and co-workers\cite{Parrinello2008}.

\begin{figure}[htb!]
\centering
\includegraphics[width=.33\columnwidth]{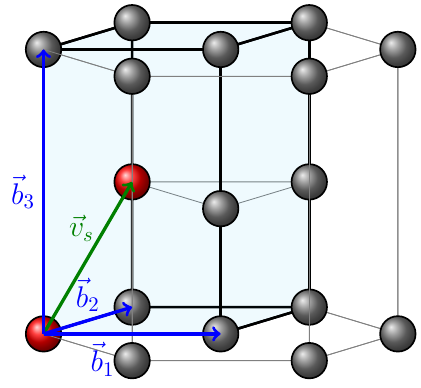}
\caption{The (distorted) cuboidal (blue lines shown on the left) and the hcp structure both with a ABABAB... sequence shown in a hexagonal unit cell with corresponding basis vectors with cell parameters $|\vec{b}_1|=a$, $|\vec{b}_2|=b$, and $|\vec{b}_3|=c$. The ratio $\gamma_{2}=c/a=\sqrt{8/3}$ together with $\angle(\vec{b}_1,\vec{b}_2)=60^\circ$ and $|\vec{b}_1|=|\vec{b}_2|=a$ leads to the optimal hcp lattice with 12 kissing spheres around a central atom. For $c=a$ and $\angle(\vec{b}_1,\vec{b}_2)=90^\circ$ we obtain the bcc lattice with 8 kissing spheres. Alternatively, we have $|\vec{b}_1|=|\vec{b}_2|=a$, $c/a=\sqrt{2}$ and $\angle(\vec{b}_1,\vec{b}_2)=90^\circ$ for the fcc lattice with 12 kissing spheres.}
\label{fig:hcp}
\end{figure}

We have recently shown\cite{Jerabek2022,RoblesNavarro2023,burrows2025a} that the Bain transformation can be effectively described within a two-parameter space $(A,r_\text{NN})$ 
, where $r_\text{NN}$ is the nearest neighbor distance in the solid and $A$ is responsible for describing the martensitic transformation path from the axial centred-cuboidal (acc) lattice ($A=1/3$), to the bcc lattice ($A=1/2$), the mean centred-cuboidal (mcc) lattice ($A=1/\sqrt2$), and finally the face-centred cubic (fcc) lattice ($A=1$) in a general body-centered tetragonal (bct) arrangement. The advantage of this choice of parameters instead of the usual bct lattice parameters $(a,c)$ is that the nearest neighbor distance changes only slightly along the Bain transformation path and a two-dimensional picture can therefore be avoided. We will show that a similar choice of parameters for a bi-lattice shown in Figure \ref{fig:hcp} can be used to describe the Burgers transformation from hcp to fcc, and demonstrate that symmetry breaking effects need to be considered in the  Burgers-Bain transformation, with the possible appearance of a bcc phase representing a local minimum along the Burgers path. Furthermore, the observed symmetry breaking results in a connection between the Burgers and the Bain path. The formalism can be used for future density functional studies and the activation energy obtained can be seen as an upper bound to the true (and possibly more complex) minimum energy path requiring supercell treatments and consideration of Barlow packings.

In the coming section we detail the theory of the Burgers-Bain hcp$\leftrightarrow$fcc$\leftrightarrow$bcc transformation and develop the corresponding lattice sums for inverse power potentials in terms of fast converging Bessel sums to capture important long-range effects. We then apply these lattice sums to various $(n,m)$ Lennard-Jones potentials and discuss the results for the transition paths. We will show that for such potentials, starting from the hcp phase, the minimum energy transition path leads to fcc rather than directly to the bcc phase within the parameter space chosen, except when we allow for symmetry breaking within an extended parameter space for the lattice constants for soft LJ potentials, that is in the case where bcc becomes a metastable minimum. As a special case we discuss solid argon where we also apply density functional calculations together with the variable cell nudge elastic band algorithm to map out the MEP. A conclusion and future perspectives on this topic is given in the final section.

\section{Theory}

To correctly describe the Burgers-Bain martensitic hcp$\leftrightarrow$fcc/bcc phase transition,\cite{Bain1924,Burgers1934} we define the cuboidal and hexagonal lattices within the smallest common unit cell, that is, a monoclinic cell with two specified layers A and B as shown in Figure \ref{fig:hcp}. We then introduce a common set of basis vectors containing parameters that transform the lattices smoothly into each other.

\subsection{Lattice vectors and properties}

\subsubsection{The hexagonal close packed structure}
\label{sec:properties}

The hcp phase is a bi-lattice \cite{Stillinger2019} that requires two hexagonal Bravais lattices to describe the close-packed structure, or alternatively one hexagonal unit cell with one atom positioned inside specified by its Wyckoff positions with fractional coordinates $(0,0,0)$ and $(\frac{1}{3},\frac{1}{3},\frac{1}{2})$. It belongs together with all the other Barlow lattices to the more general class of multi-lattices \cite{Stillinger2019}.

To define the hcp lattice, we start by considering the underlying hexagonal Bravais lattice ($h$) with the accompanying basis vectors,
\begin{equation} \label{eq:latvec}
\vec{b}_1^\top = a\left( 1, 0, 0  \right) , \quad
\vec{b}_2^\top = b\left(\frac{1}{2},  \frac{\sqrt{3}}{2}, 0 \right) ,\quad
\vec{b}_3^\top = c\left( 0, 0,  1   \right)
\end{equation}
with the lattice parameters $|\vec{b}_1|=a$, $|\vec{b}_2|=b$, $|\vec{b}_3|=c$, $\angle(\vec{b}_1,\vec{b}_2)=60^{\circ}$ and $\angle(\vec{b}_1,\vec{b}_3)=\angle(\vec{b}_2,\vec{b}_3)=90^\circ$. These basis vectors build the A layers in Fig. \ref{fig:hcp}. The parameters $a$, $b$ and $c$, allow for flexibility in the cell dimensions, and the hexagonal unit cell is recovered when $a=b$. We define the generator matrix $B$ (also called the cell tensor) with consisting of these three vectors as
\begin{equation}
B_h = \begin{pmatrix}
\vec{b}_1^\top \\
\vec{b}_2^\top \\
\vec{b}_3^\top 
\end{pmatrix}
= a
\begin{pmatrix}
1  & 0  &  0 \\
\frac{\gamma_{1}}{2} &\frac{\sqrt{3}\gamma_{1}}{2} &  0 \\
0 & 0  &  \gamma_{2}
\end{pmatrix}
\end{equation}
where we use the notation $\gamma_{1}=\frac{b}{a}$ and $\gamma_{2}=\frac{c}{a}$. This leads to the following positive definite symmetric Gram matrix 
\begin{equation}
\left( G_{ij} \right)_h = \left( \vec{b}_i^\top \vec{b}_j \right)= a^2
\begin{pmatrix}
1 & \frac{\gamma_{1}}{2}  &  0 \\
\frac{\gamma_{1}}{2} & \gamma_{1}^{2} &  0 \\
0 & 0  & \gamma_2^2 
\end{pmatrix}
\end{equation}
or $G=BB^\top$, with $\text{det}G=\frac{3}{4}a^{6}\gamma_{1}^{2}\gamma_{2}^{2}=\frac{3}{4}a^{2}b^{2}c^{2}>0$. From an arbitrarily chosen atom at the origin, all points in the hexagonal lattice (A layers only) are described by 
\begin{equation}
\vec{r}^h_1(\vec{i}) = B_h^\top \vec{i} = i_1\vec{b}_1+ i_2\vec{b}_2+ i_3\vec{b}_3 \\
\end{equation}
with $\vec{i}\in\mathbb{Z}^3$. Their distances to the origin are given by the quadratic form 
\begin{equation}
|\vec{r}^h_1(\vec{i})| = \sqrt{\vec{i}^\top G_h \vec{i} } = a \sqrt{i_1^2 + \gamma_{1}i_1i_2 + \gamma_{1}^{2}i_2^2 +\gamma_{2}^2 i_3^2} \,.
\end{equation}
The volume of the unit cell is determined through the Gram of $B$ matrix
\begin{equation}
V(a,\gamma_{1},\gamma_{2})={\textrm{det}}B_{h}=\sqrt{{\textrm{det}}G_{h}}= \frac{\sqrt{3}}{2}a^{3}\gamma_{1}\gamma_{2}
\label{eq:vol} 
\end{equation}
which is just the product of the diagonal elements of the $B$-matrix as this matrix has been set is in triangular form eliminating unwanted rotations of the unit cell.  The nearest neighbor distance is given by
\begin{equation}
r^h_{\textrm{NN}}(a,b,c)=\textrm{min}\{r^h_1(\vec{i}) \} = {\textrm{min}}\left\{ a,b,c\right\} \,.
\end{equation}
and the kissing number for a lattice $\mathcal{L}$ is defined as
\begin{equation}
\label{eq:kiss}
\kappa(\mathcal{L})=\#\{\vec{r}(\vec{i})\in \mathcal{L} ~|~ |\vec{r}(\vec{i})|=r_\text{NN}(\mathcal{L})\}
\end{equation}
where $\#$ stands for the count (the number of distances). For the hcp structure with ideal value of $\gamma_{\textrm{hcp}} = \sqrt{ \tfrac{8}{3}}$ the kissing number is $\kappa=12$. Similarly, we have $\kappa=12$ for fcc and $\kappa=8$ for bcc.

We now introduce the second layer, i.e. the B-layer as shown in Figure \ref{fig:hcp}, to complete the hcp structure. If we assume, for example, that the sphere of the second layer lies above the centroid of the triangle defined by the vectors $\vec{b}_1$ and $\vec{b}_2$, then the B-layer is shifted by a vector 
of $\vec{v}_h^\top=\frac{a}{2} \left( \frac{2+\gamma_{1}}{3}, \frac{\gamma_{1}}{\sqrt{3}}, \gamma_{2} \right)$ with respect to the lattice vectors given in \eqref{eq:latvec}, such that the position of any atom in the B layers is given by
\begin{align}
\label{eq:shift}
\vec{r}^h_2(\vec{i}) &=  B_h^\top \vec{i} +\vec{v}_h \;,
\end{align}
Note that the generator matrix transforms the fractional into the cartesian coordinates for atoms located inside the unit cell, and vice versa through the inverse $B^{-1}$. As a side note we could also have taken the center of the inscribing circle of the triangle defined by the vectors $\vec{b}_1$ and $\vec{b}_2$, but this is not important in our analysis as we shall see.

We call the set $\{ \vec{b}_i,\vec{v}_h \}$ generalized lattice vectors. They include the Wyckoff positions within a unit cell. The resulting set of distance vectors $\{ \vec{r}^h_1(\vec{i}), \vec{r}^h_2(\vec{i})\}$ then produce all points in 3D space for the hcp bi-lattice. For the minimum distance $r^h_{\textrm{NN}}$ in an hcp bi-lattice we now have,
\begin{equation}
r^h_{\textrm{NN}}(a,b,c,|\vec{v}_h|)={\textrm{min}}\left\{ a,b,c,|\vec{v}_h| \right\}=a\, {\textrm{min}}\left\{ 1,\gamma_{1},\gamma_2,\frac{|\vec{v}_h|}{a} \right\} \,.
\label{eq:nearneigh}
\end{equation}

\subsubsection{The cubic structures}

From Figure \ref{fig:hcp} we see how the hcp and cubic structures are related and can be smoothly transformed into each other, which is what Burgers had in mind in his original 1934 paper\cite{Burgers1934}. We therefore make the transformation from cubic to the hexagonal lattices through an underlying monoclinic unit cell, as defined above for the hexagonal lattice. The description of the simple cubic ($c$) lattice in terms of a primitive monoclinic ($mP$) can be achieved by introducing the following basis vectors
\begin{equation} \label{eq:latvec1}
\vec{b'}_{1}^\top = a  \left( 1, 0, 0  \right) , \quad
\vec{b'}_2^\top = a \left( 0, \gamma_{1}, 0  \right)  ,\quad
\vec{b'}_3^\top = a \left( 0, 0,  \gamma_{2}   \right) \,.
\end{equation}
For the cubic structure we have $\angle(\vec{b}_{1},\vec{b}_{2})=\angle(\vec{b}_{1},\vec{b}_{3})=\angle(\vec{b}_{2},\vec{b}_{3})=90^{\circ}$ and the simple cubic unit cell is obtained when $\gamma_{1}=\gamma_{2}=1$. We conveniently keep $\gamma_{1}$ and $\gamma_{2}$ in the definition of the generator matrix for the cubic lattice as it becomes important later-on for the definition of the bcc and fcc lattices,
\begin{equation}\label{eq:CubicLattice}
B_c = a
\begin{pmatrix}
1 & 0  &  0 \\
0 & \gamma_{1} &  0 \\
0 & 0  &  \gamma_{2}
\end{pmatrix}
\end{equation}
We can already see the familiarity between the two matrices $B_h$ and $B_c$. This gives the following Gram matrix for $B_c$,
\begin{equation}
G_c =  a^2
\begin{pmatrix}
1 & 0  &  0 \\
0 & \gamma_{1}^{2} &  0 \\
0 & 0  & \gamma_{2}^{2} 
\end{pmatrix}
\end{equation}
with $\text{det}(G)=a^{6}\gamma_{1}^{2}\gamma_{2}^{2}>0$. 
The volume is therefore $V(a,\gamma_{1},\gamma_{2})=a^3\gamma_{1}\gamma_{2}=abc$, or for the ideal bcc lattice $V=a^3$. The distances to all atoms within the A layers are given by $\vec{r}_{1}^{c}(\vec{i}) =  B_c^\top \vec{i} $, and we have the expression in terms of the quadratic form $(\vec{i}^\top G_c \vec{i} )$,
\begin{equation}
|\vec{r}^c_1(\vec{i})| = \left(\vec{i}^\top G_c \vec{i} \right)^\frac{1}{2} = a \sqrt{i_{1}^{2} + \gamma_{1}^{2}i_{2}^{2} + \gamma_{2}^{2} i_{3}^{2}} \,.
\end{equation}
We now address again the B-layer in Figure \ref{fig:hcp} for the bcc lattice by using the shift vector $\vec{v}_c^\top=\frac{a}{2} \left( 1 , \gamma_{1}, \gamma_{2} \right)$ with length $|\vec{v}_c|=\frac{a}{2}\sqrt{1+\gamma_{1}^{2}+\gamma_{2}^{2}}$.
The distances are given by $\vec{r}^c_2(\vec{i}) = B_c^\top \vec{i} +\vec{v}_c$. Again, the set of both vectors $\{ \vec{r}^c_1(\vec{i}), \vec{r}^c_2(\vec{i})\}$ produce all points in 3D space for the bcc lattice. So far we have a three-parameter space for the distances $\vec{r_k}(a,\gamma_1,\gamma_2)$ for the two lattices hcp and bcc. They will become variable parameters when we discuss the Lennard-Jones potential further below. 

There is one important fact to note. In the anticipated hcp$\leftrightarrow$bcc transformation we start with an hcp multi-lattice using the hexagonal primitive cell as the underlying unit cell which transforms along a certain path into a cubic multi-lattice. This cubic unit cell is not the primitive bcc cell. In fact, the smallest distance in the bcc cell is given along the [111] direction to the center lattice point with a distance of $a_c=|\vec{v}_c|=a\sqrt{3}/2$ for $\gamma_{1}=\gamma_{2}=1$. Lennard-Jones and Ingham therefore introduced an additional multiplicative factor to the lattice for the bcc cell \cite{Jones-1925}, which we discuss further below for the more general Burgers transformation between the hcp and the cuboidal lattices. 

\subsubsection{The Burgers-Bain transformation}
\label{sec:PhaseTrans}

The connection between the hexagonal and the cubic generator matrices is given by
\begin{equation}
B_h = U_B B_c =
\begin{pmatrix}
1 & 0   &  0 \\
\frac{\gamma_{1}}{2} &\frac{\sqrt{3}}{2} &  0 \\
0 & 0  & 1 
\end{pmatrix}
B_c 
\end{equation}
This means that the distances will transform like $\vec{r}_k^h=U_B\vec{r}_k^c ~(k=1,2)$. The second set of distances, $\vec{r}_2$, also concerns the transformation of the shift vector
\begin{equation}
\vec{v'} = U_B \vec{v}_c = \frac{a}{2} \left(1,\gamma_{1}\left(\tfrac{1+\sqrt{3}}{2}\right),\gamma_2\right)^\top
\end{equation}
However, the vector $\vec{v'} $ differs markedly from $\vec{v}_h$ in its first and second component. It is therefore more suitable to introduce the shift vector directly with freely varying parameters $\beta_{1},\beta_{2}$, and $\beta_{3}$ such that 
\begin{equation}
\vec{v}_s^\top=\frac{a}{2} \left( \beta_{1} , \beta_{2}, \beta_{3} \right)
\label{eq:sf}
\end{equation}
with $\beta_{1}^{c}=\beta_{2}^{c}=1$ for the ideal cubical lattices and $\beta_{1}^{h}=1$ and $\beta_{2}^h=\frac{1}{\sqrt{3}}$ for the ideal hexagonal lattices. The length of the shift vector becomes $|\vec{v}_s|=\tfrac{a}{2}\sqrt{\beta_{1}^{2}+\beta_{2}^{2}+\beta_{3}^{2}}$. This factor becomes important when we rescale our lattice sums.

\begin{figure}[htb!]
\centering
\includegraphics[width=.66\columnwidth]{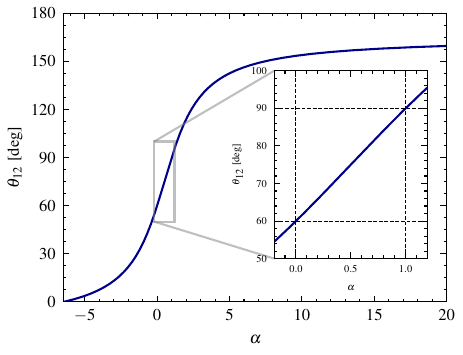}
\caption{Relationship between the parameter $\alpha$ and the angle between the two base vectors $\vec{b}_1$ and $\vec{b}_2$.}
\label{fig:angle}
\end{figure}

The Gram matrices transform as
\begin{equation}
G_h=B_h B_h^\top  = U_B B_c B_c^\top U_B^\top = U_B G_c  U_B^\top = (U_B U_B^\top) G_c =
\begin{pmatrix}
1 & \frac{\gamma_{1}}{2} &  0 \\
\frac{\gamma_{1}}{2} & \frac{\gamma_{1}^{2}+3}{4} &  0 \\
0 & 0  & 1 
\end{pmatrix}
G_c = S G_c .
\end{equation}
We can now formulate the smooth transition from the cuboidal lattice to hcp as
\begin{align}\label{linearcombination}
B&(a,\alpha,\gamma_{1},\gamma_{2})=\alpha B_c(a,\gamma_{1},\gamma_{2}) +(1-\alpha) B_h(a,\gamma_{1},\gamma_{2}) 
\nonumber
= \left\{ \alpha +(1-\alpha) U_B \right\} B_c(a,\gamma) \\
&= a
\begin{pmatrix}
1  & 0  &  0 \\
\frac{\gamma_{1}(1-\alpha)}{2} & \gamma_{1}\left[\alpha\left(1-\frac{\sqrt{3}}{2}\right)+\frac{\sqrt{3}}{2}\right] &  0 \\
0 & 0  & \gamma_{2}
\end{pmatrix}
= a
\begin{pmatrix}
1  & 0  &  0 \\
\omega_1(\alpha,\gamma_{1}) &\omega_3(\alpha,\gamma_{1}) &  0 \\
0 & 0  &  \gamma_{2}
\end{pmatrix}
\end{align}
where $\alpha$ is the structure parameter defining the type of lattice being $\alpha=0$ for the hexagonal and $\alpha=1$ for the cubical lattice, by definition, $\omega_{1}(\alpha,\gamma_{1}) = \frac{\gamma_{1}}{2}(1-\alpha)$, and $\omega_{3}(\alpha,\gamma_{1}) = \gamma_{1}\left[\alpha\left(1-\frac{\sqrt{3}}{2}\right)+\frac{\sqrt{3}}{2}\right]$. Note that in this form, the length of the vector $\vec{b}_{2}$ changes during the transition at fixed $b$ (in fact, $b$ is a parameter to be optimized), and its variation with respect to $\alpha$ is given by
\begin{equation}\label{b2vector}
    |\vec{b}_{2}| = b\sqrt{1-\alpha(\alpha-1)(\sqrt{3}-2)} = bf_{s}(\alpha) = a\gamma_1 f_{s}(\alpha)
\end{equation}
where $f_{s}(\alpha)$ is a scaling factor that has its minimum value $f_{s}=0.965926$ at $\alpha=0.5$. Consequently, the transformation matrix $B$ is describing the rotation and scaling of the vector $\vec{b}_{2}$. As shown in Figure \ref{fig:angle}, the angle $\angle(\vec{b}_{1},\vec{b}_{2})\equiv \theta_{12}$ approaches $0^{\circ}$ at low $\alpha$ but never reaches $180^{\circ}$. In the interval of interest, $\alpha = [0,1]$, it varies almost linearly with $\alpha$ and is thus a good measure of the angle of the base of the unit cell. In fact we have
\begin{equation}\label{eq:angle}
\cos\theta_{12}=\frac{\omega_{1}}{\sqrt{\omega_1^2+\omega_3^2}}.
\end{equation}
The linear behavior in the plot could be approximated as $\theta_{12} \approx \tfrac{(\alpha+2)\pi}{6}$ in the $[0,1]$ region.

The two sets of distance vectors are given by
\begin{align}
\label{distvec}
\vec{r}_1(\vec{i}) =  B(a,\alpha,\gamma_{1},\gamma_{2})^\top \vec{i} \quad \text{and} \quad \vec{r}_2(\vec{i}) =  B(a,\alpha,\gamma_{1},\gamma_{2})^\top \vec{i} + \vec{v}_s(a,\beta_{1},\beta_{2},\beta_{3})
\end{align}
We now have a seven-parameter space $\{p_i\}=(a,\alpha,\beta_{1},\beta_{2},\beta_{3},\gamma_{1},\gamma_{2})$ as a minimum set of parameters to describe the Burgers-Bain transformation with an underlying primitive monoclinic cell. If the middle layer sits exactly in between the two neighboring layers we expect that $\beta_3=\gamma_2$. 

The symmetric Gram matrix for $B(a,\alpha,\gamma_{1},\gamma_{2})$ becomes,
\begin{equation}\label{eq:S3}
\begin{aligned}
    G(a,\alpha,\gamma_1,\gamma_2) &= B(a,\alpha,\gamma_1,\gamma_2) B(a,\alpha,\gamma_1,\gamma_2)^\top = a^2 
\begin{pmatrix}
1  &\omega_1(\alpha,\gamma_{1})  &  0 \\
\omega_1(\alpha,\gamma_{1})  & \omega_2(\alpha,\gamma_{1})  &  0 \\
0 & 0  &  \gamma_{2}^2
\end{pmatrix} \\
&= a^2 S_3(\alpha,\gamma_1,\gamma_2)
\end{aligned}
\end{equation}
where 
\begin{equation}\label{eq:omegarel}
\omega_{2}(\alpha,\gamma_{1})=\omega_{1}(\alpha,\gamma_{1})^{2}+\omega_{3}(\alpha,\gamma_{1})^{2}
\end{equation}
For $G(a,\alpha,\gamma_1,\gamma_2)$ to be positive definite we have the condition that $\omega_3(\alpha)>0$, i.e. $\alpha>-(3+\sqrt{3})\approx -4.732$.

\begin{figure}[htbp!]
\centering
\includegraphics[width=.75\columnwidth]{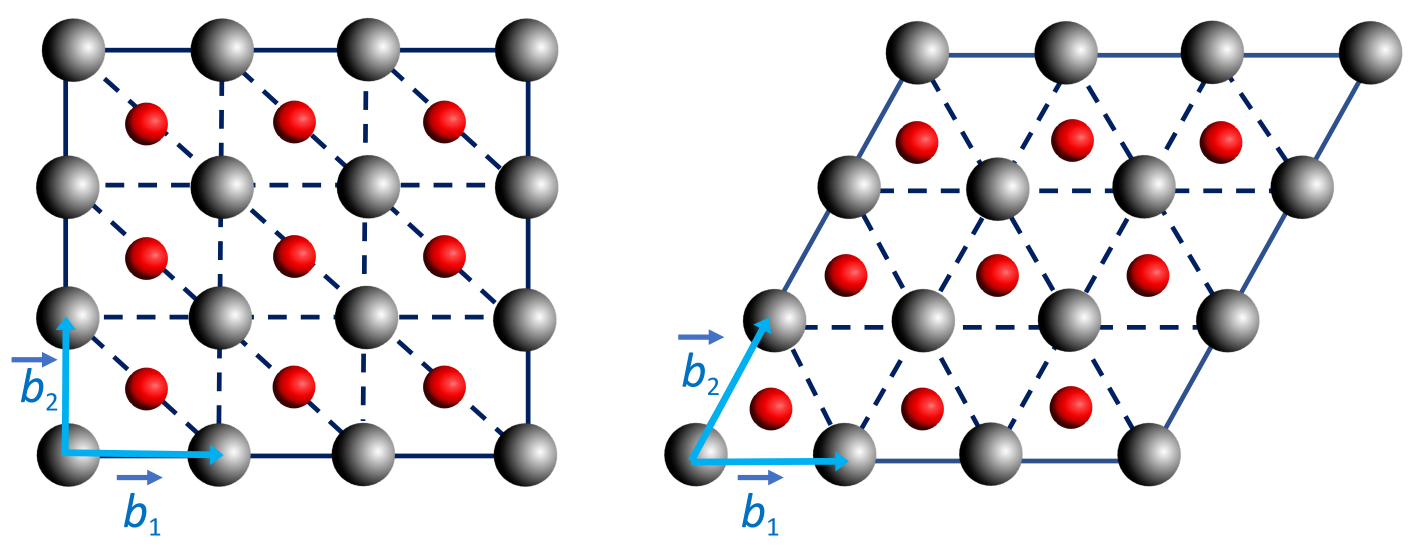}
\caption{The shearing in the A-layer (gray spheres, change of angle between $\vec{b}_1$ and $\vec{b}_2$) of the base hexagonal plane and sliding of the B-layer (red spheres, sitting at $c/2$ above the A-layer) in the Burgers transformation. Left: bcc lattice with $\angle(\vec{b}_1,\vec{b}_2)=90^\circ$, $\gamma_{2}=1$ and fractional coordinates $\vec{u}^\top=(\frac{1}{2},\frac{1}{2},\frac{1}{2})$, or fcc lattice with $\gamma_{2}=\sqrt{2}$ and $\vec{u}^\top=(\frac{1}{2},\frac{1}{2},\frac{1}{2})$. Right: hcp structure with $\angle(\vec{b}_1,\vec{b}_2)=60^\circ$, $\gamma_{2}=\sqrt{\frac{8}{3}}$ and $\vec{u}^\top=(\frac{1}{3},\frac{1}{3},\frac{1}{2})$.}
\label{fig:Burgers}
\end{figure}

It is now clear that the martensitic transition is described by a shearing of the A-layer, related mainly to the parameter $\alpha$, with a change in angle between the two basis vectors $\vec{b}_1$ and $\vec{b}_2$ together with a sliding of the B-layer in direction orthogonal to the $\vec{b}_1$ and $\vec{b}_3$ vectors and additional changes in the lattice parameters, which are summarized in Table \ref{tab:latticeparameters} for the three lattices hcp, fcc and bcc. The Burgers transformation is shown schematically in Figure \ref{fig:Burgers}. We note that the Burgers path maintains the symmetry of a primitive monoclinic unit cell (space group $mP$).

\begin{center}
\begin{table}[hbtp!] 
\setlength{\tabcolsep}{8pt}
\begin{tabular}{ |l|r|r|r|r| } 
 \hline
 Parameter                          & hcp                   & fcc           & bcc(Bain)     & bcc(Burgers)\\ 
 \hline 
$\alpha$                            & $0$                   & $1$           & $1$           & $\frac{1}{3}$ \\
$\gamma_1$                          & $1$                   & $1$           & $1$           & 1 \\
$\gamma_2$                          & $\sqrt{\frac{8}{3}}$  & $\sqrt{2}$    & $1$           & $\sqrt{\frac{8}{3}}$ \\
$\beta_1$                           & $1$                   & $1$           & $1$           & $\frac{4}{3}$\\
$\beta_2$                           & $\frac{1}{\sqrt{3}}$  & $1$         & $1$           & $\frac{2\sqrt{2}}{3}$\\
$\beta_3$                           & $\sqrt{\frac{8}{3}}$  & $\sqrt{2}$    & $1$           & $\sqrt{\frac{8}{3}}$ \\
$\angle(\vec{b}_{1},\vec{b}_{2})$   & 60$^\circ$            & 90$^\circ$    & 90$^\circ$    & arcos$\left( \frac{1}{3}\right)$ \\
$\omega_1(\alpha,\gamma_1)$         & $\frac{1}{2}$         & 0             & 0             & $\frac{1}{3}$\\
$\omega_2(\alpha,\gamma_1)$         & 1                     & 1             & 1             & $1$\\
$\omega_3(\alpha,\gamma_1)$         & $\frac{\sqrt{3}}{2}$  & 1             & 1             & $\frac{2\sqrt{2}}{3}$ \\
$V$                                 & $\sqrt{2}$            & $\sqrt{2}$    & $\frac{8}{3\sqrt{3}}$& $\frac{8}{3\sqrt{3}}$ \\
$\rho$                              & $\frac{\pi}{3\sqrt{2}}$& $\frac{\pi}{3\sqrt{2}}$& $\frac{\pi\sqrt{3}}{8}$& $\frac{\pi\sqrt{3}}{8}$ \\
\hline
\end{tabular}
\caption{Ideal lattice parameters using for the three lattices hcp, fcc and bcc with lattice constant $a$, $b$ and $c$ as shown in Figure \ref{fig:hcp}. The values for $\omega_1$, $\omega_2$, $\omega_3$ and the hard-sphere volumes $V$ for hcp and fcc and for bcc (setting $a=1$) and corresponding packing densities are also shown. The values for the bcc structure occurring in the vicinity of the Burgers path are also shown using Model 2.}
\label{tab:latticeparameters}
\end{table}
\end{center}

The specific choice for the $\alpha$-dependence of the basis vector $\vec{b}_2$ in \eqref{b2vector} originates directly from the linear combination between the two $B$-matrices given in eq.\eqref{linearcombination}. In the following we will call this model for the Burgers transformation \textit{Model 1}. As it turns out, the minimum energy path for the Burgers transformation within Model 1 fulfills the condition $\gamma_1f_S=1$, that is for the whole path with $\alpha\in[0,1]$ we have the condition that $|\vec{b}_1|=|\vec{b}_2|$ for the two basis vectors of the hexagonal sheet. This saves one parameter ($\gamma_1$) in our optimization procedure.
Alternatively, we could adopt from the beginning on the condition that $|\vec{b}_1|=|\vec{b}_2|$ for $\gamma_1=1$. This implies that 
\begin{equation}\label{eq:omega12cond}
\omega_3=\sqrt{1-\omega_1^2}=\frac{1}{2}\sqrt{4-(1-\alpha)^2}
\end{equation}
and $\omega_2=1$, keeping the original definition for $\omega_1$. This changes the $B$-matrix to
\begin{align}\label{secondchoice}
B(a,\alpha,\gamma_2)&=a(\vec{b}_1,\vec{b}_2,\vec{b}_3)^\top=a 
\begin{pmatrix}
1  & 0  &  0 \\
\frac{1}{2}(1-\alpha) &\frac{1}{2}\sqrt{(1+\alpha)(3-\alpha)} &  0 \\
0 & 0  &  \gamma_2
\end{pmatrix}
\end{align}
We call this choice \textit{Model 2}. In this model the angle $\angle(\vec{b}_{1},\vec{b}_{2})= \theta_{12}$ is easily obtained from the parameter $\alpha$,
\begin{equation}\label{eq:angle1}
\cos\theta_{12}=\frac{\omega_1}{\sqrt{\omega_2}}=\omega_{1}=\frac{1}{2}(1-\alpha).
\end{equation}
and we have a nice linear relationship as shown in Figure \ref{fig:angle}. We also give the inverse $B$-matrix for model 2 required to obtain the fractional coordinates through $\vec{u}=B^{-1}\vec{v}_s$,
\begin{align}\label{secondchoice}
B^{-1}(a,\alpha,\gamma_2)&=a^{-1} 
\begin{pmatrix}
1  & \frac{(\alpha-1)}{\sqrt{(1+\alpha)(3-\alpha)}}  &  0 \\
0 &\frac{2}{\sqrt{(1+\alpha)(3-\alpha)}} &  0 \\
0 & 0  &  \gamma_2^{-1}
\end{pmatrix}
\end{align}

As we must get the same cohesive energies at a specific angle $\theta_{12}$, we can use eq.\eqref{eq:angle1} to derive the relationship between the $\alpha$-values in the two different models, i.e.
\begin{equation}\label{eq:transform1}
\alpha_\text{M2}=1-\frac{1-\alpha_\text{M1}}{\sqrt{1-2C\alpha_\text{M1}(1-\alpha_\text{M1})}}
\end{equation}
with $C=(1-\frac{\sqrt{3}}{2})$. The back-transformation requires the solution of a quadratic equation for which we get the following expression (taking the negative sign for the square-root expression in the quadratic equation),
\begin{equation}\label{eq:transform2}
\alpha_\text{M1}=\frac{1-C(1-\alpha_\text{M2})^2 - \sqrt{(1-\alpha_\text{M2})^2(1+C(1-\alpha_\text{M2})^2(C-2))}}{1-2C(1-\alpha_\text{M2})^2}
\end{equation}
Substituting $\alpha_\text{M1}$ by $\alpha_\text{M2}$ in \eqref{linearcombination} changes the Burgers transformation into a non-linear transformation of the $B$-matrices. However, for the reduction of the parameter space the $B$ matrix \eqref{secondchoice} is quite useful as $\gamma_1=1$ if $|\vec{b}_{1}|=|\vec{b}_{2}|$ and any symmetry breaking through $|\vec{b}_1|\ne|\vec{b}_2|$ will be described directly by $\gamma_1\ne 1$. It is clear that $\alpha_\text{M1}=\alpha_\text{M2}$ for $\alpha_\text{M1}=0$ and  $\alpha_\text{M1}=1$. The maximum deviation between the two $\alpha$ values is $\alpha_\text{M1}-\alpha_\text{M2}=0.02078464$ with $\alpha_\text{M1}=0.33846180$ and $\alpha_\text{M2}=0.31767716$ in the interval $\alpha_\text{M1}\in[0,1]$. Note that the bcc structure found at $\alpha_{\text{M2}}=1/3$ (see Table \ref{tab:latticeparameters}) corresponds to $\alpha_{\text{M1}}=0.35411421\dots$, highlighting the advantage of Model 2 locating the bcc structure at a rational value of the parameter $\alpha$.

The volume of the $\alpha$-dependent unit cell then is,
\begin{align}\label{eq:vol1}
V(a,\alpha,\gamma_1,\gamma_2)=\sqrt{\text{det}G(a,\alpha,\gamma_1,\gamma_2) }= a^3\gamma_{2}\omega_{3}(\alpha,\gamma_{1})
\end{align}
This results in a packing density compared to spheres of radius $r$,
\begin{equation}
\rho=\frac{V_{\textrm{sphere}}}{V_{\textrm{hex}}}=\frac{4\pi r^3}{3a^3\gamma_{2}\omega_3(\alpha,\gamma_{1})}
\label{eq:vol2} 
\end{equation} 
where we considered that we have two spheres in the unit cell. With the values listed in Table \ref{tab:latticeparameters} this gives $\rho_{\text{bcc},\alpha=1}=\frac{\sqrt{3}}{8}\pi$ (with $r=\frac{|\vec{v}_c|}{2}=\frac{\sqrt{3}}{4}a$), 
$\rho_\text{fcc}=\frac{\sqrt{2}}{6}\pi$ ($r=\frac{1}{2}a$), 
and $\rho_\text{hcp}=\frac{\sqrt{2}}{6}\pi$ ($r=\frac{1}{2}a$).

\begin{figure}[htb!]
\centering
\includegraphics[width=.66\linewidth]{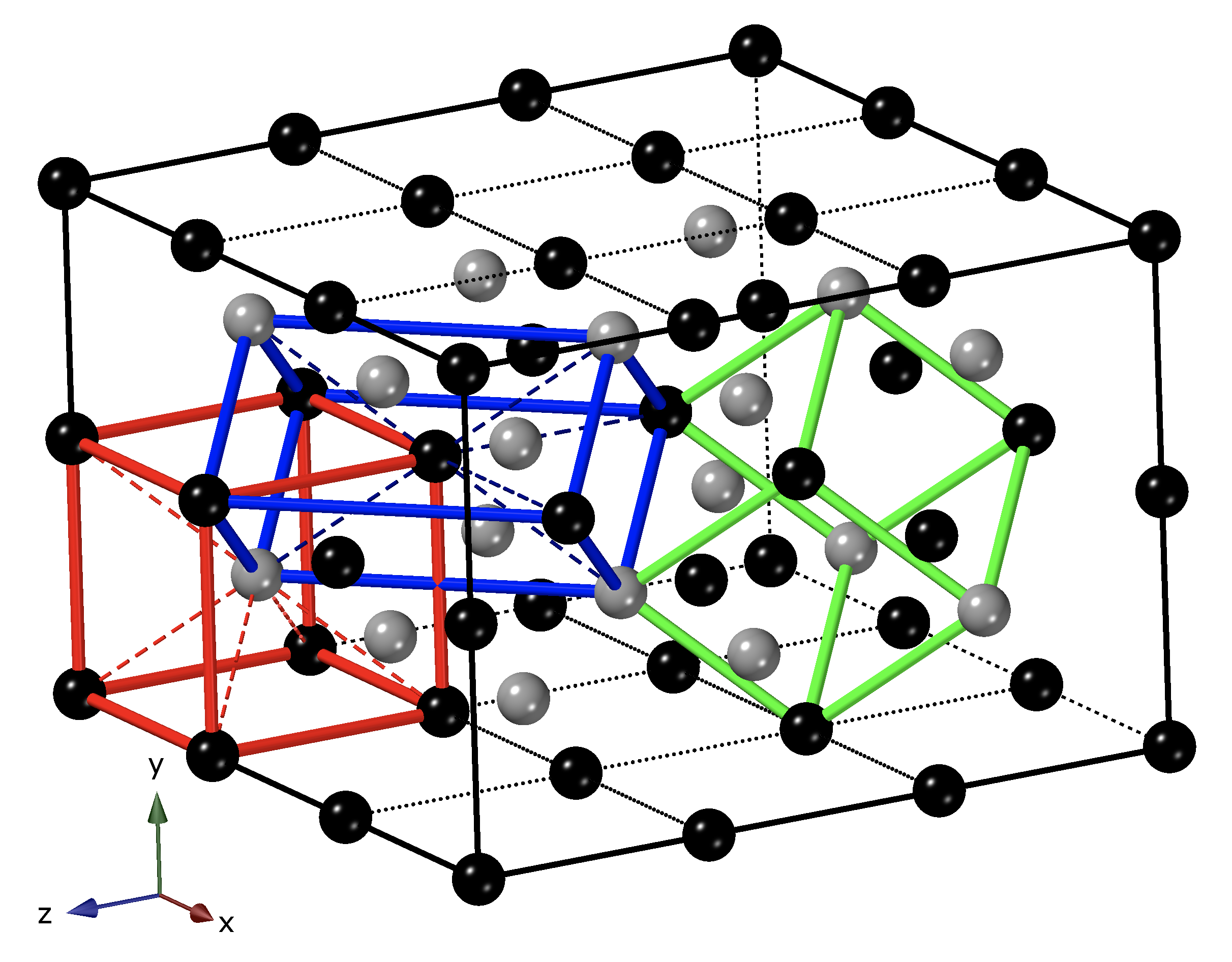}
\caption{bcc phase shown in a $(3\times 2\times 3)$ super-cell. Left: the conventional bcc bi-lattice shown in red color (left) with the body-centered atom in grey; middle: the bi-lattice shown in blue with unit cell identical to the one defined in Fig.\ref{fig:hcp} with $(\angle(\vec{b}_{1},\vec{b}_{2})=70.5287794^\text{o}=180^\text{o}-\theta_T)$, where $\theta_T$ is the tetrahedral angle; right: the rhombohedral primitive bcc cell of equal lengths shown in green (containing no central atom), with a ratio for the length of the lattice vectors compared to the conventional unit cell of $\frac{\sqrt{3}}{2}$. The angles between the lattice vectors are $180^\text{o}-\theta_T$. Note that the primitive bcc cell has half the volume of the conventional one.}
\label{fig:StellOct}
\end{figure}
Table \ref{tab:latticeparameters} shows that at specific lattice parameters at $\alpha=\frac{1}{3}$ (Model 2) an additional bcc structure appears along the Burgers path situated in-between the hcp ($\alpha=0$) and fcc ($\alpha=1$) structures. The corresponding lattice vectors are $\vec{b}_1^\top = a(1,0,0), ~\vec{b}_2^\top = a(\frac{1}{3},\frac{2\sqrt{2}}{3},0)$, and $\vec{b}_3^\top = a(0,0,\sqrt{\frac{8}{3}})$. Notice that the angle $\angle(\vec{b}_{1},\vec{b}_{2})=\arccos\left( \frac{1}{3}\right)=180^\text{o}-\theta_T$, where $\theta_T=109.4712206\dots$ is the tetrahedral angle. The corresponding shift vector is $\vec{v}_s^\top=a ( \frac{2}{3} , \frac{\sqrt{2}}{3}, \sqrt{\frac{2}{3}})$, or expressed in terms of fractional coordinates $u$, i.e. $\vec{u}=B^{-1}\vec{v}_s= ( \frac{1}{2}, \frac{1}{2}, \frac{1}{2})^\top$. Hence, the atom located inside the unit cell is exactly body-centered with a distance from the origin of $\frac{2a}{\sqrt{3}}$. This is just another unit cell definition for the bcc structure as shown (in blue color) in Figure \ref{fig:StellOct}. The primitive rhombohedral cell for bcc (in green color) with lattice parameters $a_1=b_1=c_1=a, \alpha_1=\beta_1=\gamma_1=\theta_T$, and the conventional unit cell (in red color) with lattice parameters $a_2=b_2=c_2=\frac{2a}{\sqrt{3}}, \alpha_2=\beta_2=\gamma_2=90^\text{o}$ are also shown in Figure \ref{fig:StellOct}. In the interval $\alpha\in[0,1]$ we only have two occurrences of a bcc structure, i.e. at $\alpha=\frac{1}{3}$ and $\alpha=1$, where the first one is convenient to discuss distortions along the Burgers path to either hcp or fcc, and the second one for cuboidal distortions to fcc along the Bain path.

We briefly summarize. For the Burgers transformation we reduced the, in principle, nine-parameter space (6 lattice constants for the Bravais lattice plus 3 parameters for the position of the middle layer) for a bi-lattice to a seven-parameter space $\{p_i\}=(a,\alpha,\gamma_1,\gamma_2,\beta_1,\beta_2,\beta_3)$, where $\alpha\in (-1,3)$ is the important structure parameter (our reaction coordinate) describing the Burgers-Bain phase transition. All other parameters $a,\gamma_1,\gamma_2,\beta_1,\beta_2,\beta_3$ have in principle to be optimized along the transformation path. We expect approximately the following parameter range for $\alpha\in[0,1]$: $a\in [a_\text{hcp},a_\text{bcc}]$, $\gamma_{1}\approx 1.0$, $\gamma_{2}\in \left[1,\sqrt{\frac{8}{3}}\right]$, $\beta_{1}\approx 1.0$, $\beta_{2}\in \left[\frac{1}{\sqrt{3}},1\right]$, $\beta_{3}\approx\gamma_{2}$. For the Bain transformation the originally introduced distortion parameter $A$ in Ref. \onlinecite{burrows2025a} is simply related to the parameter $\gamma_{2}$ by $A=\gamma_{2}^2/2$, and similarly the nearest neighbor distance in the cuboidal lattice is given by $r_{NN}=\frac{a}{2}\sqrt{2+\gamma_{2}^{2}}$.\cite{Jerabek2022, RoblesNavarro2023} This implies that at $\alpha=1$ the transformation from fcc to bcc is described mainly through the parameter $\gamma_{2}$. An additional bcc structure appears along the Burgers path at $\alpha=\frac{1}{3}$, which will be discussed in detail for the case of a LJ potential.

\subsection{Quadratic forms and functions for the lattice sums}
\label{sec:LatticeSums}

Lattice sums have a long history and are usually based on quadratic functions. For inverse power potentials $V(r)=r^{-2s}$ such as the Lennard-Jones potential or the Madelung constant for Coulomb potentials, these are related to the so-called Epstein zeta function.\cite{borwein-2013,madelung1918,Jones-1925} For the cuboidal Bain transformation the associated lattice sums were already described in our previous papers \cite{burrows2021b,burrows2025a}. For the more general Burgers-Bain transformation we need to introduce two lattice sums for all the distances in a bi-lattice. 

We first introduce the quadratic form $Q_{A}$ involving only atoms in the base A-layer,
\begin{equation}
\label{eq:QA}
\begin{aligned}
    Q_{A}(\vec{i},\alpha,\gamma_1,\gamma_2) &=\vec{i}^\top S_3(\alpha,\gamma_1,\gamma_2)\vec{i}=a^{-2}\vec{i}^\top G(a,\alpha,\gamma_1,\gamma_2) \vec{i} \\
    &= i_1^{2}+2\omega_1(\alpha,\gamma_{1})i_1i_2+\omega_2(\alpha,\gamma_{1}) i_2^{2}+\gamma_{2}^{2} i_3^{2}
\end{aligned}
\end{equation}
with the associated lattice sum
\begin{equation}
H^h_A(s,\alpha,\gamma_1,\gamma_2)={\sum_{\vec{i}\in\mathbb{Z}^3}}^{'} \frac{1}{(\vec{i}^\top S_3(\alpha,\gamma_1,\gamma_2)\vec{i})^s}
\label{eq:HA}
\end{equation}
The prime notation at the sum indicates that we avoid the term $i_1=i_2=i_3$ in the summation. This lattice sum can be re-expressed in terms of a fast converging Bessel functions \cite{terras-1973,burrows-2020} described in detail in Appendix \ref{Besselexpansions}.

For the second lattice sum we need the quadratic function for the distances between the A- and B-layer lattice points derived from eq.\eqref{distvec},
	\begin{align}\label{eq:QB}
		Q_{B}&(\vec{i},\{ p_i\})= 
        a^{-2} \left( \vec{i}^\top B(a,\alpha,\gamma_1,\gamma_2)  + \vec{v}_s^\top(\beta_1,\beta_2,\gamma_{2})\right) \left(B(a,\alpha,\gamma_1,\gamma_2)^\top \vec{i} + \vec{v}_s(\beta_1,\beta_2,\gamma_{2})\right) \\
    \nonumber
		&= a^{-2} \left[  \vec{i}^\top G(a,\alpha,\gamma_1,\gamma_2)\vec{i} + 2 \vec{i}^\top B(a,\alpha,\gamma_1,\gamma_2) \vec{v}_s(\beta_1,\beta_2,\gamma_{2}) + |\vec{v}_s(\beta_1,\beta_2,\gamma_{2})|^2\right] \\
		&=\left(i_1 + \omega_1(\alpha,\gamma_{1})i_{2} + \frac{\beta_{1}}{2}\right)^{2} + \left(\omega_{3}(\alpha,\gamma_{1})i_{2}+\frac{\beta_{2}}{2}\right)^{2} + \left(\gamma_{2}i_{3}+\frac{\beta_{3}}{2}\right)^2
	\end{align}
where $\omega_1(\alpha,\gamma_{1})$ and $\omega_3(\alpha,\gamma_{1})$ have already been defined. The corresponding lattice sum becomes
\begin{equation}
H^{h}_{B}(s,\{ p_i\})=\sum_{\vec{i}\in\mathbb{Z}^3} \frac{1}{Q_{B}(\vec{i},\alpha,\gamma_1,\gamma_2,\beta_1,\beta_2,\beta_3)^{s}}
\end{equation}
Both lattice sums are absolute convergent for finite $s>\frac{3}{2}$ with a simple pole at $s=\frac{3}{2}$, but can be analytically continued for $s<\frac{3}{2}$. However, for $s\rightarrow\infty$ the lattice sum $H^{h}_{B}\rightarrow\infty$ for the bcc lattice. This can be understood from the fact that some terms in $Q_{B}(\vec{i}$ for $(i_1,i_2,i_3)\in[-1,1]$ can become smaller than 1 in our definition of the lattice sum. This divergence of the lattice sum for $s\rightarrow\infty$ will be compensated by the optimized lattice constant $a$ obtained from a Lennard-Jones potential as discussed further below, but this results in a $(0\times\infty)$ case for $s\rightarrow\infty$. It is therefore far more convenient to avoid such divergences by re-scaling the lattice constant such that $r_\text{bc}=af(\beta_1,\beta_2,\beta_{3})$, and $r_\text{bc}=|\vec{v}_s|$ is the length from the atom at the base layer A to the nearest body-centered atom in layer B. $f(\beta_1,\beta_2,\beta_{3})$ becomes part of the lattice sum assuring convergence to a finite value for $s\rightarrow\infty$. We now redefine the two lattice sums such that they behave well in the limit $s\rightarrow \infty$,
\begin{align}
L^h_A(s,\{ p_i\})=f(\beta_1,\beta_2,\beta_{3})^{2s}H^h_A(s,\alpha,\gamma_1,\gamma_2)
=f(\beta_1,\beta_2,\beta_{3})^{2s}{\sum_{\vec{i}\in\mathbb{Z}^3}}^{'} \frac{1}{(\vec{i}^\top S_3(\alpha,\gamma_1,\gamma_2)\vec{i})^s}
\end{align}
and
\begin{align}
L^{h}_{B}&(s,\{ p_i\})=f(\beta_1,\beta_2,\beta_{3})^{2s} H^{h}_{B}(s,\{ p_i\})
=f(\beta_1,\beta_2,\beta_{3})^{2s}\sum_{\vec{i}\in\mathbb{Z}^3} \frac{1}{Q_{B}(\vec{i},\{ p_i\})^{s}}
\end{align}
with
\begin{equation}\label{eq:factor}
f(\beta_1,\beta_2,\beta_{3})=\frac{r_\text{bc}}{a}=\frac{1}{2}\sqrt{\beta_{1}^{2}+\beta_{2}^{2}+\beta_{3}^{2}}
\end{equation}
This ensures that the bcc lattice sum is identical to the one defined originally by Lennard-Jones and Ingham \cite{Jones-1925}. The total lattice sum is then given by
\begin{align}
\label{eq:TLS}
L^h(s,\{ p_i\})=L^h_A(s,\{ p_i\})+L^h_B(s,\{ p_i\})\\
H^h(s,\{ p_i\})=H^h_A(s,\{ p_i\})+H^h_B(s,\{ p_i\})
\end{align}
Both lattice sums can be expressed in terms of fast converging series involving Bessel functions as derived in Appendix \ref{Besselexpansions}. We note that it is sufficient to take just the sum of the two layer contributions, e.g.  $L^h_A+L^h_B$, as in our definition of the bi-lattice each atom is equivalent to all the others, i.e. they have the same surrounding. This can also be seen from a more general formula for Barlow packings.\cite{Schwerdtfeger-2024b} 

\subsection{Lattice sums for the special cases of hcp, fcc and bcc} 

Using the parameters in Table \ref{tab:latticeparameters} for the hcp structure we get $f(\beta_1,\beta_2,\beta_3)=1$, and for the lattice sums
\begin{align}
L^\text{hcp}_A &= {\sum_{\vec{i}\in\mathbb{Z}^3}}^{'} \left( i_1^{2}+i_1i_2+ i_2^{2}+\tfrac{8}{3} i_3^{2} \right)^{-s} \\
\nonumber
L^\text{hcp}_B &= \sum_{\vec{i}\in\mathbb{Z}^3} \left( (i_1+\tfrac{1}{2}i_2+\tfrac{1}{2})^2 + (\tfrac{\sqrt{3}}{2}i_2+\tfrac{1}{2\sqrt{3}})^2 + \tfrac{8}{3} (i_3+\tfrac{1}{2})^2 \right)^{-s}\\
\nonumber
&= \sum_{\vec{i}\in\mathbb{Z}^3}  \left( (i_1 +\tfrac{1}{3})^2+(i_2 +\tfrac{1}{3})^2+(i_1 +\tfrac{1}{3})(i_2 +\tfrac{1}{3})+ \tfrac{8}{3} (i_3+\tfrac{1}{2})^2 \right)^{-s}
\end{align}
These are exactly the lattice sums as shown in Refs. \onlinecite{bell1976rare,burrows-2020,Cooper2023}. For bcc we get $f(\beta_{1},\beta_{2},\beta_{3})=\sqrt{3}/2$ and
\begin{align}
L^\text{bcc}_A &= \left( \frac{\sqrt{3}}{2} \right)^{2s} {\sum_{\vec{i}\in\mathbb{Z}^3}}^{'} \left( i_1^{2}+ i_2^{2}+i_3^{2} \right)^{-s} \\
\nonumber
L^\text{bcc}_B &=  3^s \sum_{\vec{i}\in\mathbb{Z}^3} \left((2i_1+1)^2 + (2i_2+1)^2 +  (2i_3+1)^2 \right)^{-s}
\end{align}
These are identical to the ones given by Lennard-Jones and Ingham \cite{Jones-1925} and used in Ref. \onlinecite{burrows-2020} for various Bessel function expansions. 

For fcc we get $f(\beta_{1},\beta_{2},\beta_{3})=1$ and
\begin{align} \label{eq:LJIfcc}
L^\text{fcc}_A &= {\sum_{\vec{i}\in\mathbb{Z}^3}}^{'} \left( i_1^{2}+ i_2^{2}+2i_3^{2} \right)^{-s} \\
\nonumber
L^\text{fcc}_B &=  \sum_{\vec{i}\in\mathbb{Z}^3} \left((i_1+\tfrac{1}{2})^2 + (i_2+\tfrac{1}{2})^2 +  2(i_3+\tfrac{1}{2})^2 \right)^{-s}
\end{align}
This can be brought into a more familiar form discussed in Ref. \onlinecite{burrows-2020} (see Appendix \ref{fcclattice}),
\begin{equation} \label{eq:LJIfcc1}
L^\text{fcc}_A + L^\text{fcc}_B = 2^{s-1}{\sum_{\vec{i}\in\mathbb{Z}^3}}^{'} 
\left[ 1+(-1)^{i_1+i_2+i_3}\right] \left( i_1^2 + i_2^2 + i_3^2 \right)^{-s} \\
\end{equation}

\subsection{The Lennard-Jones cohesive energies for the lattices along the fcc to hcp transition path} 

The cohesive energy for the hexagonal-cuboidal structures for a general $(n,m)$ LJ potential can be expressed in terms of lattice sums and is given by the expression\cite{Jones-1925,Schwerdtfeger-2006,Smits2021}
\begin{align} \label{eq:cohLJhcp}
E_{\text{LJ}}^{\text{coh}}(n,m,\{ p_i\})&= \frac{\epsilon nm}{2(n-m)} \; \left[ \frac{H^h(\frac{n}{2},\{ p_i\})}{n}\left( \frac{r_e}{a} \right)^{n} - \frac{H^h(\frac{m}{2},\{ p_i\})}{m} \left( \frac{r_e}{a} \right)^{m} \right]\\
&= \frac{\epsilon nm}{2(n-m)} \; \left[ \frac{L^h(\frac{n}{2},\{ p_i\})}{n}\left( \frac{r_e}{r_\text{bc}} \right)^{n} - \frac{L^h(\frac{m}{2},\{ p_i\})}{m} \left( \frac{r_e}{r_\text{bc}} \right)^{m} \right] 
\nonumber
\end{align}
where the distance from the base to the body-centered atom is related to the lattice constant $r_\text{bc}= af(\beta_1,\beta_2,\beta_{3})$ as already mentioned. Note that the factor $f(\beta_1,\beta_2,\beta_3)$ in eq.\eqref{eq:factor} cancels out when we with the lattice sums $H\rightarrow L$ and the distances $r_\text{bc}\rightarrow a$ at the same time. The expression for the lattice sum $L^h(s,\{ p_i\})$ is taken from Eq. \eqref{eq:TLS} and the corresponding Bessel function expansions we used in our work are given in Appendix \ref{Besselexpansions}. As these Bessel sum expansions are fast converging series, they can easily be obtained to arbitrary computer precision, i.e. to double precision accuracy within a few seconds of computer time.

In order to discuss the behavior for the LJ potential with varying parameters $p_i$ we treat the lattice constant $a$ independently from all the other parameters to be optimized. For this, we follow the procedure in Ref. \onlinecite{burrows2021b} and get from $\partial E_{\textrm{LJ}}^{\textrm{coh}}/ \partial a = 0$ the minimum lattice parameter,
\begin{equation} \label{eq:RLJnmmin}
 a^*_{\text{min}}=\frac{a_{\text{min}}(\alpha,\beta_1,\beta_2,\beta_3,\gamma_1,\gamma_2)}{r_e}=\frac{1}{f(\beta_1,\beta_2,\beta_{3})}\left( \frac{L^h(\frac{n}{2},\alpha,\beta_1,\beta_2,\beta_3,\gamma_1,\gamma_2)}{L^h(\frac{m}{2},\alpha,\beta_1,\beta_2,\beta_3,\gamma_1,\gamma_2)}\right)^{\tfrac{1}{n-m}} \,,
\end{equation}
and the * indicates that reduced (or dimensionless) units are used. We can then evaluate the cohesive energy at $ a^*_{\text{min}}$ and get
\begin{equation} \label{eq:ELJnmmin}
E_{nm}^*=E_{\text{LJ}}^{\text{coh}}(n,m,a^*_\text{min},\alpha,\beta_1,\beta_2,\beta_3,\gamma_1,\gamma_2)/\epsilon=-\frac{1}{2}\left[ \frac{L^h(\frac{m}{2},\alpha,\beta_1,\beta_2,\beta_3,\gamma_1,\gamma_2)^n}{L^h(\frac{n}{2},\alpha,\beta_1,\beta_2,\beta_3,\gamma_1,\gamma_2)^m}\right]^{\tfrac{1}{n-m}}
\end{equation}
We therefore have to deal only with the six-parameter space $\alpha,\beta_1,\beta_2,\beta_3,\gamma_1,\gamma_2$ for fixed exponents $(n,m)$. As $\alpha$ is fixed to map out the Burgers path, we only have to optimize $E_{nm}^*$ with respect to both sets $\{\beta_1,\beta_2,\beta_3\}$ and $\{\gamma_1,\gamma_2\}$ using a 5D Newton-Raphson procedure as outlined in the supporting information. We chose the interval $\alpha\in[-0.2,1.2]$ in steps of $\Delta\alpha=0.05$ for the Burgers path. For the following we omit the * notation, i.e. all quantities are in reduced (dimensionless) units unless otherwise stated. 

The validity of using \eqref{eq:ELJnmmin} instead of \eqref{eq:cohLJhcp} has been checked by computations which show that we obtain exactly the same results for the extreme points (maxima and minima) and the minimum energy Burgers path. This is perhaps not surprising as we have $\partial E_{nm} / \partial a=0$ and $\partial E_{nm} / \partial \gamma_1 = \partial E_{nm} / \partial \gamma_2 = 0$ for the minimum energy path. Thus, by performing the derivative $\partial E_{nm} / \partial a=0$ and considering $\partial L^h(\frac{n}{2},\alpha,\beta_1,\beta_2,\beta_3,\gamma_1,\gamma_2) / \partial a=0$ in \eqref{eq:cohLJhcp} we can easily prove by substituting $\partial a = -a \gamma_i^{-1} \partial \gamma_i$ that \eqref{eq:ELJnmmin} is valid. For the area around the Burgers path where this condition does not exactly hold, but we obtain rather small deviations from the exact solution \eqref{eq:cohLJhcp}. Here, the situation is similar to the hcp case as discussed before.\cite{Cooper2023}

\subsection{The Burgers-Bain transformation within the kissing hard-sphere model}
\label{sec:KHS}
For large exponents, the $(n,m)$-LJ potential approaches the kissing hard-sphere limit (KHS) as defined in eq.\eqref{eq:KHS}. Within the KHS limit we optimize the number of contacts between the spheres (atoms). It is therefore illustrative to discuss the Burgers/Bain type phase transition within the KHS limit first.

\begin{figure}[htpb!]
\centering
\includegraphics[width=\textwidth]{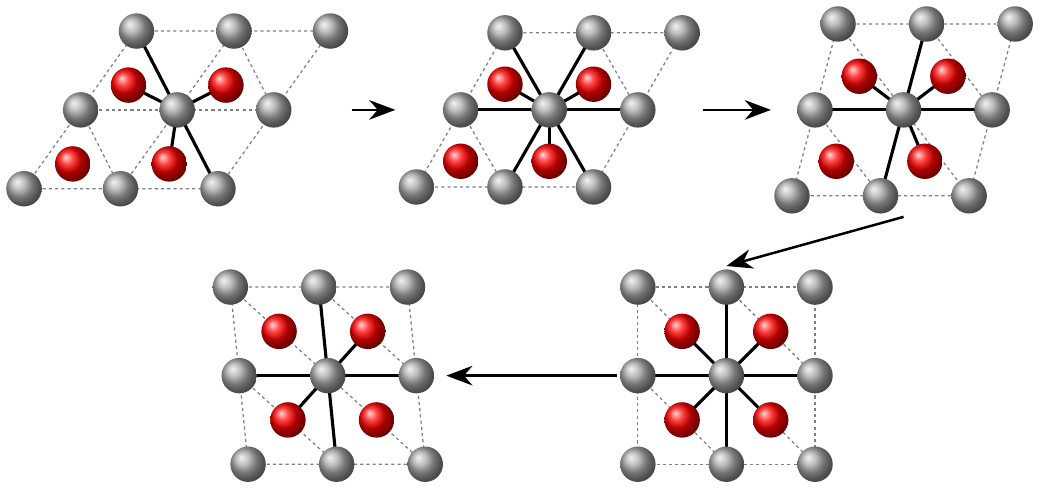}
    \caption{Burgers transformation using a kissing hard-sphere model potential along different values of $\alpha$ from  $-0.2\rightarrow0.0\rightarrow0.5\rightarrow1.0\rightarrow1.2$. The continuous black lines show the bonds from the atom at the origin to its nearest neighbors (kissing spheres). The red atoms lie in the Wyckoff positions of the unit cell and form layers above and below (B layer) the layer of gray atoms (A layer). Including the bonds for the lower B layer we count the number of solid lines and get the kissing numbers as shown in \eqref{eq:KHSE}. }
\label{fig:KHSBurgers}
\end{figure}

For the Burgers transformation the KHS cohesive energy, $E_\text{KHS}$, and kissing numbers $\kappa$ are given as
\begin{align}
\label{eq:KHSE}
E_\text{KHS}
&= \begin{cases}
-4 \quad \text{for}\quad  \alpha<0 \quad \text{and} \quad\alpha>1, \quad \kappa=8 \\
-6  \quad \text{for} \quad \alpha=0 \quad \text{and} \quad \alpha=1, \quad \kappa=12\\
-5  \quad \text{for} \quad 0<\alpha<1, \quad \kappa=10.
\end{cases}
\end{align}
with $\alpha=\alpha_\text{M1}$ or $\alpha=\alpha_\text{M2}$. $E_\text{KHS}$ is now a step function, i.e. we do not have a single point maximum for the energy anymore. In this Burgers transformation, the central atom in the middle layer is located at the bisecting line of the two lattice vectors $\vec{b}_1$ and $\vec{b}_2$ of length $a=1$ (half unit sphere of radius $R=\frac{1}{2})$, i.e. the atom moves up- or downwards at half-distance ($\frac{1}{2}|\vec{b}_1|=\frac{a}{2}$) along the Burgers path. This is illustrated in Figure \ref{fig:KHSBurgers}. The parameters change accordingly along the path and we have the boundary conditions $\beta_1=1$, $\gamma_1f_S=1$, $\beta_{3}=\gamma_{2}$ and $r_\text{bc}=1$.

We may naively assume a linear behavior for $\gamma_2(\alpha_\text{M1})$ and $\beta_2(\alpha_\text{M1})$, but this does not correctly describe the movement of the hard sphere in the second layer while the angle between $\vec{b}_1$ and $\vec{b}_2$ changes. We therefore need to derive analytical formulae for both $\gamma_2(\alpha_\text{M1})$ and $\beta_2(\alpha_\text{M1})$. For this we conveniently choose Model 2 as $\alpha_\text{M2}$ has a simple relationship with the angle between the two vectors $\vec{b}_1$ and $\vec{b}_2$ (cf. eq.\eqref{eq:angle1}).

\begin{figure}[htpb!]
\centering
\includegraphics[scale=1]{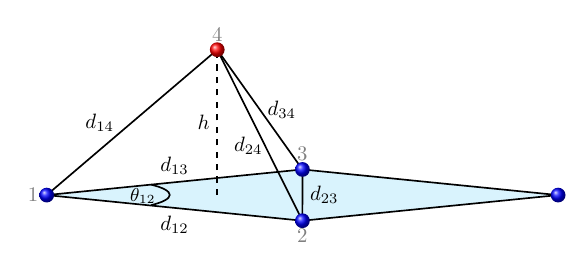}
    \caption{The movement of the second layer atom (red color) with respect to the base layer in the Burgers transformation with the parameters for the KHS limit $d_{12}=d_{13}=d_{14}=d_{24}=d_{34}=1$, $d_{23}=2\text{sin}\frac{\theta_{12}}{2}$ with $\theta_{12}\in[60,90]$ degrees, height $h=\frac{\gamma_2}{2}$ and $s=\frac{\beta_2}{2}$ described by an irregular tetrahedron. Note, the kissing number of each sphere is 10.}
\label{fig:pyramid}
\end{figure}
In order to derive the exact movement of a hard sphere of radius $a/2$ within the second (middle) layer relative to the layers above and below with changing base angle $\theta_{12}$ from 60 to 90 degrees, we use a well known relation for the volume $V_T$ of an irregular tetrahedron,
\begin{equation}
V_T=\frac{1}{3}A_Th = \frac{1}{12\sqrt{2}}\sqrt{D_\text{CM}}
\end{equation}
where $A_T=\frac{1}{2}\sin{\theta_{12}}$ is the area of the base triangle between points 1, 2 and 3 as shown in Figure \ref{fig:pyramid} and $D_\text{CM}$ is the Cayley-Menger determinant for an irregular polyhedron.\cite{Sommerville2020} The lengths of the edges for the hard-sphere model are fixed at $d_{12}=d_{13}=a=1$, $d_{14}=r_\text{bc}=1$, $d_{24}=d_{34}=1$, and we have $d_{23}=2\text{sin}\frac{\theta_{12}}{2}$. As $h=\gamma_2/2$ we obtain
\begin{equation}
\gamma_2=\frac{1}{2\sqrt{2}}\frac{\sqrt{D_\text{CM}}}{A}
\end{equation}
The $(5\times 5)$ Cayley-Menger determinant can easily be evaluated from the distance conditions above, and we have $D_\text{CM}=2d_{23}^2(3-d_{23}^2)$. After applying some basic trigonometry, this leads to
\begin{equation}
\gamma_2=\frac{d_{23}\sqrt{3-d_{23}^2}}{2A}=
\frac{2\sin{\frac{\theta_{12}}{2}} \sqrt{3-4\sin{\frac{\theta_{12}^2}{2}}}}
{\sin{\theta_{12}}}
=\sqrt{\frac{2+2\cos{\theta_{12}}(1-2\cos{\theta_{12}})}{1-\cos^2{\theta_{12}}}}
\end{equation}
From eq.\eqref{eq:angle} we finally get
\begin{equation}
\gamma_2(\alpha_\text{M2})=\sqrt{\frac{8+4\alpha_\text{M2}(1-\alpha_\text{M2})}{4-(1-\alpha_\text{M2})^2}}=2\sqrt{\frac{2-\alpha_\text{M2}}{3-\alpha_\text{M2}}}
\end{equation}
$\gamma_2$ is a monotonically decreasing function in the interval $\alpha_\text{M2}\in[0,1]$. Moreover, $\gamma_2$ changes from $\gamma_2=\sqrt{\frac{8}{3}}$ at $\alpha_\text{M2}=0$ to $\gamma_2=\sqrt{2}$ at $\alpha_\text{M2}=1$. 
In other words, the KHS model directly connects the hcp with the fcc structure (in our specific unit cell treatment), unlike the original proposed transformation by Burgers which is between hcp and bcc.\cite{Burgers1934} 

Now that we have an expression for $\gamma_2$, we can give an analytical formula for our unit cell volume along the Burgers path \eqref{eq:vol} for the KHS limit,
\begin{equation}
V(\alpha_\text{M2})=\gamma_2\omega_3=\sqrt{2+\alpha_\text{M2}(1-\alpha_\text{M2})}
\end{equation}
which is an astonishingly simple function in $\alpha_\text{M2}$. This function has a maximum at $\alpha_\text{M2}=\frac{1}{2}$ with a volume of exactly $V_\text{max}=\frac{3}{2}$. From Table \ref{tab:latticeparameters} we see that $V_\text{bcc}>V_\text{max}$. The corresponding packing density becomes
\begin{equation}
\rho(\alpha_\text{M2})=\frac{\pi}{3\sqrt{2+\alpha_\text{M2}(1-\alpha_\text{M2})}}
\end{equation}
which at maximum volume becomes $\rho(V=\frac{3}{2})=2\pi/9$. The ratio to the hard-sphere packing at fcc minimum is therefore  $\rho(V=\frac{3}{2})/\rho(V=\sqrt{2})=\frac{2\sqrt{2}}{3}=0.942809...$. This value is significantly above the ratio near melting ($\rho=0.739$) given by Mau and Huse.\cite{Mau1999}

From our expression of $\gamma_2$, we can easily derive an analytical expression for $\beta_2$ for the hard-sphere model using simple geometry,
\begin{equation}\label{eq:KHSbeta2}
\beta_2(\alpha_\text{M2})=\sqrt{\frac{4-(1-\alpha_\text{M2})(\alpha_\text{M2}+3)}{4-(1-\alpha_\text{M2})^2}} = \sqrt{\frac{1+\alpha_{\text{M2}}}{3-\alpha_{\text{M2}}}}
\end{equation}
$\beta_2$ is a monotonically increasing smooth function in $\alpha_\text{M2}\in[0,1]$. We will see later on that this implies no symmetry breaking effects within this model.

We need to mention the bcc structure which can appear on the Burgers path at $\alpha=\frac{1}{3}$ at the right conditions for the lattice parameters. Within the KHS model at $\alpha=\frac{1}{3}$ we get $\gamma_2=\sqrt{\frac{5}{2}}=1.5811388\dots$, $\beta_1=1$ and $\beta_2=\frac{1}{\sqrt{2}}=0.70710678\dots$. The $\gamma_2$ value is perhaps closest to the one associated with the bcc structure ($\gamma_2=1.632993$ in our unit cell definition, but both the KHS values for $\beta_1$ and $\beta_2$ deviate substantially from the ones defined by the bcc structure, see Table \ref{tab:latticeparameters}). This clearly shows that the KHS model bypasses the bcc structure along the Burgers path, a general feature we also expect for harder LJ potentials close to the KHS limit. This is comes at no surprise as the kissing number is constant at $\kappa=10$ for the Burgers path in the interval $\alpha\in(0,1)$ compared to the bcc structure, where $\kappa$ is reduced to 8.

As we will discuss also the fcc$\rightarrow$bcc transformation in section \ref{SymmBreak2} further down, 
we mention the Bain phase transition as a special case of our Burgers-Bain transformation by setting $\alpha=1$ and varying $\gamma_2\in[1,\sqrt{2}]$. In this transformation the 8 atoms at the edges of the cuboidal cell move slightly out as the unit cell gets compressed from fcc to bcc and we have $a>1$. The kissing number changes from 12 to 8, i.e. the cuboidal structure for $\gamma_2<\sqrt{2}$ lies energetically always above fcc or hcp and even above the hard-sphere Burgers transition path. As we have $a>r_\text{bc}=1$ for $\gamma_2<\sqrt{2}$ and $\alpha=1$ within the KHS model, we can easily derive that for the lattice constant $a$ we have
\begin{equation}
a(\gamma_2)=\frac{2}{\sqrt{2+\gamma_2^2}}
\end{equation}
As $\omega_3(\alpha=1)=1$ we get for the volume of the Bain path of a cuboidal lattice
\begin{equation}
V_\text{cub}(\gamma_2)=\frac{8\gamma_2}{(2+\gamma_2^2)^\frac{3}{2}}
\end{equation}
that is the volume is monotonically increasing from fcc to bcc within the KHS model. We can easily verify that for the appropriate $\gamma_2$ values we get the results as shown in Table \ref{tab:latticeparameters}. We can now compare the KHS model with the results obtained from $(n,m)$-LJ potentials.

\subsection{Density functional Calculations using the Variable Cell Nudge Elastic Band Method for Solid Argon}
\label{sec:DFT}

Solid argon is reasonably well described by two-body interactions\cite{Schwerdtfeger-2016} such as the (12,6)-LJ potential.\cite{Dobbs_1957,Schwerdtfeger-2006,Smits2021} Hence, in order to verify our lattice sum calculations for the hcp$\rightarrow$fcc phase transition path, we decided to carry out additional density functional theory (DFT) calculations\cite{Giannozzi_2017} using the Perdew–Burke–Ernzerhof (PBE) exchange-correlation functional\cite{PBE1996} together with an ultrasoft pseudopotential for argon\cite{Vanderbilt1990} to replace the argon core, and the PBE functional including dispersion interactions through the DFT-D3(BJ) method\cite{Grimme-2011,Grimme-D3-2010,Goerigk2011}, mapping out the Burgers path by applying the Variable Cell Nudge Elastic Band method (VC-NEB) for solid argon\cite{HANNES1998,Henkelman2000,Henkelman2000a,QIAN2013} using 25 grid points on the transition path between hcp and fcc. For this we used the same unit cell as shown in Figure \ref{fig:hcp} and optimized all variable parameters for the $B$-matrix and the shift vector to obtain the stress tensor for the path optimization. While the PBE functional may not reproduce solid state properties of argon accurately,\cite{Maerzke2009} the PBE-D3 functional performs far better, and for the transition path we assume that the errors inherent in the DFT\cite{Teale2022} or pseudopotential approximation\cite{SchwerdtfegerPP2011} should cancel out for energy differences to some extent. We also included the local density approximation (LDA)\cite{Jones1989} as a limiting case, as it is well known that this functional substantially overestimates the binding energy of Ar$_2$,\cite{Truhlar2006} and therefore of solid argon as well.

\section{Results and Discussions}

\subsection{The hcp$\leftrightarrow$fcc Burgers transformation in a restricted four-parameter space}

For the following we adopt Model 2 for convenience, i.e. $\alpha\equiv\alpha_\text{M2}$ and $\gamma_1=1$. We first consider the smaller four-parameter space $(a,\alpha,\beta_2,\gamma_2)$ by setting $\beta_1=1$ and $\beta_3=\gamma_2$ in agreement with the hard-sphere model. Only a few combinations for the exponentials $(n,m)$ of the LJ potential are selected here, sufficient to discuss the most important features of a generalized LJ potential. Furthermore, we keep the discussion in this section rather short. More details can be found in the supplementary information, and preliminary results for the (12,6)-LJ potential were already published by our group\cite{RoblesNavarro2025}. Furthermore, extension of this small parameter space results in symmetry breaking effects with a bifurcation point occurring including the appearance of an additional minimum for the bcc structure, which will be discussed in detail in sections \ref{SymmBreak} and \ref{SymmBreak2}.

\begin{figure}[htb]
\centering
\includegraphics[width=.66\linewidth]{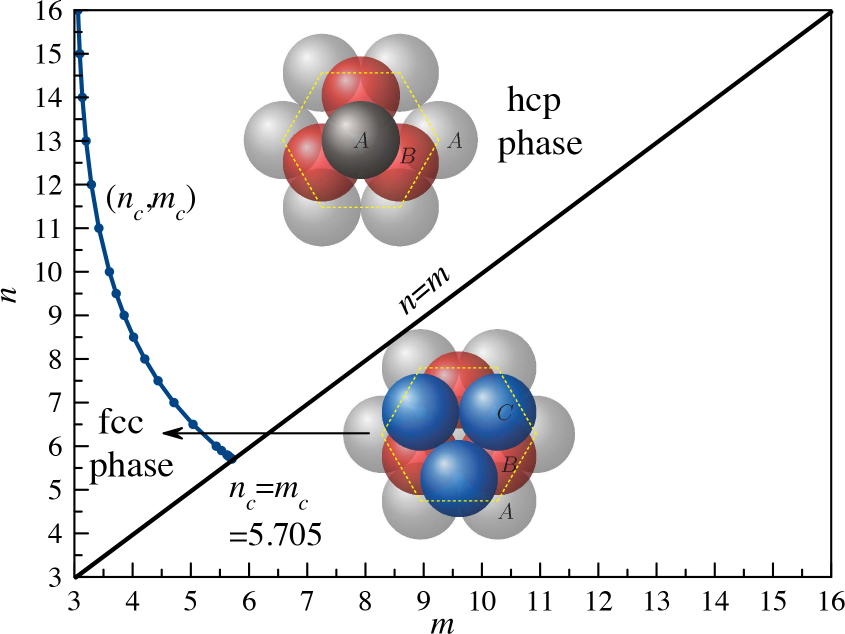}
\caption{hcp/fcc phase transition line $(n_c,m_c)$ for $(n,m)$-LJ potentials fulfilling the condition $\Delta E_{nm}=E_{nm}^\text{hcp}-E_{nm}^\text{fcc}=0$.}
\label{fig:fcchcp}
\end{figure}

The optimized lattice parameters and corresponding energies are summarized in Table \ref{tab:properties}. As can be seen, the energy difference between fcc and hcp is very small for all LJ potentials applied, e.g. $\Delta E_{nm} = E_{nm}^\text{hcp}-E_{nm}^\text{fcc}= 8.703\times$10$^{-4}$ for the (12,6)-LJ potential and $-1.6705\times$10$^{-3}$ for the (6,4)-LJ potential. This is already well documented in the literature.\cite{Kibara1952,Wallace1965,Niebel-1974,Waal1991,LotrichSzalewicz1997a,Krainyukova2011,Schwerdtfeger-2016,Smits2021} Figure \ref{fig:fcchcp} shows the hcp-fcc phase transition line between the two phases for a $(n,m)$-LJ potential where we have $\Delta E_{nm}=0$. Only for very soft long-range potentials becomes the fcc phase more stable compared to hcp. The critical exponents $n_{c},m_{c}$ for the phase transition line can be approximated by the following functional form,
\begin{equation}
    m_{c} = 3 + k_{1}e^{-k_{2}n_{c}} + \frac{c_{1}}{n_{c}} + \frac{c_{2}}{n_{c}^{2}} + \frac{c_{3}}{n_{c}^{3}} \quad \text{with} \quad n_c>5.705
\end{equation}
where $k_{1}=3.322473067\times10^{8}$, $k_{2}=5.95017115$, $c_{1}=-9.23169654$, $c_{2}=1.73880812\times10^{2}$ and $c_{3}=-1.89668849\times10^{2}$ with a coefficient of determination of $R^{2}=0.9998$. For example, for $n_c=12$ we get $m_c=3.291725434064$, which implies that for the $(12,6)$-LJ potential the hcp phase lies energetically below the fcc phase. It was recently shown\cite{Schwerdtfeger-2016} that, for a rare-gas solid like argon, phonon dispersion is required for the stabilization of the fcc over the hcp phase at 0K.

\begin{table}[hbtp!] 
\setlength{\tabcolsep}{4pt}
\begin{tabular}{ |l|r|r|r|r|r|r| } 
 \hline
 Parameter                             & $(6,4)$   & $(8,6)$   & $(12,4)$  & $(12,6)$  & $(30,6)$  & $(30,12)$\\ 
 \hline 
\textit{minima} &&&&&&\\
$a^\text{hcp}(\alpha=0)$               & 0.75523963& 0.94115696& 0.91206706& 0.97127386& 0.99229244& 0.99939864\\
$r_\text{bc}^\text{hcp}(\alpha=0)$     & 0.75533453& 0.94108456& 0.91203445& 0.97118272& 0.99225828& 0.99938521\\
$E_{nm}^\text{hcp}(\alpha=0)$          &-38.9325327&-10.4019067&-18.30985366&-8.61107046&-7.57167198&-6.11058661\\
$\delta_{nm}^\text{hcp}(\alpha=0)$     & 0.00030779&-0.00018844&-0.00008759&-0.00022986&-0.00008431&-0.00003291\\
$a^\text{fcc}(\alpha=1)$               & 0.75527318& 0.94112001& 0.91205036& 0.97123369& 0.99227815& 0.99939381 \\
$\Delta E_{nm}^\text{fcc}(\alpha=1)$   &-0.00167052& 0.00065426& 0.00053138& 0.00087030& 0.00063934& 0.00034707\\
$a^\text{cub2}(\alpha=1)$              & 0.84952560& 1.08593491& 1.06151441& 1.14235260& 1.20025739& 1.21772324\\
$r_\text{bc}^\text{cub2}(\alpha=1)$    & 0.73571075& 0.92022759& 0.89393191& 0.95447928& 0.98606235& 0.99654594\\
$\gamma_2^\text{cub2}(\alpha=1)$       & 1         & 0.93401642& 0.91472620& 0.89022278& 0.83649731& 0.82395744\\
$\Delta E_{nm}^\text{cub2}(\alpha=1)$  & 0.29641379& 0.24620888& 0.45584326& 0.34906107& 0.58211318& 0.76315575\\
\hline
\textit{TS Burgers}                    &           &           &           &           &           &\\
$\alpha^\#$                            & 0.50124827& 0.50070620& 0.50055491& 0.50036605& 0.50007315& 0.49998534\\
$\theta_{12}$                          &75.55941782&75.54338146&75.53890556&75.53331802&75.52465212&75.52205406\\
$a^\#$                                 & 0.73561911& 0.92214149& 0.89552993& 0.95622314& 0.98638714& 0.99664997\\
$r_\text{bc}^\#$                       & 0.76148854& 0.94610839& 0.91637703& 0.97463898& 0.99337015& 0.99965780 \\
$\beta_2^\#$                           & 0.84563200& 0.82802883& 0.82241205& 0.81445956& 0.78931350& 0.78087408 \\
$\gamma_2^\#$                          & 1.60349255& 1.58902301& 1.58494120& 1.57867371& 1.56007043& 1.55383900 \\
$\Delta E_{nm}(\alpha^\#)$             & 0.63868037& 0.34878984& 0.59775652& 0.41488952& 0.59694720& 0.76650812 \\
$\Delta V [\%]$                        & 1.45105875& 2.35271163& 2.73068082& 3.15510867& 4.92065604& 5.51004680\\
$\alpha_k$                             & 0.52877   & 0.51028   & 0.50911   & 0.50408   & 0.49990   & 0.49905   \\
\hline
\textit{TS Bain}                       &           &           &           &           &           &\\
$a^\#$                                 & 0.84587206& 1.06148905& 1.03057025& 1.09911882& 1.13483878& 1.14700157\\
$r_\text{bc}^\#$                       & 0.73573887& 0.91927648& 0.89250002& 0.95186482& 0.95186482& 0.99333250\\
$\Delta E_{nm}(\alpha^\#)$             & 0.29644955& 0.24972893& 0.47009499& 0.37377855& 0.77263663& 1.17257580 \\
\hline
\end{tabular}
\caption{Optimized parameters for minima and transition states (TS) for various $(n,m)$-LJ potentials (Model 2) setting $\beta_1=1$, $\beta_3=\gamma_2$ and $\gamma_1=1$. TS properties for the Burgers ($\partial E_{nm} / \partial \alpha =0 $) and Bain paths ($\partial E_{nm} / \partial \gamma_2 =0 $) are indicated by a $\#$ symbol. For the minimum structures we get exactly $\alpha=0$ and $\beta_2=1/\sqrt{3}$ for hcp and $\alpha=1$, $\beta_2=1$ and $\gamma_2=\sqrt{2}$ for fcc. For hcp $(\alpha=0)$ we obtain a slight deviation from the ideal $\gamma_2$ value with $\delta_{nm}=\gamma_2(n,m)-\sqrt{8/3}$. cub2 indicates the second cuboidal structure after fcc, e.g. bcc if we have the ideal values of $\alpha=1$, $\beta_2=1$ and $\gamma_2=1$. For fcc $(\alpha=1)$ we have $a_\text{min}^\text{fcc}=r_\text{bc,min}^\text{fcc}$. $\Delta E_{nm}(\alpha^\#)$ for the Burgers path is taken relative to the hcp structure. The volume increase $\Delta V [\%]$ at the TS of the Burgers path relative to the fcc structure is defined as $100\{V(\alpha^\#)-V^\text{fcc}\}/V^\text{fcc}$. $\alpha_k$ defines the point on the Burger's path where the kissing number changes spontaneously from $\kappa$=4 ($\alpha<\alpha_k$) to 2 ($\alpha<\alpha_k$). For the Bain path TS we have $\alpha=1$ and $\beta_2=1$, which is located at $\gamma_2=1$ except for the (6,4)-LJ potential for which we get $\gamma_2^\#=$1.01301670. $\Delta E_{nm}(\alpha^\#)$ values for the Bain path are relative to the fcc structure.}
\label{tab:properties}
\end{table}

\begin{figure}[htb!]
\centering
\includegraphics[width=.66\columnwidth]{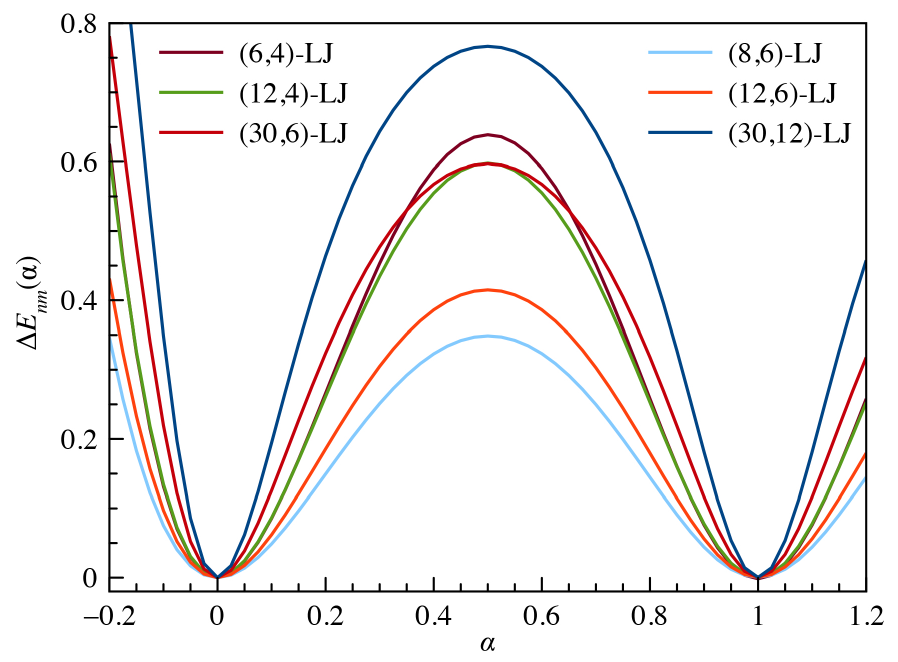}
\caption{The cohesive energy difference $\Delta E_{nm}(\alpha) = E_{nm}(\alpha)-E_{nm}^{\text{hcp}}(\alpha=0)$ for the Burgers path as a function of the distortion parameter $\alpha$ for various selected $(n,m)$-LJ potentials using Model 2 and $\beta_1=1$.}
\label{fig:Ecoh}
\end{figure}

To obtain the minimum energy path, for which we have the condition $\partial E_\text{coh}(\alpha)/\partial x_i=0$ for $x_i\in\{a,\beta_2,\gamma_2\}$, we changed the parameter $\alpha$ in forward direction from $-0.2$ to $+1.2$ in steps of 0.025 optimizing all other lattice parameters. The results show that we automatically end up with the fcc structure at $\alpha=1$ and $\gamma_{2}=\sqrt{2}$ on that path for all LJ potentials considered here, in perfect agreement with the KHS model (cf. Figure \ref{fig:Ecoh}). The reverse path starting from the bcc structure at $\alpha=1$ and $\gamma_2=1$ goes steeply uphill.

An analysis of the minimum distances reveals a kissing number of $\kappa=12$ for the two points $\alpha=0, \gamma_2=\sqrt{\frac{8}{3}}$ and $\alpha=1, \gamma_2=\sqrt{2}$ as one expects. However, around the transition state we find $\kappa=4$ for the interval $\alpha\in (0,\alpha_k)$ with shortest distances in the hexagonal layer, at critical value $\alpha_k$ we find  $\kappa=6$ picking up two short distances to the central atom, and we have $\kappa=2$ for the interval $\alpha\in (\alpha_k,1)$ with shortest distances to the central atom as listed in Table \ref{tab:properties} for various LJ potentials. We note that $\alpha_k$ is close to $\frac{1}{2}$. The optimized lattice parameters $\beta_2(\alpha)$ and $\gamma_2\alpha)$ follow qualitatively the KHS model, see supporting information.

The cohesive energies along the Burgers path for different $(n,m)$-LJ potentials are shown in Figure \ref{fig:Ecoh}. The transition state is close to the midpoint of $\alpha=1/2$ for all LJ potentials considered here as one would expect. For example, for the (12,6)-LJ potential we find the transition state at $\alpha^\#=0.50037$ with an energy difference of $\Delta E_{12,6}(\alpha^\#) = E_{12,6}(\alpha^\#)-E_{12,6}^{\text{hcp}}=0.41490$, see Table \ref{tab:properties}. An important finding here is that within this parameter space the energy and its first and second derivatives with respect to the lattice parameters show a smooth functional behavior. In the following section we will show that this is an artifact of limiting our parameter space according to the KHS limit. 

The Bain transformation from fcc to bcc is well understood because of its simplicity in the crystal transformation.\cite{burrows2021b,Schwerdtfeger2022jpc,Underwood_2015,RoblesNavarro2025} For recent applications to group 1 and 11 metals we refer to refs.\onlinecite{Jerabek2022,RoblesNavarro2023}. In our limited parameter space model, the activation energies for the Bain phase transition are usually below the ones for the Burgers path (cf. Table \ref{tab:properties}), except for very repulsive walls. For example, the highest barrier for the Bain path is obtained by the most repulsive $(30,12)$-LJ potential as expected from the KHS model, which puts bcc energetically above the Burgers path. In addition, the MD simulations by Bingxi Li et al.\cite{Bingxi2017} for solid argon show no transition from hcp or fcc to the bcc phase. We note that the hcp to fcc transition has been modeled before for LJ potentials by Jackson et al\cite{Ackland2002} using a fixed cutoff using the lattice-switch Monte Carlo method.

\subsection{The Burgers phase transition along a symmetry broken minimum energy path}
\label{SymmBreak}

So far we adopted a fixed value of $\beta_1=1$ as derived from the KHS model resulting in a smooth Burgers-Bain transformation. This initial choice to save computer time leads tp bypassing the bcc structure at $\alpha=\frac{1}{3}$ and $\beta_1=\frac{4}{3}$ along the Burgers path. For example, for the (12,6)-LJ potential we obtain an optimized value of $\beta_2=0.73409675$ at $\alpha=\frac{1}{3}$ substantially different to $\beta_2=2\sqrt{2}/3=0.94280904$ expected for the bcc structure. The question therefore arises how valid the choice of parameters from the KHS approximation for LJ potentials really is to correctly describe the hcp transition to the cuboidal phase. We therefore decided to add the $\beta_1$ lattice constant to our variable parameter space. We therefore restricted our lattice sum treatment for the Burgers path from in principle nine (six for the $B$-matrix and three for the shift vector) to seven varying lattice parameters, with the two remaining fixed ones being the two angles $\angle(\vec{b_1},\vec{b_3})$ and $\angle(\vec{b_2},\vec{b_3})$ kept at 90$^{\circ}$. This not only saves computer time, but has the advantage that we are able to derive efficient expansions of the required lattice sums in terms of Bessel functions, see Appendix \ref{Besselexpansions}. Otherwise one has to use different techniques for the efficient treatment of the Epstein zeta function, see for example Buchheit and Busse\cite{buchheit2025epsteinzeta}. However, from preliminary optimization runs it turns out that we can impose two further restrictions, i.e. $(i)$~ $|\vec{b}_1|=|\vec{b}_2|$ resulting in $\gamma_1f_S=1$ for Model 1 and $\gamma_1=1$ for Model 2, and  $(ii)$~ $\beta_3/\gamma_2=1$ as we observe no symmetry breaking in direction of the $c-$axis within the range of $\gamma_2$ values applied (if $\gamma_2$ becomes large one will eventually observe a Peierls distortion). Hence, for the Burgers path, our in general nine-dimensional optimization problem reduces to five, that is $x_i\in\{a,\alpha,\gamma_2,\beta_1,\beta_2\}$. This space of freely varying parameters is larger than used in previous work, cf. Ref. \onlinecite{Natarajan2019}, and therefore allows for a more accurate treatment of the Burgers path within the LJ model. As we will see in the next section, there is an exact relation between the two parameters $\beta_1(\alpha)$ and $\beta_2(\alpha)$ in the interval $\alpha\in[0,1]$, which would reduce our variable parameter space even further to four. As we shall see, the parameter $\beta_1$ is responsible for the appearance of a bifurcation point and symmetry breaking effect in the middle layer movement, leading to a significant lowering in energy especially for the soft LJ potentials. The optimized parameters $x_i$ for the transition state along the Burgers path are shown in Table \ref{tab:propertiestrans1}. 

\begin{center}
\begin{table}[hbt!] 
\setlength{\tabcolsep}{3pt}
    \fontsize{10}{12}\selectfont
\begin{tabular}{ |l|r|r|r|r|r|r|r| } 
 \hline
 Parameter                  & $(6,4)$ LM& $(6,4)$& $(8,6)$   & $(12,4)$  & $(12,6)$  & $(30,6)$  & $(30,12)$\\ 
\hline
$\alpha^\#$                 & 0.33333333& 0.35642304& 0.39334946& 0.41043348& 0.43318316& 0.48184258& 0.49317558\\
$\theta_{12}$               &70.52877937&71.22888103&72.34256361&72.85545878&73.53626429&74.98459698&75.32047856\\
$a^\#$                      & 0.73571075& 0.73573887& 0.92025077& 0.89394548& 0.95452850& 0.98609243& 0.99655952\\
$r_\text{bc}^\#$            & 0.84952560& 0.84587207& 1.01384023& 0.97080269& 1.01649542& 1.00388530& 1.00331254\\
$\beta_1^\#$                & 1.33333333& 1.14969060& 1.21375677& 1.17935344& 1.13411523& 1.03648224& 1.01364020\\
$\beta_2^\#$                & 0.94280904& 0.73573887& 0.88740177& 0.87037568& 0.84744319& 0.79509947& 0.78230716\\
$\gamma_2^\#$               & 1.63299316& 1.62590804& 1.61067804& 1.60279158& 1.59117300& 1.56178561& 1.55400238\\
$\Delta E_{nm}(\alpha^\#)$  & 0.29641379& 0.29644955& 0.24610211& 0.45568950& 0.34858732& 0.58118291& 0.76251538\\
$\Delta V [\%]$             & 0.62414868& 0.61309906& 1.46557560& 1.97613085& 2.42740664& 3.70201581& 5.39593019\\
\hline
\end{tabular}
\caption{Optimized parameters for the Burgers path transition state (TS) for various $(n,m)$-LJ potentials. See Table \ref{tab:properties} for a detailed description of the parameters. $\Delta E_{nm}(\alpha^\#)$ is taken relative to the hcp structure. The volume increase $\Delta V [\%]$ at the transition point of the Burgers path relative to the fcc structure is defined as $100\{V(\alpha^\#)-V^\text{fcc}\}/V^\text{fcc}$. For the (6,4) potential we report both the local minimum (LM) and the second transition state closer to the fcc structure (see text for details).}
\label{tab:propertiestrans1}
\end{table}
\end{center}

\begin{figure}[b!]
\centering
\includegraphics[width=.66\columnwidth]{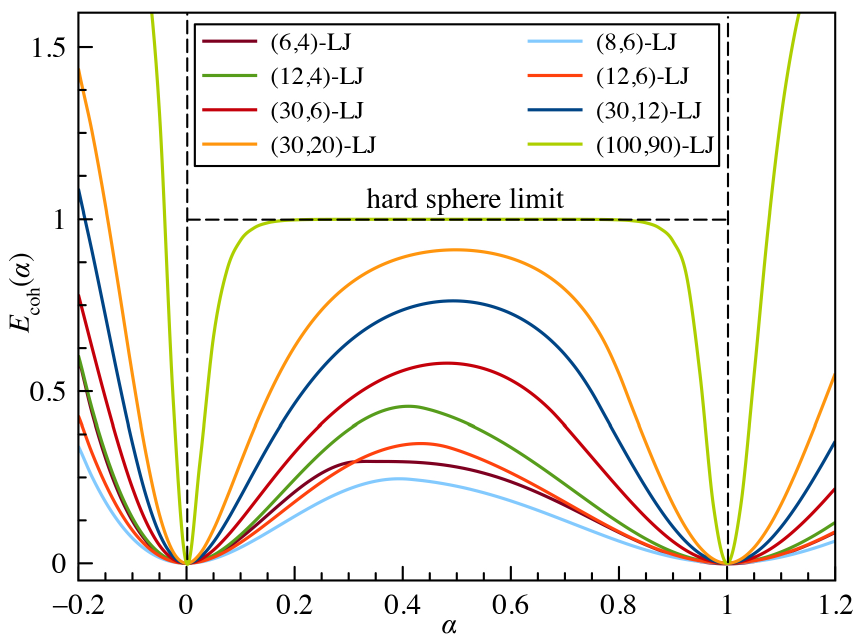}
\caption{The cohesive energy difference $\Delta E_{nm}(\alpha) = E_{nm}(\alpha)-E_{nm}^{\text{hcp}}(\alpha=0)$ for the symmetry-broken Burgers path as a function of the distortion parameter $\alpha$ for various selected $(n,m)$-LJ potentials using Model 2.}
\label{fig:Ecohsymmbreak}
\end{figure}

The minimum cohesive energy path $E_\text{coh}(\alpha)$ for the Burgers transformation within our extended five parameter space is shown in Figure \ref{fig:Ecohsymmbreak}. We see that the transition state is now significantly lowered especially for the softer LJ potentials compared to our previous results where we imposed the hard-sphere limit restriction of $\beta_1=1$. For example, for the (12,6)-LJ potential the transition state is slightly shifted towards the hcp structure ($\Delta\alpha^\#=-0.06718$) while the energy is lowered by $\Delta E_\text{coh}=-0.06630$ compared to the case where $\beta_1=1$. This is even more so for the soft (6,4)-LJ potential, i.e. $\Delta\alpha^\#=-0.18023$ and $\Delta E_\text{coh}=-0.34222$ (here we took the second transition state on the hcp$\rightarrow$fcc path, see discussion below). The increase in volume from the minimum structure to the transition state is also  reduced significantly. The KHS limit is an upper bound for the change in cohesive energy, and calculations show that the (100,90)-LJ potential is already very close to this limit with larger deviations only observed in the vicinity of the two limiting points of hcp ($\alpha=0$) and fcc ($\alpha=1$).

In order to analyse the influence of the $\beta_1$ parameter even further, we note that the kissing number changes from $\kappa(\alpha<0)=2\rightarrow\kappa(\alpha=0)=12$ (ideal hcp) $\rightarrow\kappa(0<\alpha<\alpha_k)=4\rightarrow\kappa(\alpha_k)=8\rightarrow\kappa(\alpha_k<\alpha<1)=4\rightarrow\kappa(\alpha=1)=12\rightarrow\kappa(\alpha>1)=4$ where $\alpha_k$ is close to the transition state. This implies a larger kissing number after the transition state towards fcc compared to the previous symmetric model where we fixed $\beta_1=1$, which rationalizes the significant energy lowering of the transition state. Moreover, we see a change in the shortest distances along the Burgers path, i.e. from hcp to the transition state the shortest distances are those within the hexagonal layers, while after the transition state towards the fcc structure the shortest connections are from the hexagonal to the middle layer atoms.

\begin{figure}[htb!]
\centering
a)\includegraphics[width=.45\linewidth]{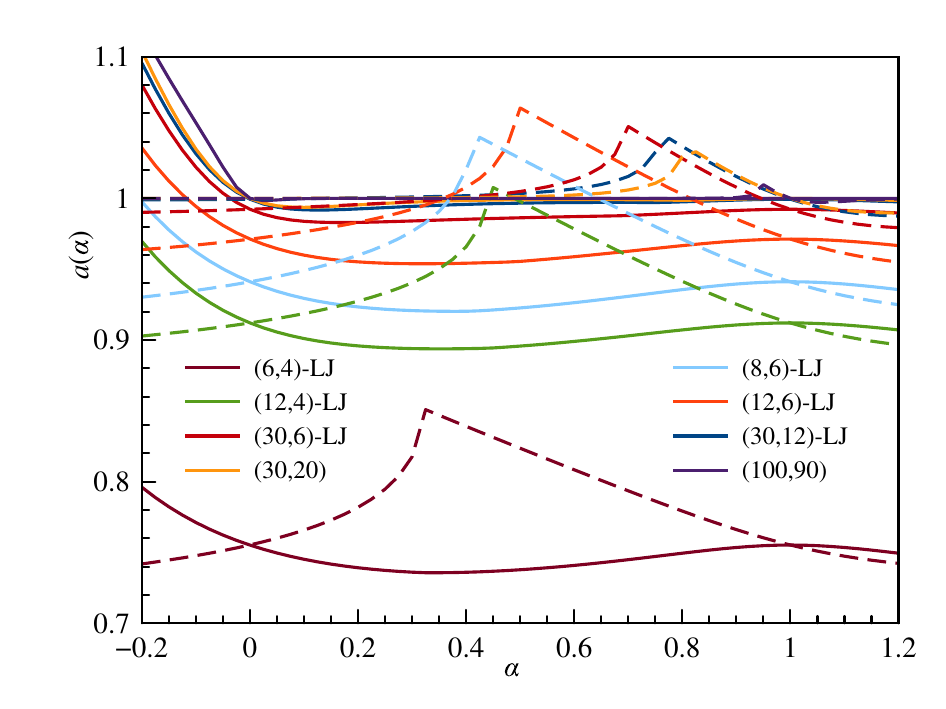}
b)\includegraphics[width=.45\linewidth]{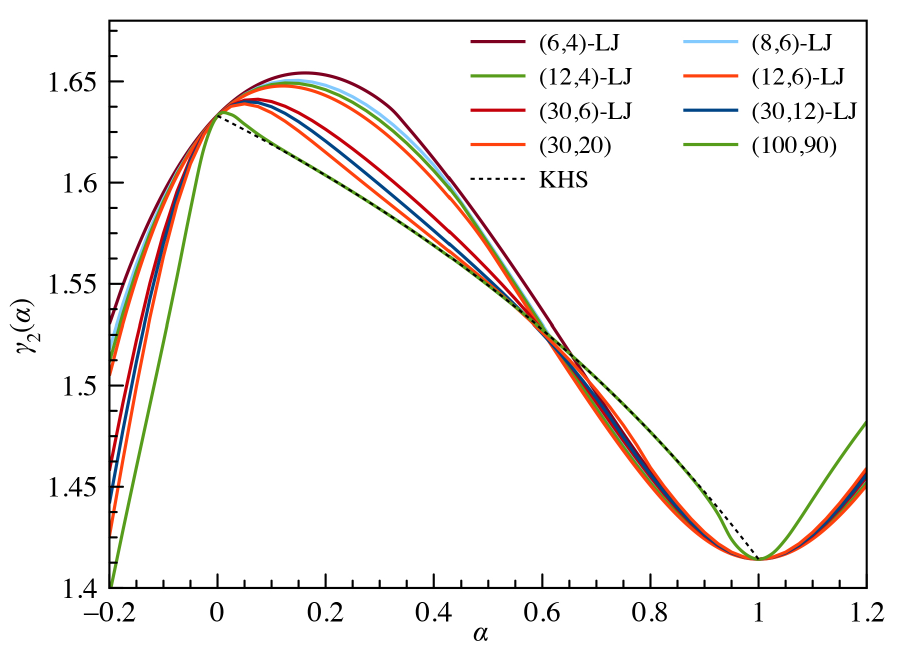}
c)\includegraphics[width=.45\linewidth]{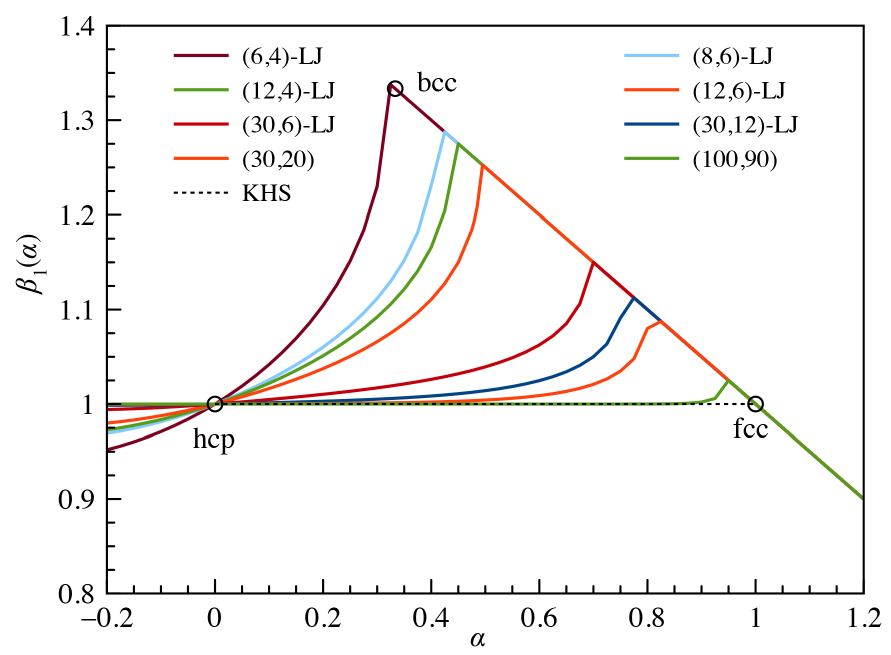}
d)\includegraphics[width=.45\linewidth]{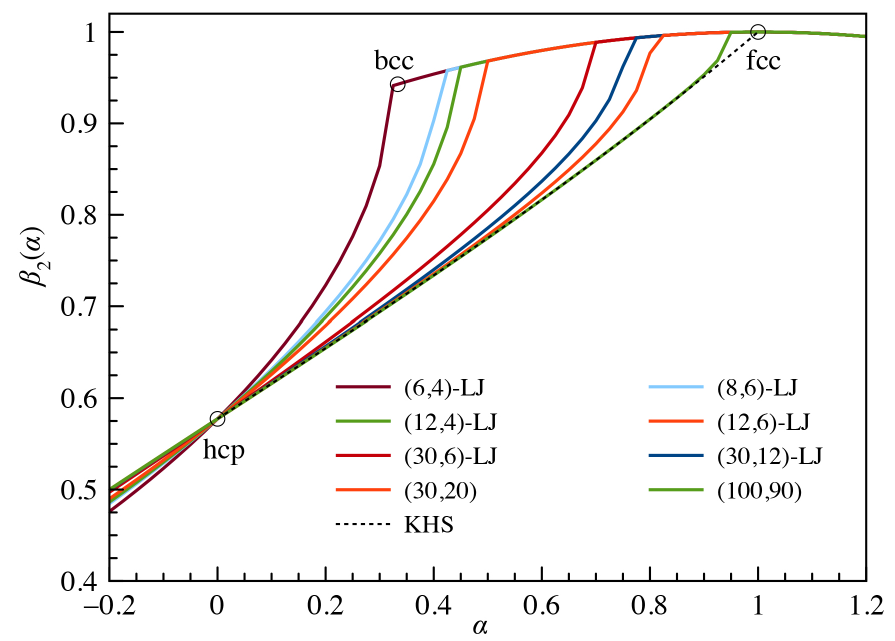}
\caption{Dependence of the optimized parameters $a,\gamma_2,\beta_1,\beta_2$ on $\alpha$ along the symmetry-broken Burgers path for various selected $(n,m)$-LJ potentials using Model 2. For the lattic constant $a$, the solid lines show the distances between nearest neighbors in the base layer ($a=|\vec{b}_1|$), while the dashed lines show the optimized distances from the base to the body-centered atom ($r_\text{bc}=|\vec{v}_s|$), see Figure \ref{fig:hcp}. The KHS limits are also shown as described in detail in section \ref{sec:KHS}. In c) and d) the three different structures, hcp, fcc and bcc, are indicated by black circles.}
\label{fig:fcchcpsymmbreak}
\end{figure}
The lattice parameters shown in Figure \ref{fig:fcchcpsymmbreak} clearly demonstrate that we break symmetry along the Burgers path as compared to the symmetric path obtained for fixed $\beta_1=1$. This distinct symmetry breaking effect leads to a critical point $\alpha_c$ along the path where the parameters $\beta_1$ and $\beta_2$ change abruptly while the cohesive energy $E_\text{coh}(\alpha)$ still behaves like a smooth function. The distance to the middle layer $r_\text{bc}$ also changes abruptly at the critical point due to its dependence on $\beta_1$ and $\beta_2$, see eq.\eqref{eq:sf}. At the critical point $\alpha_c$ we have the following conditions for the optimized parameters describing the movement of the middle layer (in the following we discuss the optimized parameters along the minimum energy path $E_\text{coh}(\alpha)$, i.e. we have the condition $\partial E_\text{coh}(\alpha)/\partial \beta_1=0$ and $\partial E_\text{coh}(\alpha)/\partial \beta_2=0$),
\begin{equation} \label{eq:discont1}
\lim_{\alpha\rightarrow \alpha_c} \beta_1(\alpha)=\lim_{\alpha_c\leftarrow \alpha} \beta_1(\alpha) \quad \text{and} \quad
\lim_{\alpha\rightarrow \alpha_c} \beta_2(\alpha)=\lim_{\alpha_c\leftarrow \alpha} \beta_2(\alpha)
\end{equation}
but
\begin{equation} \label{eq:discont2}
\lim_{\alpha\rightarrow \alpha_c} \partial \beta_1(\alpha)/\partial\alpha\ne\lim_{\alpha_c\leftarrow \alpha} \partial \beta_1(\alpha)/\partial\alpha \quad \text{and} \quad
\lim_{\alpha\rightarrow \alpha_c} \partial \beta_2(\alpha)/\partial\alpha\ne\lim_{\alpha_c\leftarrow \alpha} \partial \beta_2(\alpha)/\partial\alpha 
\end{equation}
where the right arrow indicates that we move in forward direction, from hcp to fcc, and the left arrow for the reverse path from fcc to hcp. In fact, in forward direction the functions $\beta_1(\alpha)$ and $\beta_2(\alpha)$, $\alpha<\alpha_c$, terminate by joining another curve coming from the reverse direction which continues beyond $\alpha_c$ with increasing cohesive energy, in fact towards the bcc structure at $\alpha=\frac{1}{3}$ as discussed in more detail below. This implies that we have a bifurcation point along our minimum energy path for both $\beta_1(\alpha)$ and $\beta_2(\alpha)$ as shown in Figure \ref{fig:fcchcpsymmbreak}. We refer to the two directions for the minimum energy path as left (forward) and right (backward) directions in the following. The second derivative of $\beta_1(\alpha)$ and $\beta_2(\alpha)$ with respect to $\alpha$ therefore shows a jump discontinuity. However, the ratio $\beta_1(\alpha)/\beta_2(\alpha)$ behaves smoothly as discussed in the next section.

\begin{figure}[htb!]
\centering
\includegraphics[width=.66\linewidth]{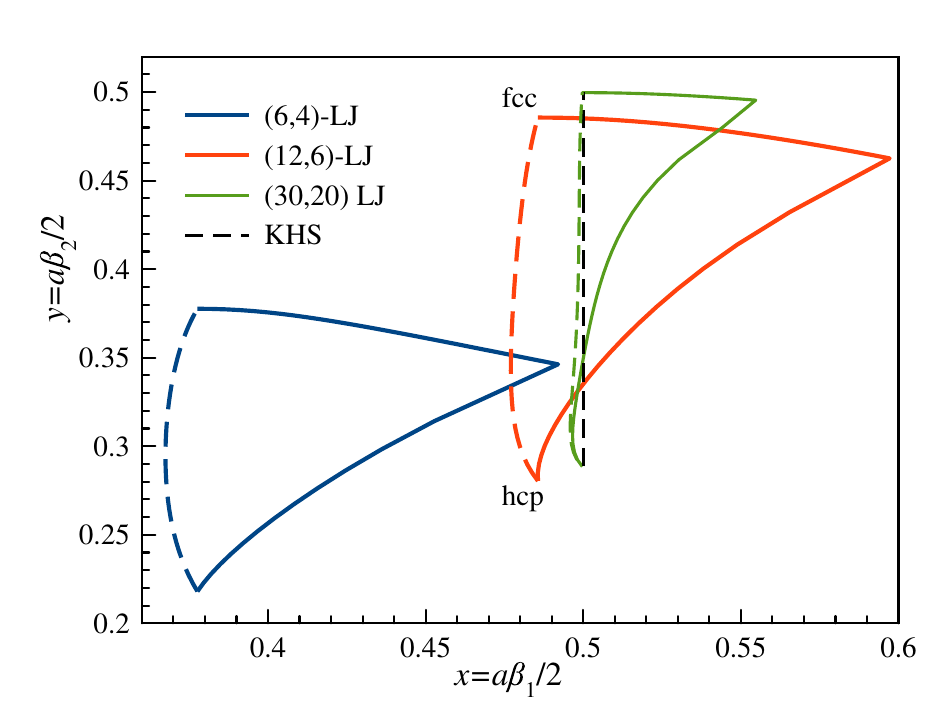}
\caption{Movement of central atom within the $(x,y)$-plane, spanned by the vectors $\vec{b}_1$ and $\vec{b}_2$, for three selected LJ potential. For the hard-sphere limit we have $\beta_1=1$ and $\beta_2=1/\sqrt{3}$ for hcp and $\beta_2=1$ for fcc. The dashed lines are from calculations setting $\beta_1=1$.}
\label{fig:Movementsymmbreak}
\end{figure}
The movement of the atom in the middle layer in the $(x,y)=(\vec{b}_1,\vec{b}_2)$ plane is shown in Figure \ref{fig:Movementsymmbreak}. In this symmetry broken Burgers path, the central atom in the middle layer is moving away from the origin to the right up to the critical point where it starts to move back again. We also show the line for the KHS model and the one for the smooth transition with fixed $\beta_1=1$, the latter not being the correct minimum energy path in our extended parameter space. With increasing LJ exponents we converge towards the KHS limit.

It is computationally challenging to accurately obtain the bifurcation (critical) point $\alpha_c$ as we do not have smooth functions of $\beta_1(\alpha)$, $\beta_2(\alpha)$ or $r_\text{bc}(\alpha)=|\vec{v}_s(\alpha)|$ anymore. Moreover, it is difficult to converge into the left (forward) solution near the critical point, while the right (backward) solution causes no particular challenge. Nevertheless, as it turns out, the second derivatives $\partial^2E_\text{coh}(\alpha)/\partial \beta_1^2$ and $\partial^2E_\text{coh}(\alpha)/\partial \beta_2^2$ have an almost linear behavior for both the left and right solutions and can therefore be used to determine the critical point where the two solutions have identical values. The lattice parameters and energies at the critical point obtained by this method are listed in Table \ref{tab:propertiescrit} for a few selected LJ potentials. We note that $\alpha_c$ increases with increasing hardness of the potential, that is the bifurcation point moves towards the fcc minimum by increasing the exponents of the LJ potential, and for the KHS model we have $\alpha_c=1$ where the bifurcation point becomes identical to the fcc minimum.
\begin{center}
\begin{table}[hbtp] 
\setlength{\tabcolsep}{4pt}
\begin{tabular}{ |l|r|r|r|r|r|r| } 
 \hline
 Parameter                  & $(6,4)$   & $(8,6)$   & $(12,4)$  & $(12,6)$  & $(30,6)$  & $(30,12)$\\ 
\hline
$\alpha_c$                  & 0.321022  & 0.409953  & 0.441264  & 0.491655  & 0.700331  & 0.756979\\
$\theta_{12}$               &70.154256  &72.841043  &73.777520  &75.275443  &81.382657  &83.020728\\
$a_c$                       & 0.735736  & 0.920300  & 0.894201  & 0.955241  & 0.965239  & 0.997027\\
$r_{\text{bc},c}$             & 0.851483  & 1.046068  & 1.009332  & 1.065546  & 1.050608  & 1.046645\\
$\beta_{1,c}$                 & 1.339486  & 1.295023  & 1.279364  & 1.254168  & 1.149835  & 1.121482\\
$\beta_{2,c}$                 & 0.940608  & 0.955490  & 0.960181  & 0.967155  & 0.988711  & 0.992565\\
$\gamma_{2,c}$                & 1.636646  & 1.605597  & 1.592986  & 1.571241  & 1.561786  & 1.471437\\
$\Delta E_{nm}(\alpha_c)$   & 0.296457  & 0.245010  & 0.449619  & 0.334208  & 0.399023  & 0.477019\\
\hline
\end{tabular}
\caption{Optimized parameters for the Burgers path critical point (CP) for various $(n,m)$-LJ potentials using Model 2. See Table \ref{tab:properties} for details. The parameter $\alpha_c$ defines the location of the critical point.}
\label{tab:propertiescrit}
\end{table}
\end{center}

Concerning cohesive energy derivatives with respect to the lattice parameters, we have for the two critical parameters the condition
\begin{equation} \label{eq:discont2a}
\lim_{\alpha\rightarrow \alpha^c} \partial^2 E_\text{coh}(\alpha)/\partial\beta_1^2=\lim_{\alpha_c\leftarrow \alpha} \partial^2 E_\text{coh}(\alpha)/\partial\beta_1^2
\end{equation}
and
\begin{equation} \label{eq:discont2b}
\lim_{\alpha\rightarrow \alpha^c} \partial^2 E_\text{coh}(\alpha)/\partial\beta_2^2=\lim_{\alpha_c\leftarrow \alpha} \partial^2 E_\text{coh}(\alpha)/\partial\beta_2^2
\end{equation}
with an abrupt change (kink) at $\alpha_c$ along the minimum energy path. Both the left and right second derivative curves with respect to the transition parameter $\alpha$ are very close to linearity (at least to numerical accuracy) and it is therefore difficult to predict accurately the higher order derivatives which will be very small for these two parameters. Ehrenfest's definition of a $n$th-order phase transition is that the free energy exhibits a discontinuity (jump discontinuity, singularities etc.) in the $n$th-order derivative with respect any thermodynamic variables like the pressure or temperature at a certain critical point.\cite{Ehrenfest1933,Jaeger1998} Any discontinuity in the lattice parameters for the middle layer may lead to a discontinuity in the Taylor expansion of the total energy in Cartesian space and therefore to a discontinuity in the phonon spectrum. Indeed, using DFT for several transition metals, Feng and Widom\cite{Feng2018} observed a $\Lambda$-point phonon instability along the Burgers path from bcc to hcp related to the slide deformation of layer B (see Figure \ref{fig:hcp}). They chose a three-parameter space for their definition of a bi-lattice directly connecting the bcc with the hcp phase as compared to our choice which includes hcp and all all cuboidal structures. At the point of instability the lattice reduces symmetry from a cubic to an orthorhombic phase moving towards the hcp structure. This symmetry breaking is clearly related to what we are observing here using LJ potentials. Similarly, Casperson and Carter used a three-parameter space but different definition for the Burgers transformation for their DFT calculations of metallic lithium\cite{Caspersen2005}. The calculations were carried out under finite pressure, but they did not observe any symmetry breaking effects most likely due to the fact that the cohesive energy is rather insensitive to such effects. 
We finally mention that the discontinuity in the second derivatives will lead to an abrupt change in the entropy however small this effect might be. One can therefore only speculate that the Burgers transition is a first-order phase transition. However, it would be very difficult if not impossible to detect this discontinuity either by experiment or by molecular dynamics simulations.

\subsection{The bcc structure as an intermediate between the fcc and hcp phase}
\label{SymmBreak2}

The question arises if there is a direct minimum energy path from hcp to bcc as suggested for example by Carter and co-workers for iron or lithium \cite{Carter2008a,Caspersen2005} and by several other authors\cite{Straub1971,Bassett1987,LiWang1999,Cayron2016,Natarajan2019,Feng2018}, or if the fcc phase is required as an intermediate\cite{lu2014,RoblesNavarro2025}. In other words, we might have missed another transition path. Moreover, it was already shown in 1940 that the elastic stiffness tensor for cubic crystals fulfills Born's second stability criterium, $C_{11}-C_{12}>0$, only for very soft LJ potentials for the bcc structure, e.g. for exponents such as $(n,m)=(6,4)$ of $(5,4)$, otherwise one observes a distortion of the rhombohedral bcc primitive cell\cite{Born_1940,Misra_1940} as shown in Figure \ref{fig:StellOct}, (loosely termed rhombohedral distortion in the literature). In contrast, the fcc phase for a $(n,m)$-LJ potential is always stable (local minimum). Surprisingly, the exact minimum energy path and mechanism of the rhombohedral distortion towards presumably either the fcc or the hcp phase has never been investigated in great detail apart from the conventional cuboidal distortion towards the fcc phase along a simple Bain path\cite{burrows2021b,Jerabek2022,RoblesNavarro2023,RoblesNavarro2025}.

\begin{figure}[htb]
\centering
\includegraphics[width=.65\linewidth]{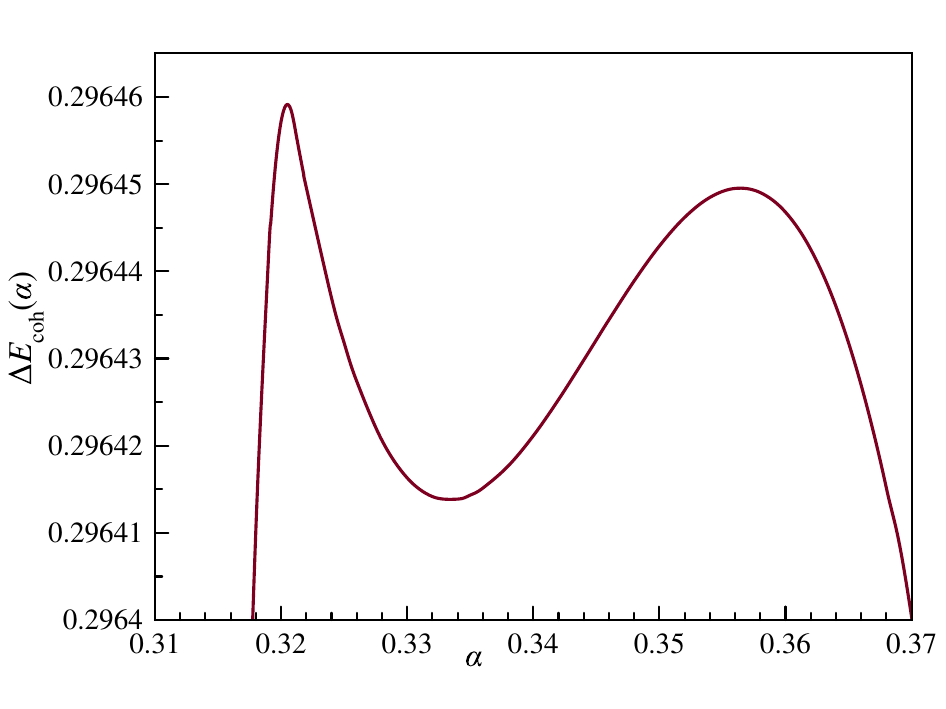}
\caption{A closer look into the minimum energy path with the two transition states and the bcc minimum for the (6,4)-LJ potential with a minimum occurring at the bcc structure with $\alpha=\frac{1}{3}$.}
\label{fig:64transition}
\end{figure}

Figure \ref{fig:64transition} shows an interesting feature along the Burgers path in that we obtain an additional (metastable) minimum  for the soft (6,4)-LJ potential. This is not the case for any of the other harder potentials studied here such as the usual (12,6)-LJ potential. The highest energy point for the (6,4)-LJ potential is now located at the critical point $\alpha_c$ where a bifurcation takes place, with a second transition state following at $\alpha^\#$ (compare the values in Tables \ref{tab:propertiestrans1} and \ref{tab:propertiescrit}). Between both maxima we have a minimum located exactly at the lattice parameters satisfying the condition for a bcc phase (cf. Table \ref{tab:latticeparameters}), e.g. $\alpha=\frac{1}{3}$, albeit the energy differences between the extremal points around the bcc phase are very small and the bcc phase is therefore easily overlooked in the Burgers phase transition. At this metastable bcc minimum we observe an increase in the kissing number from four to eight, which obviously implies increased stability for softer potentials (in the KHS limit we have $\kappa=10$ for the whole of the Burgers path). Hence, if the bcc structure becomes a minimum along the Burgers path, we have a direct path from hcp to bcc and further to fcc. An unstable bcc structure will distort to either the hcp or the fcc structure, and the Burgers path will bypass bcc. In fact, bcc remains an extremum (zero gradient for all lattice parameters) for all LJ potentials considered, but eventually turns into a second-order saddle point if the exponents $(n,m)$ of the LJ potential are increased\cite{burrows2021b}. A more detailed analysis reveals that in this case the negative eigenvalues of the Hessian with respect to our parameter space shows a distortion within the two-parameter space $(\alpha,\beta_1)$.  

The question arises of how to describe the mechanism of the structural distortion of the bcc phase. Figure \ref{fig:fcchcpsymmbreak} shows that bcc lies on the straight $\beta_1(\alpha_c)$ line towards the fcc structure containing all the bifurcation points for the $(n,m)$-LJ potentials, i.e. we have
\begin{equation} \label{eq:linearright}
\beta_1(\alpha)=\frac{1}{2}\left(3-\alpha\right)
\end{equation}
with small deviations possibly due to numerical noise in our data (with a standard deviation of $1.6\times10^{-5}$). The Hessian for the cohesive energy shows that the modes along the lattice parameters $\alpha$ and $\beta$ are strongly coupled with large off-diagonal elements $\partial^2E_\text{coh}/{\partial \alpha \partial \beta_1}$ and $\partial^2E_\text{coh}/\partial \alpha^2\approx \partial^2E_\text{coh}/\partial \beta_1^2$ along the bifurcation line. Solving the eigenvalue problem for the Hessian containing only the second derivatives with respect to $\alpha$ and $\beta_1$ one can easily show that the existence of a minimum for bcc solely depends on the sign of $\partial^2 E_\text{coh}/\partial \alpha^2$ (or $\partial^2 E_\text{coh}/\partial \beta_1^2$), which is positive for the (6,4)-LJ potential as seen in Figure \ref{fig:64transition}.

For the second lattice parameter $\beta_2(\alpha)$ we get a simple analytical expression derived in appendix \ref{BurgersBain}, 
\begin{equation}\label{eq:beta2alpha}
\beta_2(\alpha) =  \sqrt{1-\tfrac{1}{4}(1-\alpha)^2}
\end{equation}
This formula is substantially different to one derived for the KHS limit, eq.\eqref{eq:KHSbeta2}, except at the end point at $\alpha=1$. This implies that for a specific value of the bifurcation point $\alpha_c$ for a $(n,m)$-LJ potential we can determine exactly the corresponding $\beta_1$ and $\beta_2$ values. Unfortunately, $\alpha_c$ for a specific $(n,m)$-LJ potential has to be determined numerically from the lattice sums. Furthermore, the ratio $\beta_1(\alpha)/\beta_2(\alpha)$ obtained from the two eqs. \eqref{eq:linearright} and \eqref{eq:beta2alpha} is valid for the whole range of $\alpha\in[0,1]$ according to our calculations. Hence, this implies that the variable parameter space can be further reduced by using this simple relation. Furthermore, it implies that any symmetry breaking in $\beta_1$ leads to a symmetry breaking in $\beta_2$.

While rhombohedral primitive cells can be found for both fcc and bcc, for the whole cuboidal transformation we can define the basis vectors in the following way (see appendix \ref{BurgersBain}),
\begin{align}\label{eq:pvectors}
\nonumber
\vec{p}_1&=\frac{a_\text{cub}}{2}(1,-1,-\gamma_\text{cub})^\top\\
\vec{p}_2&=\frac{a_\text{cub}}{2}(1,1,-\gamma_\text{cub})^\top\\
\nonumber
\vec{p}_3&=\frac{a_\text{cub}}{2}(1,1,\gamma_\text{cub})^\top
\end{align}
This implies that for the bcc$\rightarrow$fcc transformation, $\gamma_\text{cub}=1\rightarrow\sqrt{2}$, we have equidistant basis vectors which transform as $|\vec{p}_i|=\frac{a_\text{cub}}{2}\sqrt{2+\gamma_\text{cub}^2}$, i.e. they are increasing from bcc to fcc as pointed out before\cite{burrows2025a}. The rhombohedral distortion of the bcc structure is then driven by the change in angles between these vectors, i.e. we have $\angle(\vec{p}_1,\vec{p}_2)=180^\text{o}-\theta_T\rightarrow 60^\text{o}$, $\angle(\vec{p}_1,\vec{p}_3)=\theta_T\rightarrow 120^\text{o}$, and $\angle(\vec{p}_2,\vec{p}_3)=180^\text{o}-\theta_T\rightarrow 90^\text{o}$, where $\theta_T=109.47122063^\text{o}$ is the tetrahedral angle. We can see that this leads to a primitive fcc cell different to the normal primitive rhombohedral cell used in the literature where the corresponding angles are set to $60^\text{o}$. The important result here is that the rhombohedral distortion of the bcc phase is along a Bain path and therefore identical to a cuboidal distortion along the Bain path(!). In addition, at the bifurcation point $\alpha_c$ along the Burgers distortion path towards hcp at $\alpha<\alpha_c$ the lattice becomes a true bi-lattice which cannot be described anymore by a single Bravais lattice without adding another atom inside the unit cell. Such topological changes are typical for bifurcation points. Hence, if we follow along this $bcc\rightarrow hcp$ path, the bcc distortion should perhaps not be described anymore as a simple rhombohedral distortion. In appendix \ref{BurgersBain} we demonstrate the Burgers path on the linear $\beta_1(\alpha)$ line containing all the bifurcation points for the different $(n,m)$-LJ potentials is identical to the original Bain path as discussed in ref. \onlinecite{burrows2025a}. This agree with the analysis of Feng and Widom using density functional theory for several transition metals\cite{Feng2018}. Moreover, the bifurcation point of symmetry breaking becomes exactly the point where we move from a simple lattice into a bi-lattice. We expect that this point of symmetry breaking becomes identical with the bcc structure when the bcc minimum vanishes, i.e. $\partial^2 E_\text{coh}/\partial \gamma_c^2=0$ at $\gamma_c=1$, which needs to be further investigated.

It is perhaps useful to summarize our important findings in this section:\\
1) In our extended parameter space we observe a distinct symmetry breaking effect with a bifurcation point appearing at $\alpha=\alpha_c$, where the remaining hcp$\rightarrow$fcc transition joins the traditional Bain path for $\alpha>\alpha_c$, and with the possibility of going through a metastable bcc minimum for very soft LJ-potentials. This explains the phonon instability observed by Feng and Widom\cite{Feng2018}.\\
2) For the remaining Bain path we obtain analytical relationships between the lattice parameters of a cuboidal transformation and our Burgers transformation for $\alpha>\alpha_c$, see appendix \ref{app:bcc}.\\
3) The rhombohedral distortion of the bcc phase in our LJ model is along the path to fcc until one reaches the bifurcation point where distortion into hcp can happen and the lattice changes into a bi-lattice. If bcc is a metastable minimum we see both distortion paths.

\subsection{The case of solid argon}
\label{Argon}

\begin{figure}[htb!]
\centering
\includegraphics[width=.62\columnwidth]{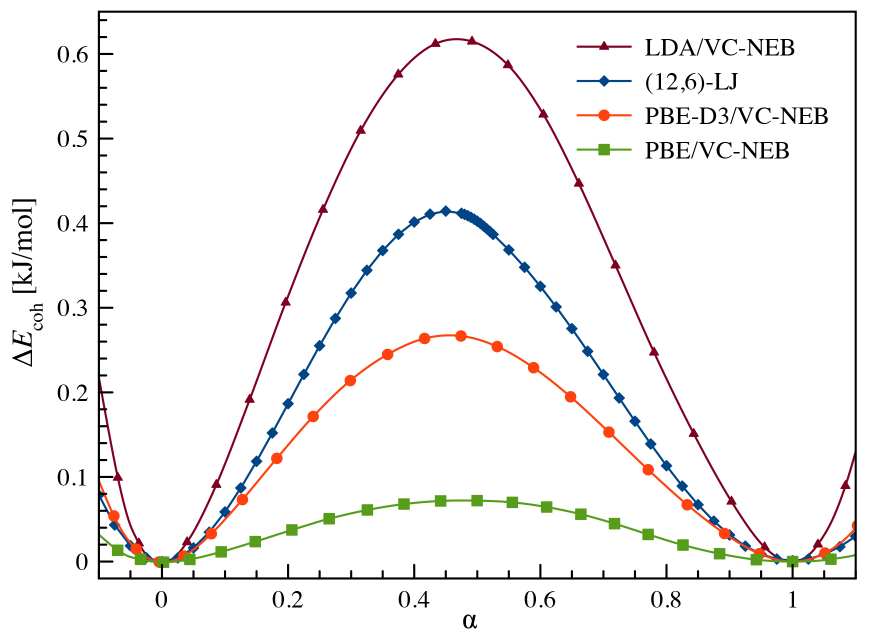}
\caption{Comparison between the (12,6)-LJ (including symmetry breaking effects) and DFT calculations (LDA, PBE and PBE-D3) using the VC-NEB method for the Burgers path from hcp to fcc.}
\label{fig:VCNEBDFT}
\end{figure}

To give our results a physical meaning for real solids, we consider solid argon by multiplying the activation energy for the (12,6)-LJ potential with the most accurate available dissociation energy of $\epsilon=99.351\pm 0.32$ cm$^{-1}$ for Ar$_2$ ($1.189\pm 0.038$ kJ/mol) obtained from relativistic coupled-cluster calculations by Patkowski and Szalewicz.\cite{PatkowskiSzalewicz2010} This results in an activation energy for the Burgers transition state of 34.6 cm$^{-1}$ (0.41 kJ/mol) relative to the hcp structure. Calculations performed with the VC-NEB method using DFT with both the PBE and the PBE-D3 functional automatically results in a Burgers path for the hcp$\rightarrow$fcc transformation in solid argon within our usual definition of the unit cell. However, the barrier is lower compared to the (12,6)-LJ result, i.e. 22.4 cm$^{-1}$ (0.267 kJ/mol) for PBE-D3, 6.0 cm$^{-1}$ (0.072 kJ/mol) for PBE, and 51.6 cm$^{-1}$ (0.618 kJ/mol) for LDA. This compares to the experimentally known cohesive energy of fcc argon of 7722(11) J/mol,\cite{Schwalbe-1977}, i.e. the barrier height for PBE-D3 is only about 3.5\% of the fcc cohesive energy. For a more precise notion, the cohesive energy of both fcc and hcp structures with the LDA functional is $13.45$ kJ/mol, while for PBE is $2.23$ kJ/mol. The effect of dispersion can be seen through the PBE-D3 cohesive energy of fcc and hcp being $8.33$ and $8.32$ kJ/mol, respectively, which is in good agreement with the value obtained in Ref. \onlinecite{Schwerdtfeger-2016}. This means that the actual barrier heights with the different density functionals are less than $5\%$ of the total cohesive energy of each phase. Furthermore, the activation energy is smaller than the zero-point vibrational contribution for fcc argon which is 67.4 cm$^{-1}$ (0.806 kJ/mol).\cite{Schwerdtfeger-2016} Comparing PBE with PBE-D3 as shown in Figure \ref{fig:VCNEBDFT}, we see that the barrier height in the Burgers transformation is strongly influenced by the addition of Grimme's dispersion term. It is clear that by increasing the cohesive energy and decreasing the nearest neighbor distances the barrier height increases. It comes therefore at no surprise that LDA overestimates the phase transition barrier substantially. The sensitivity of reaction barriers to the density functional approximation applied in molecular systems is well documented\cite{Cohen2012}. Concerning the location of the transition state, it is very close to the midpoint of the path in agreement with the (12,6)-LJ potential, i.e. $\alpha=0.456$ for PBE-D3, $\alpha=0.478$ for PBE and $\alpha=0.467$ for LDA. As argon is relatively well described by the (12,6)-LJ potential, we bypass the bcc structure along the Burgers path. For example, at the PBE-D3 level of theory we get at $\alpha=\frac{1}{3}$ the deviations from the ideal bcc lattice parameters (see Table \ref{tab:latticeparameters}) $\Delta \beta_1 = \beta_1 - \beta_1^\text{bcc}= 1.590074 - \frac{4}{3} = 0.256741$, $\Delta \beta_2 = 1.105303 - \frac{2\sqrt{2}}{3} = 0.162494$, $\Delta \beta_3 = 1.597552 - \sqrt{\frac{8}{3}} = -0.035441$, and $\Delta \gamma_2 = -0.035441$, and we have $\beta_3\approx\gamma_2$.

We can also compare with recent molecular dynamics (MD) simulation by Bingxi Li et al.\cite{Bingxi2017} using a (12,6)-LJ potential and 18,000 Ar atoms in a box adjusted to a temperature of $T$=40 K and pressure of $P$= 1 bar. They calculated a barrier height for the enthalpy of $\Delta H^\#=16.0\pm 1.0$ cm$^{-1}$ (per atom) which lies in-between our PBE and PBE-D3 result. Even if we adjust for the more accurate dissociation energy of Patkowski and Szalewicz\cite{PatkowskiSzalewicz2010} then given in their paper, we get $\Delta H^\#=19.2\pm 1.1$ cm$^{-1}$. However, at 40 K the kinetic energy per atom $E_\text{kin}=\frac{3}{2}k_\text{B}T$ translates into 41.7 cm$^{-1}$ per atom, which should be seen as an upper limit, and our values are smaller. Moreover, zero-point vibrational effects will also lower the activation energy. It is therefore difficult to compare the two very different (static vs. dynamic) methods. Furthermore, their MD simulations show an accumulation of defects, stacking disorders and growth of a less ordered structures towards the transition state. They also suggested three different phase transition paths. It is, however, difficult to provide a simple picture of the phase transition from such MD simulations, even more so if the simulation cell does not reflect the change in the lattice constants and surface effects can become important without proper periodic boundary conditions. On the contrary, the Burgers path provides a good upper bound for the activation energy in the hcp$\leftrightarrow$fcc transformation, as other hypothetical minimum energy paths lying below our ideal Burgers path on a high-dimensional energy hypersurface require a large super-cell treatment. Interestingly, the volume expansion at the transition state obtained by Bingxi Li et al.\cite{Bingxi2017} of about 2.6\% and very close to our value of 2.4\% with the (12,6)-LJ potential.

\section{Conclusions}

We investigated in detail the Burgers-Bain minimum energy path for the hcp$\leftrightarrow$fcc$\leftrightarrow$bcc phase transition for a number of representative $(n,m)$-LJ potentials using exact lattice summations. For this we expressed the lattice sums for a bi-lattice, connecting hcp with the cuboidal lattices such as fcc and bcc through a single lattice parameter $\alpha\in[0,1]$, in terms of a fast converging Bessel sum expansions yielding computer precision to cohesive energies. Our first chosen simple phase transition model based on the KHS model contained only four parameters describing the change in the base lattice parameters $a$ and $c=a\gamma_2$, the shear force acting on the hexagonal base plane through the parameter $\alpha$, and the sliding force of the middle layer through one variable parameter $\beta_2$. This limited choice of parameters made it possible to gain a first detailed insight into the Burgers-Bain phase transition\cite{RoblesNavarro2025}. However, upon extending this small parameter space, we observed a distinct symmetry breaking effect due to a single lattice parameter $\beta_1$, causing a non-smooth phase transition at a bifurcation point at critical value $\alpha_c$ where the Burgers path joins the Bain path. The bifurcation point shifts to larger $\alpha_c$ values for increasingly harder LJ potentials. For our softest (6,4)-LJ potential we observe a metastable minimum belonging to the bcc structure. This demonstrates that one has to be very careful in chosing a suitable definition for a bi-lattice. For the LJ potentials we were able to derive useful analytical formulas for the lattice parameters including the KHS model where the bifurcation point shifts into the fcc minimum. From this, we obtained a more complete picture of the rhombohedral bcc distortion towards either the fcc (Bain path) or hcp structure (Burgers path). 

The true (hypothetical) minimum energy path might well lie energetically below the ideal Burgers-Bain path involving a different mechanism for the sliding of the hexagonal planes\cite{Ackland2002}. This would require a super-cell treatment and a much larger set of variable lattice parameters as we detected no additional path in our unit cell definition. For example, we can transform from hcp to fcc through Barlow packing\cite{Barlow1883,Cooper2024,Schwerdtfeger-2024b} arrangements, for example by changing the packing sequence for six hexagonal layers in the following way: hcp$(ABABAB$)$\rightarrow (ABCBAB) \rightarrow (ABCBAC) \rightarrow$fcc$(ABCABC)$. This specific example involves one double shift of hexagonal layers in a Barlow packing of size $N=6$. This path may well be lower than the Burgers transformation and some hexagonal layers such as $BAB$ remain more or less intact with kissing number 12. The minimum energy path for such a mechanism still needs to be explored; however,  we expect an increase in the angle from the original value of 60$^o$ between the two basis vectors $\angle(\vec{b}_1,\vec{b}_2)$ defining the hexagonal plane, as this is the case in the simple Burgers transformation. Moreover, shifting hexagonal layers leads from hcp to fcc through different Barlow structures, and not directly to bcc. Concerning the Bain path, the equivalence between the Bain and Zener transformation was already shown by Rifkin\cite{rifkin1984}. Other phase transitions like the Nishiyama-Wassermann transformation\cite{nishiyama2012martensitic,Sandoval_2009} still need to be explored using exact lattice summations for inverse power potentials and a suitable choice of lattice vectors. We are currently developing a more general theory on solid-state phase transitions using exact lattice summations for many-body inverse power potentials such as the LJ potential or the Axilrod-Teller-Muto three-body term\cite{AxilrodTeller1943,Muto1943}. Such many-body effects play a crucial role in stabilizing the bcc over the fcc phase\cite{burrows2025a}.

We believe that our results obtained from using LJ potentials as a simple model systems will guide future applications using more elaborate quantum theoretical methods such as DFT. For more realistic materials where strong many-body interactions\cite{Schwerdtfeger-2006} are encountered as for metallic systems,\cite{Wentzcovitch1988,Chen1988,Ekman1998,Friak2001} the bcc phase might well be the global minimum and directly connected to the hcp phase, as it is for example indicated by calculations using density functional theory employing the nudged elastic band algorithm for mapping out the minimum energy path for bulk lithium, iron or silver nanowires.\cite{Caspersen2005,Nguyen2018,Sun2022} We point out that a more advanced MD simulation\cite{Zhao2000,Djohari2009} to obtain free energy values at finite temperatures and pressures requires again a super-cell treatment with a large set of independent parameters, which limits the accuracy in calculating the free energy. In such a case exact lattice sum treatments can be developed for LJ potentials. Our simple picture of the lattice transformation offers a useful estimate for the upper limit of the phase transition activation barrier in real systems. Finally, for argon we showed that the Burgers phase transition path is very sensitive to the density functional approximation applied and the inclusion of the dispersion force is important for such weakly interacting systems. A more detailed study for solid argon under pressure including three-body effects and different minimum energy paths is currently underway. 

{\it Acknowledgment.} We acknowledge support from the Sir Neil Waters Trust at Massey University. PS thanks Profs. Paul Indelicato at the Laboratoire Kastler Brossel and Beno\^{i}t Darqui\'{e} at the Laboratoire de Physique des Lasers, Sorbonne Université (Paris), for a visiting professorship. 

\section{Appendix}
\appendix
\setcounter{table}{0}
\renewcommand{\thetable}{A\arabic{table}}
\numberwithin{equation}{section}

\section{Bessel function expansions of the lattice sums for the Burgers path}
\label{Besselexpansions}

In the following we concentrate on the lattice sums $H^h_A$ and $H^h_B$  so we do not have to carry the factor $f(\beta_1,\beta_2,\beta_3)$ along.
\subsection{The lattice sum $H^h_A$}
\label{sec:BesselA}

In three dimensions the lattice sum defined through a quadratic form is related to the Epstein zeta function $Z$ by $H_{S_3} = 2Z_{S_3}\left(s\right)$ as discussed by Terras\cite{terras-1973}, where $S_3$ is a three-dimensional symmetric and positive definite Gram matrix. The Terras decomposition of the Epstein zeta function\cite{terras-1973} can be used for any Bravais lattice and an inverse power potential, and has been detailed in our previous paper for cubic lattices,\cite{burrows-2020} 
	\begin{align}\label{eq:bessel1}
		Z_{S_3}(s) &= \frac{\pi\Gamma(s-1)t^{1-s}}{\Gamma(s)\sqrt{\det S_{2}}}\zeta(2s-2) + \frac{p^{\frac{1}{2}-s}\Gamma(s-\frac{1}{2})}{\Gamma(s)}\sqrt{\frac{\pi}{\sigma_{22}}}\zeta(2s-1)+\frac{\zeta(2s)}{\sigma_{22}^{s}} \\
  \nonumber
		&+\frac{4\pi^{s}}{\Gamma(s)}\left(\frac{p^{1-2s}}{\sigma_{22}^{1+2s}}\right)^{1/4}\sum_{u,v\in\mathbb{N}}\left(\frac{v}{u}\right)^{s-\frac{1}{2}}\cos\left(\frac{2\pi\sigma_{12}uv}{\sigma_{22}}\right)K_{s-\frac{1}{2}}\left(2\pi uv\sqrt{\frac{p(\alpha)}{\sigma_{22}}}\right) \\
    \nonumber
		&+\frac{2\pi^{s}}{\Gamma(s)}\sqrt{\frac{t^{1-s}}{\det S_{2}}}\sum_{\substack{u\in\mathbb{N} \\ \vec{w}\in\tilde{\mathbb{Z}}^{2}\setminus \{\vec{0}_2\}}}\cos\left(2\pi u\vec{w}^\top\vec{y}\right)u^{1-s}\left(\vec{w}^\top S_{2}^{-1}\vec{w}\right)^{\frac{s-1}{2}}K_{s-1}\left(2\pi u\sqrt{t\vec{w}^\top S_{2}^{-1}\vec{w}}\right)
	\end{align}
Here $S_3$ has been block-diagonalized to obtain the $(2 \times 2)$ symmetric sub-matrix $S_2=(\sigma_{ij})$,
\begin{equation} \label{eq:Bmatrix}
S_3 = \begin{pmatrix}
         m  & \vec{z}^\top  \\
         \vec{z}  & S_{2} 
\end{pmatrix}
= \begin{pmatrix}
         1  & \vec{y}^\top  \\
         \vec{0}_2  & I_2 
\end{pmatrix}
\begin{pmatrix}
         t  & \vec{0}_2^\top  \\
         \vec{0}_2  & S_{2}
\end{pmatrix}
\begin{pmatrix}
         1  & \vec{0}_2^\top  \\
         \vec{y}  & I_2 
\end{pmatrix}
\end{equation}
and $\vec{y}$, $\vec{z}$ are simple 2-vectors. This gives the relations,
\begin{equation} \label{eq:matrix1}
\vec{z} = S_2 \vec{y} \quad \implies \quad \vec{y} = S_2^{-1} \vec{z}
\end{equation}
and
\begin{equation} \label{eq:matrix2}
	m = t + \vec{y}^\top S_2 \vec{y} \quad \implies \quad t = m - \vec{z}^\top S_2^{-1} \vec{z}
\end{equation}
with $t\ne 0$ and
\begin{equation} \label{eq:matrix3}
S_2^{-1} = {\rm det}(S_2)^{-1}
\begin{pmatrix}
         \sigma_{22}  & -\sigma_{12}  \\
         -\sigma_{12}  & \sigma_{11} 
\end{pmatrix}
\end{equation}
The other parameters and functions are defined as follows,
\begin{equation} \label{eq:t}
p=\left\{ \sigma_{11}- \frac{\sigma_{12}^2}{\sigma_{22}}  \right\}
\end{equation}
The modified Bessel function is defined as
\begin{equation} \label{eq:bessel4}
K_\nu (x)=\frac{1}{2} \int_{0}^{\infty} u^{\nu-1}{\rm exp}\left\{ -x\left( u+u^{-1} \right)/2 \right\} du \quad {\rm for} \quad |{\rm arg}(x)| < \tfrac{1}{2} \pi
\end{equation}
with $K_\nu(x)=K_{-\nu}(x)$,\cite{glasser-2012} The higher-order Bessel functions can be successively reduced to lower order Bessel functions by
\begin{equation} \label{eq:Besselrecursive}
K_\nu(x)=\frac{2(\nu-1)}{x}K_{\nu-1}(x) + K_{\nu-2}(x)
\end{equation}
and all what remains to be evaluated are the Bessel functions $K_1$, $K_0$ and $K_{\tfrac{1}{2}}$. Further, for half-integer orders of the Bessel function we can use the equation
\begin{equation} \label{eq:Besselhalf}
K_{\tfrac{1}{2}}(x)=K_{-\tfrac{1}{2}}(x)=\sqrt{\frac{\pi}{2x}}e^{-x}
\end{equation}
The relation for the $\Gamma$-functions is
\begin{equation} \label{eq:Gammarecursive}
\Gamma(x+1) = x\Gamma(x)
\end{equation}
For the symmetric matrix $S_3$ we permute the rows and columns in \eqref{eq:S3} and take 
\begin{equation}
S_3(\alpha,\boldsymbol{\gamma}) = 
\begin{pmatrix}
\gamma_{2}^{2} & 0  &  0 \\
0 &1  &\omega_1(\alpha,\gamma_{1})   \\
0 & \omega_1(\alpha,\gamma_{1})  & \omega_2(\alpha,\gamma_{1})   
\end{pmatrix}
\quad \text{and} \quad S_2(\alpha,\gamma_{1}) = 
\begin{pmatrix}
1  & \omega_1(\alpha,\gamma_{1})   \\
\omega_1(\alpha,\gamma_{1})  & \omega_2(\alpha,\gamma_{1})   
\end{pmatrix}
\end{equation}
which for $\vec{z}^\top = (0,0)$ gives $\vec{y}^\top = (0,0)$, and for the other parameters $m=t=\gamma_{2}^{2}$. Taking $\sigma_{11}=1, \sigma_{12}(\alpha,\gamma_{1})=\omega_1(\alpha,\gamma_{1})$, $\sigma_{22}(\alpha)=\omega_2(\alpha,\gamma_{1})$, $p(\alpha)=1-\frac{\omega_1(\alpha,\gamma_{1})^2}{\omega_2(\alpha,\gamma_{1})}=\frac{\omega_3^2(\alpha,\gamma_{1})}{\omega_2(\alpha,\gamma_{1})}>0$ and $\det S_2(\alpha,\gamma_{1})=\omega_2(\alpha,\gamma_{1})-\omega_1(\alpha,\gamma_{1})^2 = \omega_3(\alpha,\gamma_{1})^2 = \omega_2(\alpha,\gamma_{1})p(\alpha)>0$ for $\alpha\in [0,1]$, and we get 
\begin{align}
 \nonumber
		H&_{A}^{h}(s) = \frac{2\pi\Gamma(s-1)}{\gamma_{2}^{2s-2}\Gamma(s)\omega_{3}(\alpha,\gamma_{1})}\zeta(2s-2) + \frac{2p(\alpha)^{\frac{1}{2}-s}\Gamma(s-\frac{1}{2})}{\Gamma(s)}\sqrt{\frac{\pi}{\omega_{2}(\alpha,\gamma_{1})}}\zeta(2s-1) + \frac{2\zeta(2s)}{\omega_{2}(\alpha,\gamma_{1})^{s}}  \\
  \nonumber
		&+ \frac{8\pi^{s}\omega_{3}(\alpha,\gamma_{1})^{\tfrac{1}{2}-s}}{\Gamma(s)\sqrt{\omega_{2}(\alpha,\gamma_{1})}}\sum_{u,v\in\mathbb{N}}\left(\frac{v}{u}\right)^{s-\frac{1}{2}}\cos\left(\frac{2\pi uv\omega_{1}(\alpha,\gamma_{1})}{\omega_{2}(\alpha,\gamma_{1})}\right)K_{s-\frac{1}{2}}\left(\frac{2\pi uv\omega_{3}(\alpha,\gamma_{1})}{\omega_{2}(\alpha,\gamma_{1})}\right) \\
		&+ \frac{4\pi^{s}}{\gamma_{2}^{s-1}\Gamma(s)\omega_{3}(\alpha,\gamma_{1})}\sum_{\substack{u\in\mathbb{N} \\ \vec{w}\in\mathbb{Z}^{2}\setminus \{\vec{0}_2\}}}u^{1-s}\left(\vec{w}^\top S_{2}^{-1}\vec{w}\right)^{\frac{s-1}{2}}K_{s-1}\left(2\pi\gamma_{2} u\sqrt{\vec{w}^\top S_{2}^{-1}\vec{w}}\right)
\end{align}
and for the inverse $S_2$ matrix we have
\begin{equation} \label{eq:matrix4}
S_2^{-1} = \frac{1}{\omega_3^2(\alpha,\gamma_{1})}
\begin{pmatrix}
         \omega_{2}(\alpha,\gamma_{1})  & -\omega_{1}(\alpha,\gamma_{1})  \\
         -\omega_{1}(\alpha,\gamma_{1})  & 1
\end{pmatrix}
\end{equation}
The parameter $p$ does not depend on $\gamma_{1}$ and $p(\alpha)=\sin^{2}(\theta_{12})$ where $\theta_{12}$ is the angle between $\vec{b}_{1}$ and $\vec{b}_{2}$ that can be approximated as $\tfrac{\pi(2+\alpha)}{6}$ when $\alpha\in[0,1]$.)

\subsection{The lattice sum $H^h_B$}

This is the more complicated case, but we can use the expansion derived from Van der Hoff and Benson's original expression derived from a Mellin transformation and the use of theta functions,\cite{Hoff-Benson-1953}
\begin{equation}\label{eq:Hoff-Benson0}
	\begin{aligned}
		\sum_{i\in\mathbb{Z}}\left[x^{2} + (i+\zeta)^{2}\right]^{-s} &= \frac{\sqrt{\pi}\Gamma(s-\frac{1}{2})}{\Gamma(s)\left|x\right|^{2s-1}} + \frac{4\pi^{s}}{\Gamma(s)}\sum_{n\in\mathbb{N}}\left(\frac{n}{\left|x\right|}\right)^{s-\frac{1}{2}}\cos(2\pi n\zeta)K_{s-\frac{1}{2}}(2\pi n|x|)
	\end{aligned}
\end{equation}
with $\zeta \in [0,1)$. We rewrite the quadratic function \eqref{eq:QB} into
\begin{align}
\nonumber
Q_{B}(i,j,k,\alpha,\beta_1,\beta_2,\beta_3,\gamma_1,\gamma_2) &=\gamma_{2}^2 \left[ \gamma_{2}^{-2}(i+j\omega_1(\alpha,\gamma_{1})+\tfrac{\beta_{1}}{2})^2 + \gamma_{2}^{-2}(j\omega_3(\alpha,\gamma_{1})+\tfrac{\beta_{2}}{2})^2 + (k+\tfrac{\beta_3}{2\gamma_2})^2 \right] \\
&= \gamma_{2}^2 \left[  \gamma_{2}^{-2}x^2_{ij}(\alpha,\beta_1,\beta_2,\gamma_{1}) + (k+\tfrac{\beta_{3}}{2\gamma_2})^2 \right]
\end{align}
and get
\begin{equation}
	\begin{aligned}
		H_{B}^{h}&(s, \alpha,\beta_1,\beta_2,\beta_3,\gamma_1,\gamma_2) = \sum_{\vec{i}\in\tilde{\mathbb{Z}}^3}Q_{B}(\vec{i}, \alpha, \beta_1,\beta_2,\beta_3,\gamma_1,\gamma_2)^{-s} = \frac{\sqrt{\pi}\Gamma\left(s-\frac{1}{2}\right)}{\gamma_{2}\Gamma(s)}\sum_{i,j\in\mathbb{Z}}\left|x_{ij}(\alpha,\beta_1,\beta_2,\gamma_{1})\right|^{1-2s} \\
		&+ \frac{4\pi^{s}}{\gamma_{2}^{s+\frac{1}{2}}\Gamma(s)}\sum_{i,j\in\mathbb{Z}}\sum_{n\in\mathbb{N}}\cos\left(\frac{\pi n\beta_{3}}{\gamma_2} \right)\left(\frac{n}{\left|x_{ij}(\alpha,\beta_1,\beta_2,\gamma_{1})\right|}\right)^{s-\frac{1}{2}}K_{s-\frac{1}{2}}\left(\frac{2\pi n|x_{ij}(\alpha,\beta_1,\beta_2,\gamma_{1})|}{\gamma_{2}}\right)
	\end{aligned}
\end{equation}
with
\begin{align}
 x_{ij}(\alpha,\beta_1,\beta_2,\gamma_{1})^2 = (i+j\omega_1(\alpha,\gamma_{1})+\tfrac{\beta_{1}}{2})^2 + (j\omega_3(\alpha,\gamma_{1})+\tfrac{\beta_{2}}{2})^2
\end{align}
One has to avoid $x^2_{ij}=0$ for this expansion. For the approximate range of $(\alpha,\beta_1,\beta_2)$ parameters (Table \ref{tab:latticeparameters}) with $\omega_1\in[0,\frac{1}{2}]$,  $\omega_3\in[\frac{\sqrt{3}}{2}, 1]$, one can easily show that $x^2_{ij}> 0$.
This expansion reduces the problem into a fast converging triple sum involving Bessel functions, and into an additional double sum,
\begin{align}
W(\alpha,\beta_1,\beta_2,\gamma_{1})= \sum_{i,j\in\mathbb{Z}}  \left\{(i+j\omega_1(\alpha,\gamma_{1})+\tfrac{\beta_{1}}{2})^2 + (j\omega_3(\alpha,\gamma_{1})+\tfrac{\beta_{2}}{2})^2\right\}^{\frac{1}{2}-s}
 \end{align}
which however is slowly convergent for low $s$ values. We therefore use again expansion \eqref{eq:Hoff-Benson0} with $\zeta=j\omega_1(\alpha,\gamma_{1})+\tfrac{\beta_{1}}{2}$ and $x=j\omega_3(\alpha,\gamma_{1})+\tfrac{\beta_{2}}{2}$ and get
	\begin{align}
		W&(\alpha,\beta_1,\beta_2,\gamma_{1}) = \frac{\sqrt{\pi}\Gamma(s-1)}{\Gamma(s-\frac{1}{2})}\sum_{j\in\mathbb{Z}}\frac{1}{|j\omega_{3}(\alpha,\gamma_{1})+\frac{\beta_{2}}{2}|^{2s-2}} \\
  \nonumber
		&+ \frac{4\pi^{s-\frac{1}{2}}}{\Gamma(s-\frac{1}{2})}\sum_{j\in\mathbb{Z}}\sum_{n\in\mathbb{N}}\left(\frac{n}{|j\omega_{3}(\alpha,\gamma_{1})+\frac{\beta_{2}}{2}|}\right)^{s-1}\cos\left(2\pi n\left[j\omega_{1}(\alpha,\gamma_{1})+\frac{\beta_{1}}{2}\right]\right)\\
  \nonumber
  &\times K_{s-1}\left(2\pi n\left|j\omega_{3}(\alpha,\gamma_{1})+\frac{\beta_{2}}{2}\right|\right)
	\end{align}
The first sum can be turned into a Hurwitz zeta function $h(s,x)$,
\begin{align}
\sum_{j\in\mathbb{Z}}  \frac{1} { |j+\tfrac{\beta_{2}}{2\omega_3(\alpha,\gamma_{1})}|^{2s-2} } 
=  h\left(2s-2,\tfrac{\beta_{2}}{2\omega_3(\alpha,\gamma_{1})}\right) + h\left(2s-2,1-\tfrac{\beta_{2}}{2\omega_3(\alpha,\gamma_{1})}\right) 
\end{align}
with
\begin{align}
 h(s,x)= \sum_{n\in\mathbb{N}_0} \frac{1}{(n+x)^s}
\end{align}
In summary, the lattice sum $H_{B}$ is
	\begin{align}
   \nonumber
		H&_{B}^{h}(s, \alpha,\beta_1,\beta_2,\beta_3,\gamma_1,\gamma_2) = \frac{\pi\Gamma(s-1)}{\gamma_{2}\Gamma(s)\omega_{3}(\alpha,\gamma_{1})^{2s-2}}\left[h\left(2s-2, \tfrac{\beta_{2}}{2\omega_{3}(\alpha,\gamma_{1})} + h\left(2s-2, 1-\tfrac{\beta_{2}}{2\omega_{3}(\alpha,\gamma_{1})}\right)\right)\right] \\ 
  \nonumber
		&+ \frac{4\pi^{s}}{\gamma_{2}\Gamma(s)} \sum_{j\in\mathbb{Z}}\sum_{n\in\mathbb{N}}\left(\frac{n}{|j\omega_{3}(\alpha,\gamma_{1})+\frac{\beta_{2}}{2}|}\right)^{s-1}\cos\left(2\pi n\left[j\omega_{1}(\alpha,\gamma_{1})+\frac{\beta_{1}}{2}\right]\right)K_{s-1}\left(2\pi n\left|j\omega_{3}(\alpha,\gamma_{1})+\frac{\beta_{2}}{2}\right|\right) \\
		&+ \frac{4\pi^{s}}{\gamma_{2}^{s+\frac{1}{2}}\Gamma(s)} \sum_{i,j\in\mathbb{Z}}\sum_{n\in\mathbb{N}}\cos\left(\frac{\pi n\beta_{3}}{\gamma_2} \right)\left(\frac{n}{|x_{ij}(\alpha,\beta_1,\beta_2,\gamma_{1})|}\right)^{s-\frac{1}{2}}K_{s-\frac{1}{2}}\left(\frac{2\pi n|x_{ij}(\alpha,\beta_1,\beta_2,\gamma_{1})|}{\gamma_{2}}\right)
	\end{align}

\subsection{Adding both lattice sums}
We can now add both lattice sums $H^h_A$ and $H^h_B$
\begin{equation} \label{eq:besselT}
H^{h}\left( s,\alpha,\beta_1,\beta_2,\gamma_1,\gamma_2\right) = g_0\left( s,\alpha,\beta_1,\beta_2,\gamma_1,\gamma_2\right) + \sum_{n=1}^{4}g_n\left( s,\alpha,\beta_1,\beta_2,\beta_3,\gamma_1,\gamma_2\right)
\end{equation}
with
\begin{align}\label{eq:hurwitzS1}
\nonumber
g_0&\left( s,\alpha,\beta_1,\beta_2,\gamma_1,\gamma_2\right) = \frac{2\pi\Gamma(s-1)}{\gamma_{2}^{2s-2}\Gamma(s)\omega_{3}(\alpha,\gamma_{1})}\zeta(2s-2) + \frac{2p(\alpha)^{\frac{1}{2}-s}\Gamma(s-\frac{1}{2})}{\Gamma(s)}\sqrt{\frac{\pi}{\omega_{2}(\alpha,\gamma_{1})}}\zeta(2s-1)  \\
& + \frac{2\zeta(2s)}{\omega_2(\alpha,\gamma_{1})^{s}} + \frac{\pi\Gamma \left( s-1\right)}{\gamma_{2}\Gamma (s) \omega_3(\alpha,\gamma_{1})^{2s-2} } \left\{  h\left(2s-2,\tfrac{\beta_{2}}{2\omega_3(\alpha,\gamma_{1})}\right) + h\left(2s-2,1-\tfrac{\beta_{2}}{2\omega_3(\alpha,\gamma_{1})}\right) \right\}
\end{align}
where the first three terms originate from $H_{A}$ and the last one from $H_{B}$, and each of the following $g_{n}$ terms containing an infinite sum of Bessel functions such that
\begin{equation} \label{eq:besselS1}
g_1\left(s,\alpha,\gamma_1\right)=\frac{8\pi^{s}\omega_{3}(\alpha,\gamma_{1})^{\tfrac{1}{2}-s}}{\Gamma(s)\sqrt{\omega_{2}(\alpha,\gamma_{1})}}\sum_{i,j\in\mathbb{N}}\left(\frac{i}{j}\right)^{s-\frac{1}{2}}\cos\left(\frac{2\pi\omega_{1}(\alpha,\gamma_{1})ij}{\omega_{2}(\alpha,\gamma_{1})}\right)K_{s-\frac{1}{2}}\left(2\pi ij\frac{\omega_{3}(\alpha,\gamma_{1})}{\omega_{2}(\alpha,\gamma_{1})}\right)
\end{equation}
\begin{equation} \label{eq:besselS2}
g_2\left(s,\alpha,\gamma_1,\gamma_2\right)= \frac{4\pi^{s}}{\gamma_{2}^{s-1}\Gamma(s)\omega_{3}(\alpha,\gamma_{1})}\sum_{\substack{n\in\mathbb{N} \\ \vec{w}\in\tilde{\mathbb{Z}}^{2}\setminus \{\vec{0}_2\}}}\left(\frac{\vec{w}^\top S_{2}^{-1}\vec{w}}{n^2}\right)^{\frac{s-1}{2}}K_{s-1}\left(2\pi n\gamma_{2} \sqrt{\vec{w}^\top S_{2}^{-1}\vec{w}}\right)
\end{equation}
\begin{align}
\label{eq:besselS3}
g_3\left(s,\alpha,\beta_1,\beta_2,\gamma_1,\gamma_2\right)&= \frac{4\pi^{s}}{\gamma_{2}\Gamma(s)} \sum_{j\in\mathbb{Z}}\sum_{n\in\mathbb{N}}\left(\frac{n}{|j\omega_{3}(\alpha,\gamma_{1})+\frac{\beta_{2}}{2}|}\right)^{s-1}\cos\left(2\pi n\left[j\omega_{1}(\alpha)+\frac{\beta_{1}}{2}\right]\right) \quad\\
\nonumber
&\times K_{s-1}\left(2\pi n\left|j\omega_{3}(\alpha,\gamma_{1})+\frac{\beta_{2}}{2}\right|\right)
\end{align}
\begin{equation} \label{eq:besselS4}
\begin{aligned}
g_4\left( s,\alpha,\beta_1,\beta_2,\beta_3,\gamma_1,\gamma_2\right)=& \frac{4\pi^{s}}{\gamma_{2}^{s+\frac{1}{2}}\Gamma(s)} \sum_{i,j\in\mathbb{Z}}\sum_{n\in\mathbb{N}}\cos\left(\frac{\pi n\beta_{3}}{\gamma_2} \right)\left(\frac{n}{|x_{ij}(\alpha,\beta_1,\beta_2,\gamma_{1})|}\right)^{s-\frac{1}{2}}\\
&\times K_{s-\frac{1}{2}}\left(2\pi n\gamma_{2}^{-1}|x_{ij}(\alpha,\beta_1,\beta_2,\gamma_{1})|\right)
\end{aligned}
\end{equation}
with $p(\alpha,\gamma_{1})=1-\frac{\omega_1(\alpha,\gamma_{1})^2}{\omega_2(\alpha,\gamma_{1})}>0$ as defined before. Here $g_1$ and $g_2$ originate from the lattice sum $H_{A}$, and $g_3$ and $g_4$ from the lattice sum $H_{B}$.

For computational purposes we need to treat the first three Bessel sums more efficiently. For the first Bessel sum $g_1$ appearing in $H^h_A$ we use permutation symmetry between $i$ and $j$ and get,
\begin{align}\label{eq:besselS1a}
 \nonumber
		g_1\left(s,\alpha,\gamma_{1}\right) &=  \frac{8\pi^{s}\omega_{3}(\alpha,\gamma_{1})^{\tfrac{1}{2}-s}}{\Gamma(s)\sqrt{\omega_{2}(\alpha,\gamma_{1})}} \sum_{i\ge j\in\mathbb{N}}\left( 1-\tfrac{1}{2}\delta_{ij} \right)\left[\left(\frac{i}{j}\right)^{s-\frac{1}{2}}+\left(\frac{j}{i}\right)^{s-\frac{1}{2}}\right] \cos\left(\frac{2\pi\omega_{1}(\alpha,\gamma_{1})ij}{\omega_{2}(\alpha,\gamma_{1})}\right)  \\
  &\times K_{s-\frac{1}{2}}\left(2\pi ij\frac{\omega_{3}(\alpha,\gamma_{1})}{\omega_{2}(\alpha,\gamma_{1})}\right)
\end{align}
where $\delta_{ij}$ is the Kronecker symbol.  

For the triple sum $g_{2}$ we simply use the fact that $S_{2}^{-1}$ is a symmetric matrix and get,
\begin{equation}\label{eq:besselS3a}
	\begin{aligned}
	g_2\left( s,\alpha,\boldsymbol{\gamma}\right)&= \frac{4\pi^{s}}{\gamma_{2}^{s-1}\Gamma(s)\omega_{3}(\alpha,\gamma_{1})}\sum_{n\in\mathbb{N}} \sum_{\substack{\vec{w}\in\tilde{\mathbb{Z}}^{2}\setminus \{\vec{0}_2\} \\ w_1\le w_2}}
     \left( 2-\delta_{w_1w_2}\right) \left(\frac{\vec{w}^\top S_{2}^{-1}\vec{w}}{n^2}\right)^{\frac{s-1}{2}} \\
     & \times K_{s-1}\left(2\pi n\gamma_{2} \sqrt{\vec{w}^\top S_{2}^{-1}\vec{w}}\right)
	\end{aligned}
\end{equation}

The last sum $g_{3}$ and $g_{4}$ for $H^h_B$ cannot be further simplified, only for the special case $\beta_1=1$ we get for $g_3$
\begin{align}\label{eq:besselS2a}
 \nonumber
g_{3}&(s, \alpha, \beta_1,\beta_2,\gamma_1,\gamma_2) = \frac{4\pi^{s}}{\gamma_{2}\Gamma(s)}\sum_{i\in\mathbb{N}} \left( -1 \right)^i \left(\frac{2i}{\beta_{2}}\right)^{s-1}  K_{s-1}\left(\pi\beta_{2} i\right) \\
  &+ \frac{4\pi^{s}}{\gamma_{2}\Gamma(s)}\sum_{i,j\in\mathbb{N}}(-1)^i \cos\left(2\pi\omega_{1}(\alpha,\gamma_{1}) ij\right) \\
  \nonumber
&\times \sum_{k\in\{-1,+1\}}\left(\frac{i}{\left(j\omega_{3}(\alpha,\gamma_{1})+\tfrac{k\beta_{2}}{2}\right)}\right)^{s-1}K_{s-1}\left(2\pi i\left(j\omega_{3}(\alpha,\gamma_{1})+\frac{k\beta_{2}}{2}\right)\right)
	\end{align}
and we used cos$(x+n\pi)=(-1)^n$cos$(x)$ and remember that $\beta_2>0$ and $j\omega_{3}(\alpha)>\frac{1}{2}k\beta_2$. The first single sum is related to the infinite sum in \eqref{eq:Hoff-Benson0} and is one of the most common Bessel function series in the literature \cite{Fucci_2015}. In general, however, the two triple sums $g_{3}$ and $g_{4}$ are fast convergent. We note that for larger $s$ values we obtain large compensating numbers for the individual terms in the lattice sums, which offsets double precision results. In this case, taking the direct summation for the lattice sum for values $s\ge 8$ is the preferred option as outlined in the next section. Alternatively, one can switch to quadruple precision arithmetic.

\section{Direct summation}

For the lattice sum $H^h_A(s,\alpha,\gamma_1,\gamma_2)$ in Eq.\eqref{eq:HA} of the quadratic form $Q_{A}(\vec{i},\alpha,\gamma_1,\gamma_2)$ we can take the $k=0$ term out and get the following expression,
\begin{equation}
\begin{aligned}
\label{eq:HAdirect}
H^h_A(s,\alpha,\gamma_1,\gamma_2)= &{\sum_{i,j\in \mathbb{Z}}}^{'}  \left( i^{2}+2\omega_1(\alpha,\gamma_{1})ij+\omega_2(\alpha,\gamma_{1}) j^{2} \right)^{-s} \\
&+ 2\sum_{i,j\in \mathbb{Z}} \sum_{k\in \mathbb{N}} \left( i^{2}+2\omega_1(\alpha,\gamma_{1})ij+\omega_2(\alpha,\gamma_{1}) j^{2}+\gamma_{2}^{2} k^{2} \right)^{-s}
\end{aligned}
\end{equation}
For the special case of $\omega_2(\alpha)=1$ we can utilize permutation symmetry between $i$ and $j$, otherwise we have to treat the sum as it is. Notice, there are no general solutions in terms of standard functions for the double sum in \eqref{eq:HAdirect}, only for some special cases for the coefficients $\omega_1$ and $\omega_2$ analytical expressions are known \cite{Zucker-1975a}.

For the second lattice sum we can only simplify the summation over $k$ if $\beta_3/\gamma_2=1$,
\begin{equation}
\label{eq:HBdirect}
	\begin{aligned}
		H^{h}_{B}(s,\alpha,\beta_1,\beta_2,\gamma_1,\gamma_2)&=2\sum_{i,j\in \mathbb{Z}} \sum_{k\in \mathbb{N}}
  \left( (i+j\omega_1(\alpha,\gamma_{1})+\tfrac{\beta_{1}}{2})^2 + (j\omega_3(\alpha,\gamma_{1})+\tfrac{\beta_{2}}{2})^2 + \gamma_{2}^2 (k-\tfrac{1}{2})^2 \right)^{-s}
	\end{aligned}
\end{equation}
Nevertheless, for $s=4$ we reach double precision accuracy by summing over $i,j,k\in [-n_L,n_L]$ within a few seconds of computer time. In our application for the LJ potential we chose for the interval $n_L=750$ except for the hypersurface production where we adopted $n_L=500$ to save computer time.

\section{Relation between the lattice sums of the fcc lattice} 
\label{fcclattice}

Here we show how eq. \eqref{eq:LJIfcc} can be brought into the more common form \eqref{eq:LJIfcc1}.
Starting with \eqref{eq:LJIfcc} we have
\begin{align}
L_A^{\text{fcc}} + L_B^{\text{fcc}}
&= \sideset{}{'}\sum_{\vec{i}\in \mathbb{Z}^3} (i_1^2+i_2^2+2i_3^2)^{-s} 
+ {\sum_{\vec{i}\in \mathbb{Z}^3}} \left((i_1+\tfrac12)^2+(i_2+\tfrac12)^2+2(i_3+\tfrac12)^2\right)^{-s} \\
\nonumber
&= \sideset{}{'}\sum_{i_1\equiv i_2\equiv i_3 (\text{mod}\,2)} \left( \left(\tfrac{i_1}{2}\right)^2 + \left(\tfrac{i_2}{2}\right)^2 + 2\left(\tfrac{i_3}{2}\right)^2\right)^{-s} 
=2^{2s} \sideset{}{'}\sum_{i_1\equiv i_2\equiv i_3 (\text{mod}\,2)} \left( i_1^2+i_2^2+2i_3^2\right)^{-s}
\end{align}
using the standard notation for congruences. Since $i_1$ and $i_2$ have the same parity, both of $i_1-i_2$ and $i_1+i_2$ will be even. Hence, we make the change of variable $i_1-i_2 = 2x$ and $i_1+i_2=2y$, or equivalently $i_1 = x+y$ and $i_2 = x-y$. This gives
\begin{align}
L_A^{\text{fcc}} + L_B^{\text{fcc}}
&=2^{2s} \sideset{}{'}\sum_{x+y \equiv i_3 (\text{mod}\,2)} \left( 2x^2+2y^2+2i_3^2\right)^{-s} 
= 2^s \sideset{}{'}\sum_{x+y \equiv i_3 (\text{mod}\,2)} \left( x^2+y^2+i_3^2\right)^{-s}.
\end{align}
Now $x+y$ and $i_3$ will have the same parity if and only if $x+y+i_3 \equiv 0 \pmod{2}$. The indicator function for this condition is
$$
\tfrac12 \left(1+(-1)^{x+y+i_3}\right) 
= \begin{cases} 1 & \text{if $x+y+i_3 \equiv 0 \pmod{2}$,} \\ 0 & \text{if $x+y+i_3 \not\equiv 0 \pmod{2}$.} \end{cases}
$$
On applying this to the sum above we deduce
\begin{align}
L_A^{\text{fcc}} + L_B^{\text{fcc}}
= 2^s\sideset{}{'}\sum_{\vec{i}\in \mathbb{Z}^3} \tfrac12 \left(1+(-1)^{x+y+i_3}\right) \left( x^2+y^2+i_3^2\right)^{-s}
\end{align}
and this is \eqref{eq:LJIfcc1}.

\section{The Bain path as part of the symmetry-broken Burgers transformation} 
\label{BurgersBain}

In this appendix, we show that part of the Burgers path\cite{Burgers1934}  after joining the bcc$\leftrightarrow$fcc transformation path for values $\alpha>\alpha_c$, where $\alpha_c$ is the bifurcation points dependent on the exponents $(n,m)$ of the LJ potential, is identical to the classical Bain path\cite{Bain1924, burrows2025a} at $\alpha=1$ with variable $\gamma_2$. From continuum mechanics we know that the position of each material point, $\vec{x}$, within a body at a reference configuration $\mathcal{S}_{\alpha}$ can be followed along a deformation to a target configuration $\mathcal{S}_{\alpha'}$ through the displacement field $\chi(\vec{x})$, such that the new position is given by $\vec{y}=\chi(\vec{x})$. The mapping from the initial to the end point is given by the deformation gradient tensor, $F=\nabla\chi(\vec{x})$, that maps every vector from the reference configuration to the target.\cite{Gurtin2010,Abeyaratne1998} If $\mathrm{d}x$ is a line element in $\mathcal{S}_{\alpha}$, its length in $\mathcal{S}_{\alpha'}$ is therefore $|\mathrm{d}y|=\sqrt{|F\mathrm{d}x|^{2}}=\sqrt{\mathrm{d}x^{\top}C\mathrm{d}x}$, where $C=F^{\top}F$ is the symmetric positive definite Cauchy-Green tensor. Hence, $C$ has positive eigenvalues, $\lambda_{i}$, corresponding to the transformation $|F\mathrm{d}x|^{2},$ and orthonormal eigenvectors.

For a lattice $\mathcal{L}(A)=\{A\vec{i}=\sum_{n}i_{n}\vec{a}_{n}|\vec{i}\in\mathbb{Z}^{N}\}$, the transformation to a lattice $\mathcal{L}(B)$ is given by an affine map $\chi(\vec{x})=F\vec{x}+\vec{c}$, where $\vec{c}$ is a translation vector that can be set to zero. Thus, the transformation $\mathcal{L}(A)\to\mathcal{L}(B)$ is represented by $B\vec{i}=FA\vec{i}$, which leads to $F=BA^{-1}$. Applying this formalism to the current three-dimensional case of the Burgers transformation, we see that the relation between two conventional unit cells defined along the Burgers path with $\alpha$ and $\alpha'$ is given by the deformation gradient tensor,
\begin{equation}\label{eq:DeformationTensor}
    F = B_{\alpha'}B_{\alpha}^{-1} = \frac{a'}{a}
    \begin{pmatrix}
        1 & 0 & 0 \\
        \omega'_{1}-\frac{\omega_{1}\omega'_{3}}{\omega_{3}} & \frac{\omega'_{3}}{\omega_{3}} & 0 \\
        0 & 0 & \frac{\gamma'_{2}}{\gamma_{2}}
    \end{pmatrix}
\end{equation}
where $\omega_{1}=\tfrac{1-\alpha}{2}$, $\omega_{3}=\tfrac{1}{2}\sqrt{(1+\alpha)(3-\alpha)}$ and for $\alpha=\alpha'$ we trivially have $F = B_{\alpha}B_{\alpha}^{-1}=I$. The shift vector transforms under this deformation as
\begin{equation}
    \vec{v}^{~'}_{s} = \frac{a'}{2}\begin{pmatrix} \beta'_{1} \\ \beta'_{2} \\ \beta'_{3} \end{pmatrix} 
    = F\vec{v}_{s} = \frac{a'}{2}
    \begin{pmatrix}
        \beta_{1} \\
        \left[\omega'_{1}-\frac{\omega_{1}\omega'_{3}}{\omega_{3}}\right]\beta_{1} + \left[\frac{\omega'_{3}}{\omega_{3}}\right]\beta_{2} \\
        \left[\frac{\gamma'_{2}}{\gamma_{2}}\right]\beta_{3}
    \end{pmatrix}
\end{equation}
A polar decomposition can be performed on the deformation gradient tensor, such that $F=RU$, where $R$ is a rotation (orthogonal) matrix and $U$ is a symmetric positive definite matrix called stretch tensor.\cite{Gurtin2010,Abeyaratne1998} Hence, the Cauchy-Green tensor is $C=U^{\top}U$, since $\mathrm{det}R=1$. Therefore, the stretch tensor is calculated as $U=\sqrt{C}$ and its eigenvalues are $s_{i}=\sqrt{\lambda_{i}}$, corresponding to the stretches of the body along each direction in the lattice, thus capturing shape changes without considering rotations. The stretch tensor is then given by
\begin{equation}
    U = \frac{a'}{a}
    \begin{pmatrix}
    \frac{1+\eta_{1}^{2}+|\eta_{2}|}{t} & \frac{\eta_{1}\eta_{2}}{t} & 0 \\
    \frac{\eta_{1}\eta_{2}}{t} & \frac{\eta_{2}^{2}+|\eta_{2}|}{t} & 0 \\
    0 & 0 & |\eta_{3}| \\
    \end{pmatrix}
\end{equation}
where $~t=\sqrt{1+\eta_{1}^{2}+\eta_{2}^{2}+2|\eta_{2}|}$, $\eta_{1}=\omega'_{1}-\tfrac{\omega_{1}\omega'_{3}}{\omega_{3}}$, $\eta_{2}=\omega'_{3}/\omega_{3}$, and $\eta_{3}=\gamma'_{2}/\gamma_{2}$. The eigenvalues of $U$ show the scale factors along each direction are given by
\begin{equation}
    s_{1,2} = \frac{a'}{a}\sqrt{\frac{1}{2}\left[1+\eta_{1}^{2}+\eta_{2}^{2} \pm \sqrt{(1+\eta_{1}^{2}+\eta_{2}^{2})^{2} - 4\eta_{2}^{2}}\right]} \quad\text{and}\quad s_{3} = \frac{a'\gamma'_{2}}{a\gamma_{2}}
\end{equation}
Both transformations from the bcc structure at $\alpha=\frac{1}{3}$ (Burgers) and $\alpha=1$ (Bain) to fcc have the eigenvalues $s_{1}=1.2573106383758839$ and $s_{2,3}=0.8890528784535745$, which implies that both follow the Bain path but with different orientations in three-dimensional space. Note that $s_{1}=\sqrt{2}s_{2,3}$ which is characteristic of the Bain transformation. This concludes our proof.

\section{Some useful relations for the lattice parameters along the bcc-to-fcc phase transition} 
\label{app:bcc}

\begin{figure}
    \centering
    a)\includegraphics[width=0.45\textwidth]{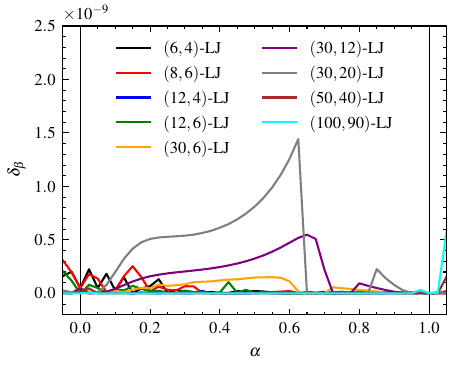}
    b)\includegraphics[width=0.45\textwidth]{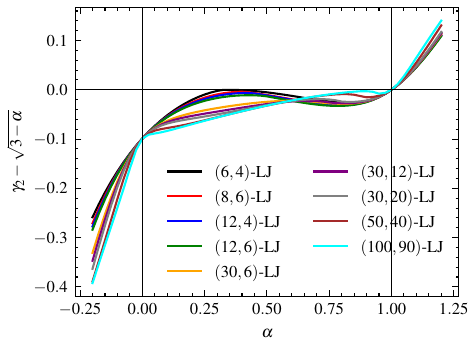}
    \caption{a) Deviation from Eq. \eqref{eq:ratiobeta} for the Lennard-Jones potentials studied here, where $\delta_{\beta}=\Big|\tfrac{\beta_{1}}{\beta_{2}}-\sqrt{\frac{3-\alpha}{1+\alpha}}\Big|$. b) Deviation from Eq. \eqref{eq:Gamma2EqualAngle} for the Lennard-Jones potentials.}
    \label{fig:DeviationBetaRatio}
\end{figure}

To discuss the rhombohedral distortion of the bcc phase to another phase as mentioned already by Max Born in 1940\cite{Born_1940,born1940}, we need to analyze the Bain part of the Burgers transformation in more detail. The transformation from our unit cell (shown in blue color in Figure \ref{fig:StellOct}), spanned by the vectors $\vec{b}_i$ (see eq.\eqref{secondchoice}) chosen for the Burgers path with an additional atom at position $\vec{v}_s$ (thus a bi-lattice by definition), to a primitive unit cell (shown green color in Figure \ref{fig:StellOct}) is only possible along the cuboidal transformation path from bcc ($\alpha=1/3$) to fcc ($\alpha=1$) (or vice versa) where the crystal is completely defined by a single unit cell. Since the unit cell, given by $B_{\text{cub}}$, defined here contains two atoms, the condition for the primitive unit cell, $B_{\text{p}}$, is that 
\begin{equation}
V_{\text{p}}=\det({B_{\text{p}}})=\det({B_{\text{cub}}})/2=V_{\text{cub}}/2 \quad .
\end{equation}
Choosing the lattice basis vectors for the primitive unit cell as $\vec{p}_{1}=\vec{v}_{s}$, $\vec{p}_{2}=\vec{b}_{2}$, and $\vec{p}_{3}=\vec{b}_{3}$, the only way to ensure that the volume per atom is preserved is by the following condition
\begin{equation}\label{eq:VolumeCondition}
    \beta_{1} - \frac{\beta_{2}\omega_{1}(\alpha)}{\omega_{3}(\alpha)} = 1
\end{equation}
where $\beta_{3}=\gamma_{2}$ as detailed below. 
From eqs.\eqref{eq:omega12cond} and \eqref{eq:angle1} we get the condition for two of the $\beta$ parameters in terms of $\alpha$,
\begin{equation}\label{eq:beta1beta2}
    \beta_1 = \frac{\beta_2(1-\alpha)}{\sqrt{4-(1-\alpha)^2}} + 1
\end{equation}
A further condition can be applied that keeps the tetragonal body-centered symmetry followed during the Bain path, namely, that the central atom is located at the center of the parallelepiped during the transformation,
\begin{equation}
\vec{v}_{s} = \frac{1}{2}\left(\vec{b}_{1}+\vec{b}_{2}+\vec{b}_{3}\right)
\end{equation}
Further, it is convenient to write the shift vector $\vec{v}_{s}$ in terms of the $\vec{b}_{i}$-basis as
\begin{equation}
    \vec{v}_{s} =\frac{1}{2}\left[\beta_{1}-\frac{\beta_{2}\omega_{1}(\alpha)}{\omega_{3}(\alpha)}\right]\vec{b}_{1} + \frac{1}{2}\left[\frac{\beta_{2}}{\omega_{3}(\alpha)}\right]\vec{b}_{2} + \frac{1}{2}\left[\frac{\beta_{3}}{\gamma_{2}}\right]\vec{b}_{3} \quad .
\end{equation}
such that in addition to eq.\eqref{eq:VolumeCondition} we get
\begin{equation}\label{eq:Centering}
    \beta_{2}=\omega_{3}(\alpha) \quad \text{and} \quad \beta_{3}=\gamma_{2}
\end{equation}
for any $\alpha\in[\frac{1}{3},1]$. From this we derive
\begin{equation}\label{eq:Centering1}
\beta_{1}= 1+\omega_{1}(\alpha) = \frac{3-\alpha}{2} \qquad .
\end{equation}
 Let $\phi_{1}$ be the angle between $\vec{b}_{1}$ and the projection of $\vec{p}_{1}$ into the $(\vec{b}_{1}\vec{b}_{2})$-plane. The constraint on the position of the second atom in the cell also implies $\phi_{1}=\theta_{12}/2$, where $\theta_{12}$ is the angle between $\vec{b}_{1}$ and $\vec{b}_{2}$. Therefore, using the half-angle formulas for cosine and sine together with Eq. \eqref{eq:angle1}, we obtain
\begin{equation}\label{eq:Centering1}
    \frac{\beta_{1}}{\sqrt{\beta_{1}^{2}+\beta_{2}^{2}}} = \sqrt{\frac{3-\alpha}{4}} \qquad\text{and}\qquad \frac{\beta_{2}}{\sqrt{\beta_{1}^{2}+\beta_{2}^{2}}} = \sqrt{\frac{1+\alpha}{4}}
\end{equation}
which then fixes the ratio to
\begin{equation}\label{eq:ratiobeta}
    \frac{\beta_{1}}{\beta_{2}}=\sqrt{\frac{3-\alpha}{1+\alpha}}
\end{equation}
This relation is in agreement with the KHS model values where we fixed $\beta_{1}=1$, see Eq. \eqref{eq:KHSbeta2}. From Figure \ref{fig:DeviationBetaRatio}a, it can be seen that the largest deviation from eq.\eqref{eq:ratiobeta} within $\alpha\in[0,1]$ is found for the $(30,20)$-LJ potential with a magnitude less than $2\times10^{-9}$ due to numerical errors. Hence we get the interesting result that Eq.\eqref{eq:ratiobeta} is valid for the whole range of $\alpha\in[0,1]$ and for any choice of LJ potentials. Moreover, any symmetry breaking effect in $\beta_1$ will lead to a symmetry breaking in $\beta_2$. And last, the derivatives of $\beta_1$ with respect to $\alpha$ are also connected to the corresponding derivatives of $\beta_2$. Also note that from Eqs. \eqref{eq:Centering} and \eqref{eq:Centering1} it follows that $\beta_{1}^{2}+\beta_{2}^{2}=2\beta_{1}$ along the Bain path for $\alpha\in[\tfrac{1}{3},1]$. 

\begin{figure}
    \centering
    \includegraphics[width=0.5\linewidth]{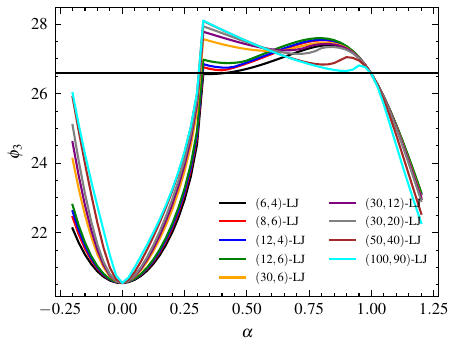}
    \caption{Angle between $\vec{b}_{3}$ and the projection of $\vec{p}_{1}$ on the plane spanned by $\vec{b}_{2}$ and $\vec{b}_{3}$. The horizontal line is $\phi_{3}=26.57^\circ$, which correspond to both fcc and bcc.}
    \label{fig:Angle_phi3}
\end{figure}

We can also derive more approximate but useful relationships for the lattice parameters. The tangent of the angle between $\vec{b}_{3}$ and the projection of $\vec{p}_{1}$ in the ($\vec{b}_{2}\vec{b}_{3}$)-plane is 
\begin{equation}
    \tan\phi_{3} = \frac{\beta_{1}\omega_{1}+\beta_{2}\omega_{3}}{\beta_{3}\gamma_{2}}
\end{equation}
The numerator in this formula is equal to $1$ for fcc and bcc at $\alpha=1$. For hcp, $\phi_3=20.56^\circ$, whereas for both fcc and bcc ($\alpha=\tfrac{1}{3}$), we get $\phi_{3}=26.57^\circ$. Using the centering conditions in Eq. \eqref{eq:Centering}, this relation simplifies to
\begin{equation}
    \tan\phi_{3} = \frac{1+\omega_{1}}{\gamma_{2}^{2}}
\end{equation}
As can be seen in Figure \ref{fig:Angle_phi3} for the different Lennard-Jones potentials, the variation of this angle is small during the cuboidal transformation between bcc ($\alpha=\tfrac{1}{3}$) and fcc. One could naively constrain this angle to be constant along the Bain transformation, such that $\tan\phi_{3}=\tfrac{1}{2}$, i.e., $\phi_{3}=26.57^{\circ}$. Therefore, the parameter $\gamma_{2}$ can be approximately related to $\alpha$ along this path by
\begin{equation}\label{eq:Gamma2EqualAngle}
    \gamma_{2}(\alpha) \approx \sqrt{3-\alpha}
\end{equation}
Keeping this angle constant means that as the base gets skewed along a variable $\alpha$ path, the height of the unit cell must compensate to keep the tilt of the body-centered atom position in the ($\vec{b}_{2},\vec{b}_{3}$) plane fixed, making all right triangles in this plane similar.

Instead of fixing the angle $\phi_3$, another possible constraint is keeping the volume fixed along the transformation. That is, for $a=1$, we have $V=\gamma_{2}\omega_{3}=K$. From Figure S3 in the Supporting Information, we see that the volume changes slightly when using a very soft potential. Since both hcp and fcc have the same volume, we can arbitrarily set $K=\sqrt{2}$. Therefore, another approximate formula for $\gamma_{2}$ appears:
\begin{equation}
    \gamma_{2}(\alpha) \approx \sqrt{\frac{8}{4-(1-\alpha)^{2}}}
\end{equation}
Since this formula for $\gamma_{2}$ is highly dependent on the choice of $K$, its usefulness could be limited, however the deviations are less than $0.15$ from hcp to fcc when $K=\sqrt{2}$. Furthermore, Figure \ref{fig:DeviationBetaRatio}b shows that the deviation of $\gamma_{2}$ from Eq. \eqref{eq:Gamma2EqualAngle} is also less than $0.10$ within the interval of the Burgers transformation. Hence, these two formulas can be used for initial guesses at optimizations with exact lattice sums or ab initio methods. 

In the case of a cuboidal transformation at constant $\alpha=1$, it is straightforward to show that $|\vec{p}_{1}|=\tfrac{a}{2}\sqrt{2+\gamma_{2}^{2}}$ leads to $\beta_{1}^{2} + \beta_{2}^{2} + \beta_{3}^{2} = 2 + \gamma_{2}^{2}$, or, due to $\gamma_{2}^{2}=\beta_{3}^{2}$,
\begin{equation}
    \beta_{1}^{2} + \beta_{2}^{2} = 2
\end{equation}
while keeping the centering conditions. By means of Eq. \eqref{eq:ratiobeta}, we see that this condition leads to $\beta_{1}(\alpha=1)=\beta_{2}(\alpha=1)=1$.

\bibliography{references}

\clearpage

\section*{Supplementary Information for Lattice Instabilities Along the Transformation from Hexagonal to Cuboidal Structures in Hard- and Soft-Sphere Models}

\title{Supplementary Information for Lattice Instabilities Along the Transformation from Hexagonal to Cuboidal Structures in Hard- and Soft-Sphere Models}

In this supplementary materials we give more details including additional figures on the Burgers path within a restricted parameter space $(a,\alpha,\beta_2,\gamma_2)$ fixing the remaining parameters to $\gamma_1=1,\beta_1=1$ and $\beta_3=\gamma_2$. We also provide details on the Newton-Raphson procedure chosen for all optimizations.

\section*{The hcp$\leftrightarrow$fcc Burgers transformation in a restricted four-parameter space}

\begin{figure}[htpb!]
\centering
\includegraphics[width=.49\columnwidth]{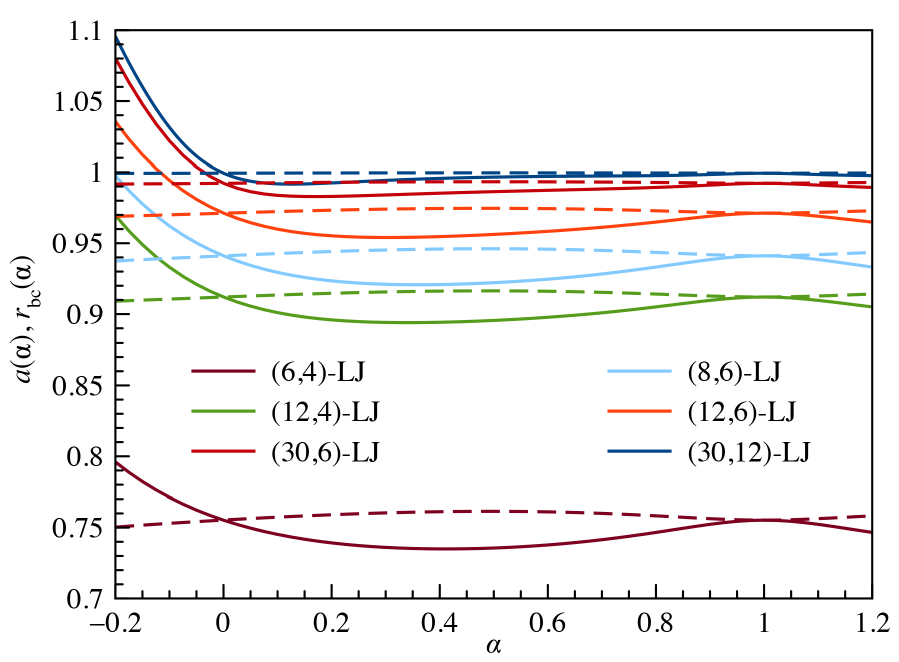}
\caption{Optimized lattice parameters $a$ and $r_\text{bc}$ as a function of the distortion parameter $\alpha$ for various selected $(n,m)$-LJ potentials using Model 2 and $\beta_1=1$. The solid lines show the distances between nearest neighbors in the base layer ($a=|\vec{b}_1|$), while the dashed lines show the optimized distances from the base to the body-centered atom ($r_\text{bc}=|\vec{v}_s|$). Note that $a=r_\text{bc}$ for the close packings at $\alpha=0$ (ideal hcp ) and $\alpha=1$ (fcc).}
\label{fig:amin}
\end{figure}

As no symmetry breaking is observed when we set $\beta_1=1$, the optimized lattice parameters $a(\alpha),\beta_2(\alpha)$ and $\gamma_2(\alpha)$ are all smooth functions in the minimum energy path parameter $\alpha$. The optimized lattice parameters $a$ for the different LJ potentials are shown in Figure \ref{fig:amin} as a function of the distortion parameter $\alpha$. For both the ideal hcp and fcc structures we have $a=r_\text{bc}$ in our definition of the unit cell because of $f(\beta_1=1.0,\beta_2,\gamma_2)=1$ (see main text). Larger deviations are only observed for the bcc structure for which we have $a/r_\text{bc}=\frac{2}{\sqrt{3}}$. For the interval $\alpha\in[0,1]$ we always have $a\le r_\text{bc}$ if we neglect the very tiny difference in $(a-r_\text{bc})$ for the nonideal hcp structure, which leads to small variations $a=r_\text{bc}\pm\epsilon$. This originates from the fact that $\gamma_2$ deviates only slightly from the ideal hard-sphere value of $\sqrt{\frac{8}{3}}$ for the hcp structure.\cite{Cooper2023} As a result of this small distortion, the kissing number is reduced from $\kappa=12$ to 6 for the (12,6)-LJ potential.  As can be further seen, the lattice parameter $a$ is more dependent on the choice of the exponents $n$ and $m$ of the LJ potential than on the distortion parameter $\alpha$. This becomes even more evident when we consider the distance $r_\text{bc}$ from the base atom at the origin to the body-centered atom, which remains almost constant over the whole range of $\alpha$ values. For the $(30,12)$-LJ potential we have $r_\text{bc}\approx 1$ being already close to the KHS model for which we have $r_\text{bc}=1$. 

\begin{figure}[htb!]
\centering
(a)\includegraphics[width=.45\columnwidth]{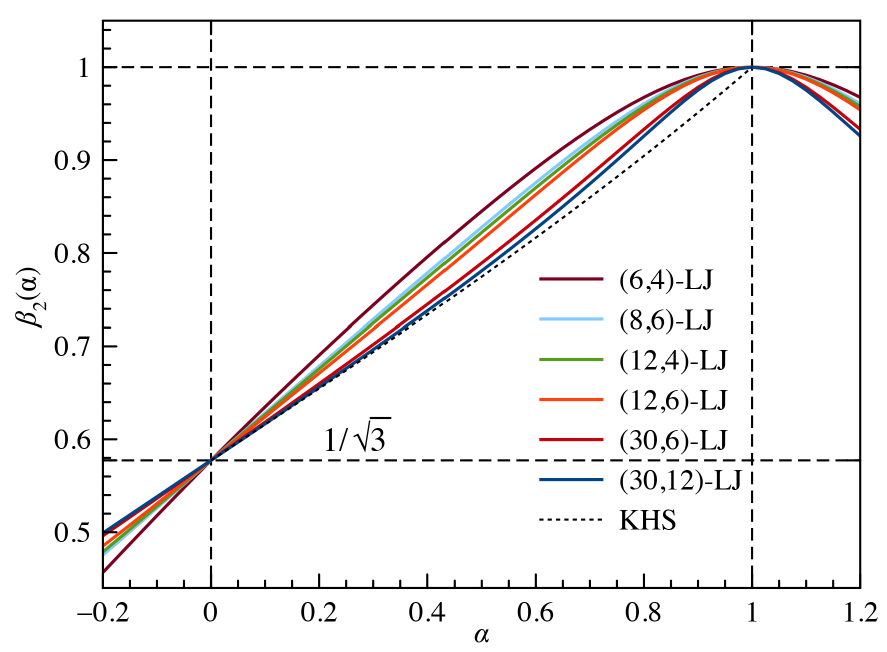}
(b)\includegraphics[width=.45\columnwidth]{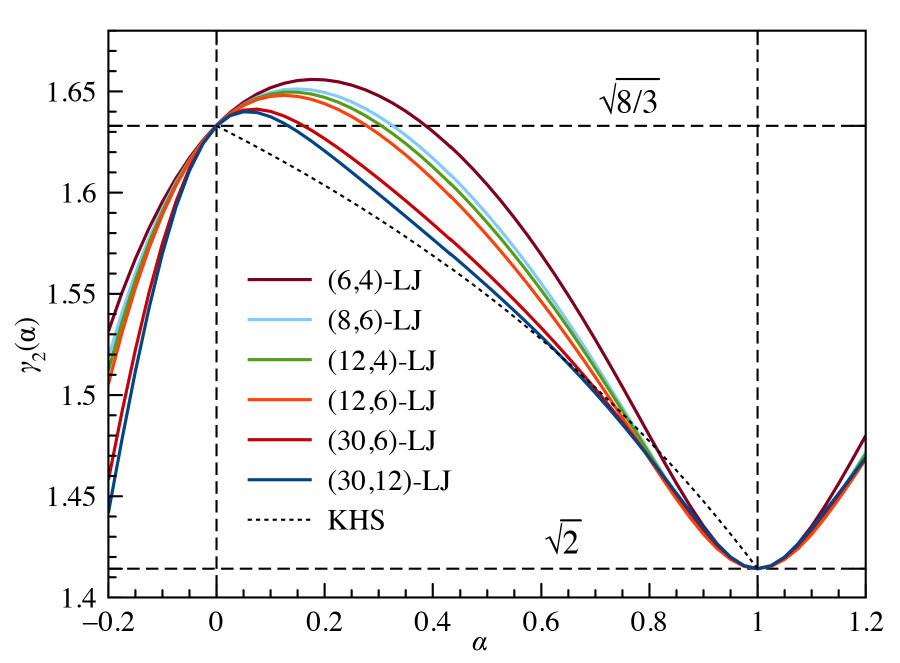}
\caption{The optimized parameter (a) $\beta_2$ and (b) $\gamma_2$ as a function of the distortion parameter $\alpha$ for various selected $(n,m)$-LJ potentials using Model 2. The KHS limit curves for $\alpha\in[0,1]$ are also shown.}
\label{fig:betagamma}
\end{figure}

The optimized lattice parameters $\beta_2$ and $\gamma_2$ are shown in Figure \ref{fig:betagamma} as a function of the distortion parameter $\alpha$. The parameter $\beta_2$, responsible for the sliding of the middle B-layer, is increasing monotonically in an almost linear fashion from $1/ \sqrt{3}$ at $\alpha=0$ to 1 at $\alpha=1$ as expected from the hard-sphere model. This implies a smooth shift of the B-layer for the Burgers transformation from hcp to fcc. The parameter $\gamma_2$, which for example describes also the Bain transformation from fcc to bcc, shows a more interesting behavior. There is a slight increase from the value of $\gamma_2\approx\sqrt{\frac{8}{3}}$ for the hcp structure with a maximum between $\alpha=0$ and $\alpha=0.2$, followed by the expected decrease to $\gamma_2=\sqrt{2}$ at $\alpha=1$ for the fcc structure along the Burgers path. This will influence both the volume $V$ and the packing density $\rho$ for spheres of radius $r$ given by
\begin{align}\label{eq:vol1}
V(a,\alpha,\gamma_1,\gamma_2)&= a^3\gamma_{2}\omega_{3}(\alpha,\gamma_{1})\\
\rho(a,\alpha,\gamma_1,\gamma_2)&=\frac{4\pi r^3}{3a^3\gamma_{2}\omega_3(\alpha,\gamma_{1})}
\label{eq:vol2} 
\end{align} 

\begin{figure}[htb!]
\centering
(a)\includegraphics[width=.45\columnwidth]{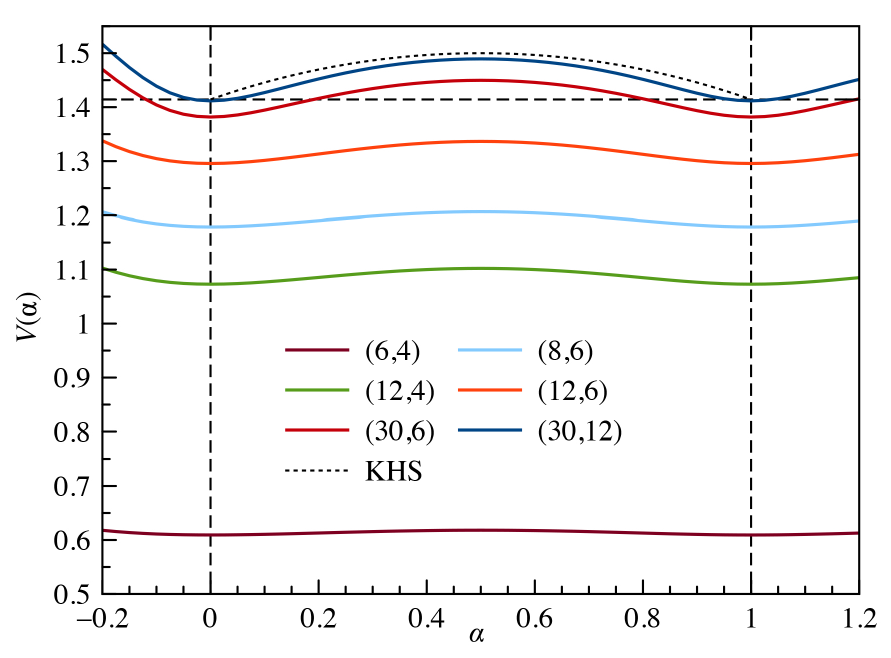}
(b)\includegraphics[width=.45\columnwidth]{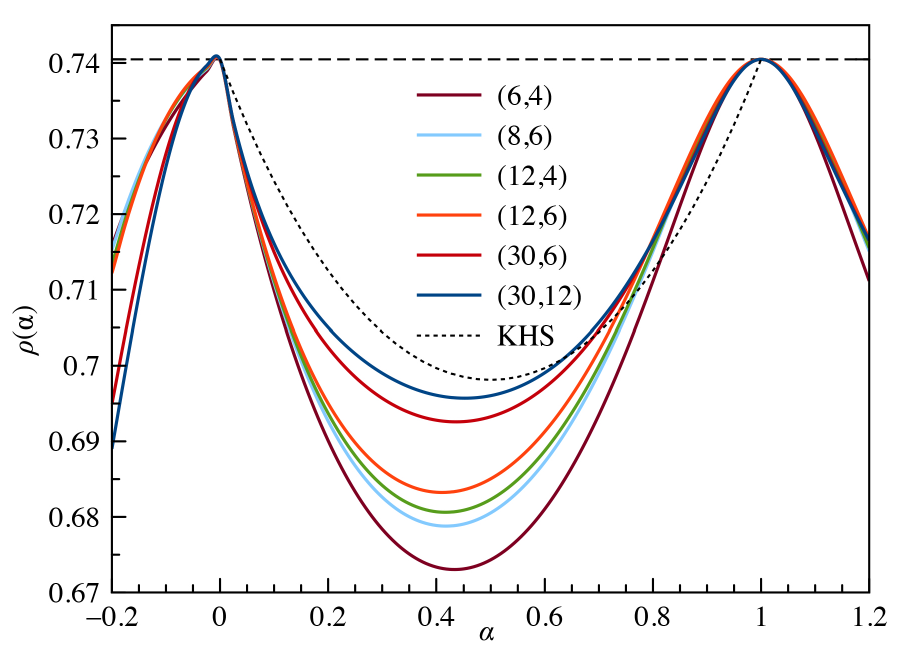}
\caption{(a) Volume $V$ from \eqref{eq:vol1} and (b) packing density $\rho$ using for the sphere radius (half the nearest neighbor distance) $r=\text{min}\{ a,a_\text{bc} \}$ in eq. \eqref{eq:vol2} for various selected $(n,m)$-LJ potentials as a function of the distortion parameter $\alpha$ using Model 2 and $\beta_1=1$. The horizontal lines indicate the hard-sphere limit for the volume, $V(\alpha=1)=\sqrt{2}$, and the maximum packing density $\rho_\text{fcc}=\rho_\text{hcp}=\frac{\sqrt{2}}{6}\pi$. The KHS limit curves are also shown for $\alpha\in[0,1]$.}
\label{fig:packingdensity}
\end{figure}

Figure \ref{fig:packingdensity}(a) shows the volume as a function of the distortion parameter $\alpha$. As one expects, an increase in volume is required for the hcp$\leftrightarrow$fcc transition along the Burgers path with a volume maximum located close to the midpoint at $\alpha=0.5$ similar to the hard-sphere model. In fact, the KHS model maximum volume of $V=\frac{3}{2}$ at $\alpha=0.5$ is an upper bound to all of the $(n,m)$-LJ potentials. As expected, the $(30,12)$-LJ potential is already close to the KHS limit.

We can now derive the packing density. We define the hard sphere radius as $r=\text{min}\{ a,a_c \}/2$ in eq.\eqref{eq:vol2}, which implies a non-overlapping sphere model with $2r \leq 1$ for the different $n,m$ combinations of the LJ potential. In other words, the sphere radius changes along the Burgers transition path and the packing density is $\rho \leq \frac{\sqrt{2}}{6}\pi$. This is shown in Figure \ref{fig:packingdensity}. As a second choice, we may allow spheres to overlap and set the radius to a fixed value of $r=\frac{1}{2}$, the radius of a unit sphere. In this case one has to account for the overlap volume to obtain the correct packing density, see discussion on this subject by Iglesias-Ham et al.\cite{iglesiasham2014} In any case, the graphs in Figure \ref{fig:packingdensity} show that the packing density is reduced along the Burgers path towards the transition state as we expect.

Because of the simplicity and smooth behavior of the lattice parameters in our reduced parameter space, we were able to scan through the $(\alpha,\gamma_2)$ parameter space by optimizing $a,\beta_2$ for fixed $\beta_1=1$ at each point to obtain a potential energy hypersurface of the form $E(\alpha,\gamma_2)$. We did not consider the (30,12)-LJ potential, as it is close to the kissing hard-sphere limit and results in steep ridges for the potential energy surface around the minimum energy path. Moreover, at larger $\gamma_2$ values one gets an early onset of symmetry breaking effects for such hard potentials such that the middle B-layer becomes energetically unstable by moving either up or down towards one of the A-layers. This movement parallel to the base hexagonal sheet would cost little energy due to the short-range nature of the potential, which makes the optimization of the $\beta_2$ parameter difficult. Furthermore, the bcc structure at $\gamma_2=1$ is energetically a maximum for this hard potential distorting either to fcc or to another minimum at $\alpha=1$, $\beta_2=1$ and $\gamma_2=0.82395744$, very close to the ideal axial centered-cuboidal (acc) lattice at $\gamma_2=\sqrt{2/3}$.\cite{Conway1994}

\begin{figure}[htb!]
\centering
\includegraphics[width=.49\columnwidth]{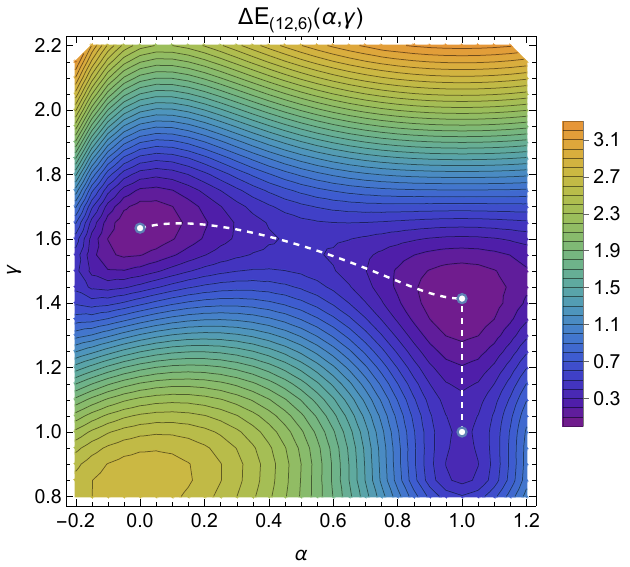}
\includegraphics[width=.49\columnwidth]{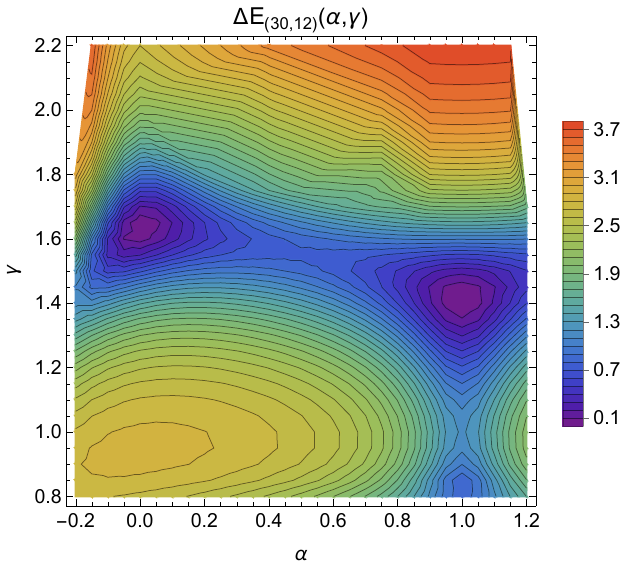}\\
\includegraphics[width=.49\columnwidth]{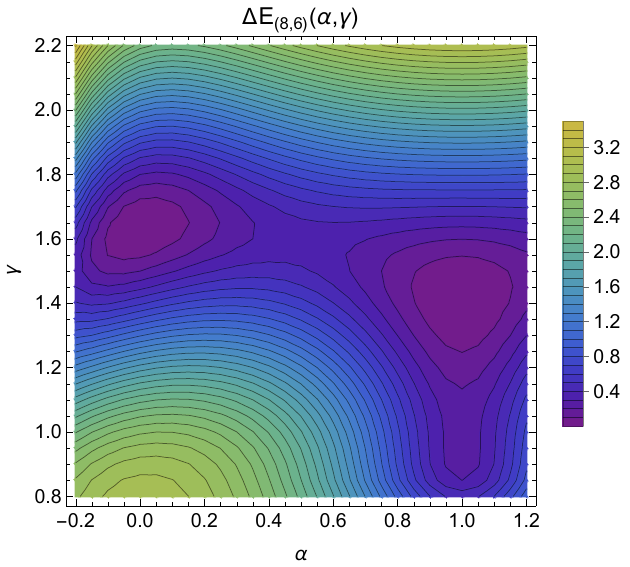}
\includegraphics[width=.49\columnwidth]{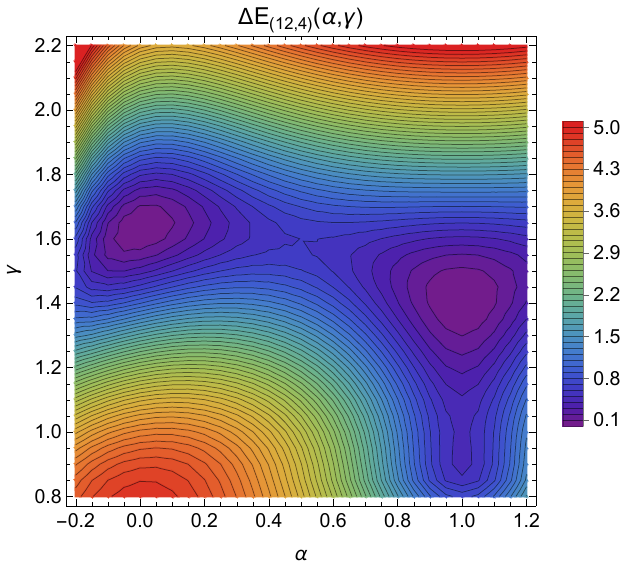}
\caption{Contour plots for the $(\alpha,\gamma_2)$ energy hypersurfaces at optimized $(a,\beta_2)$ values (at constant $\beta_1=1$) for the cohesive energy difference $\Delta E_{nm}(\alpha) = E_{nm}(\alpha)-E_{nm}^{\text{hcp}}(\alpha=0)$ of a $(12,6)$-LJ potential (top left), $(30,12)$-LJ potential (top right), $(8,6)$-LJ potential (bottom left), and $(12,4)$-LJ potential (bottom right). For the $(12,6)$-LJ the minimum Burgers path from hcp ($\alpha=0, \gamma_2=\sqrt{\frac{8}{3}}$. left minimum) to fcc ($\alpha=0, \gamma_2=\sqrt{2}$, right minimum) and the minimum Bain path at constant $\alpha=1$ from fcc to bcc ($\alpha=1, \gamma_2=1$) and beyond ($\gamma_2<1$) are shown. Note the different energy scales used as shown on the right hand side of the contour plots.}
\label{fig:hypersurface}
\end{figure}

\begin{figure}[htb!]
\centering
\includegraphics[width=.49\columnwidth]{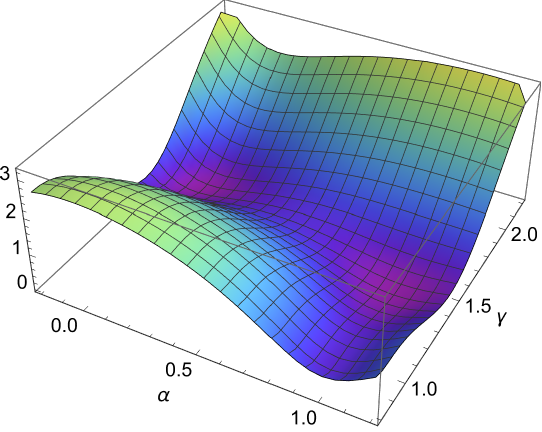}
\includegraphics[width=.49\columnwidth]{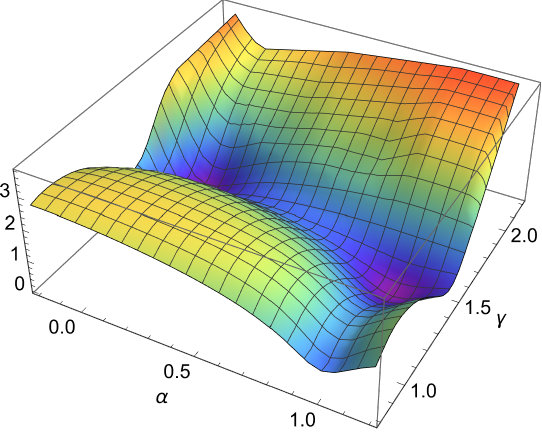}\\
\includegraphics[width=.49\columnwidth]{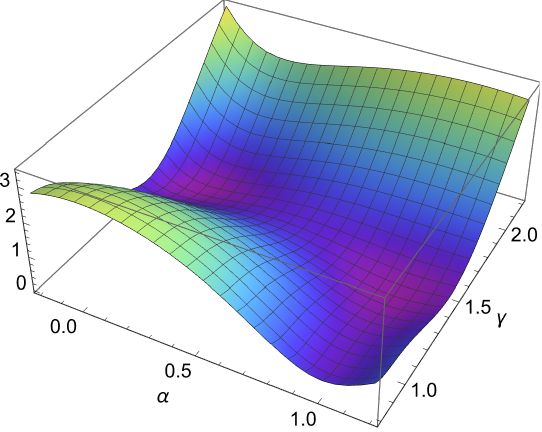}
\includegraphics[width=.49\columnwidth]{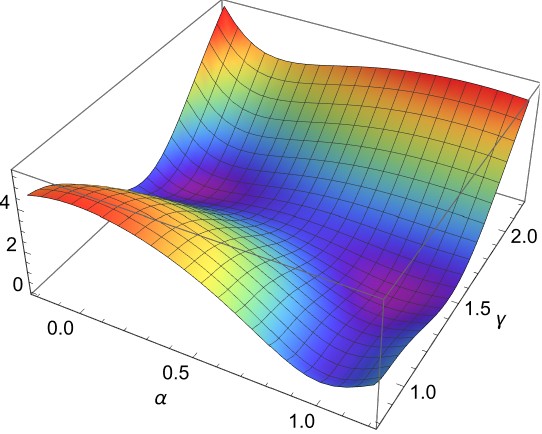}
\caption{$(\alpha,\gamma_2)$ energy hypersurfaces at optimized $(a,\beta_2)$ values (at constant $\beta_1=1$) for the cohesive energy difference $\Delta E_{nm}(\alpha) = E_{nm}(\alpha)-E_{nm}^{\text{hcp}}(\alpha=0)$ of a $(12,6)$-LJ potential (top left), $(30,12)$-LJ potential (top right), $(8,6)$-LJ potential (bottom left), and $(12,4)$-LJ potential (bottom right). See Figure \ref{fig:hypersurface} for more details.}
\label{fig:hypersurface1}
\end{figure}

The energy hypersurfaces for four different LJ potentials are shown in Figures \ref{fig:hypersurface} and \ref{fig:hypersurface1} (the 2D hypersurface for the (12,6)-LJ potential was discussed by us before\cite{RoblesNavarro2025}). We can clearly see the minimum energy path (in our limited four-parameter space neglecting rhombohedral distortions for the cuboidal structures such as bcc\cite{Born_1940,Misra_1940}) from the hcp minimum ($\alpha=0$ and $\gamma_2\approx\sqrt{\frac{8}{3}}$) on the left to the fcc minimum ($\alpha=1$ and $\gamma_2=\sqrt{2}$) on the right of the contour plot. We can also see the Bain path at constant $\alpha=1$ from fcc towards the bcc structure at $\gamma_2=1$, which represents an extremum on that path\cite{burrows2021b}. It is now clear that the hcp$\rightarrow$bcc  transition in this reduced parameter space model involves two phase transitions (if bcc represents a minimum along the Bain path), with the fcc lattice as an intermediate phase. Concerning the stability of the two different phases we already pointed out that the bcc phase lies always above the fcc structure and is metastable only for soft long-range potentials,\cite{Smits2021} otherwise it becomes unstable and distorts towards to a cubic structure structure with smaller $\gamma_2$ values \cite{burrows2021b}. It was already mentioned that the kissing number reduces to 8 for the interval $\gamma_2\in(\sqrt{2},\sqrt{\frac{2}{3}})$, which explains why this path is energetically lower compared to the Burgers path except for the potentials with large exponents $n$ and $m$. We note a maximum (second-order saddle point) around $\alpha=0.0$ and $\gamma_2=0.8$, clearly visible in Figure \ref{fig:hypersurface}. For example, for the (12,6)-LJ potential we obtain $\alpha=0$, $\beta_2=1/\sqrt{3}$, $a=0.95109476$, $\gamma_2=0.84836004$ and $\Delta E^*=2.77589575$ compared to the hcp minimum structure. These values are identical to the ones obtained before for the hcp lattice,\cite{Cooper2023} where the cohesive energy with changing $\gamma_2$ has been studied in detail.

\section*{The iterative Newton–Raphson algorithm} 
\label{NewtonRaphson}

The iterative Newton–Raphson algorithm \cite{hammerlin2012numerical} can be used to determine all critical points (except for the bifurcation point) as well as the minimum energy path (by keeping the parameter $\alpha$ fixed) for an $n$-dimensional parameter space (e.g. seven or less dimensions in our case with $\vec{x}^\top=(a,\alpha,\beta_1,\beta_2,\beta_3,\gamma_1,\gamma_2)$) to the required accuracy, 
\begin{equation}\label{eq:mincond1}
\vec{x}_{N+1}=\vec{x}_N -  \lambda \left[ (\partial_{x_i}\partial_{x_j}) E_{nm}^*(\vec{x}_N )\right]^{-1}\vec{\nabla} E_{nm}^*(\vec{x}_N )   \,,
\end{equation}
where for evaluating the gradient and Hessian we use numerical methods, i.e. the well known expressions for the first and second derivatives,\cite{abramowitzstegun2008}
\begin{equation}
\partial_x f(x)= \left\{ f(x-2h) - 8f(x-h) + 8f(x+h) -f(x+2h) \right\}/12h
\end{equation}
\begin{equation}
\partial_x^2 f(x)= \left\{ -f(x-2h) + 16f(x-h) -30f(x) + 16f(x+h) -f(x+2h) \right\}/12h^2
\end{equation}
We chose $\lambda=0.9$ and $h=1\times 10^{-4}$ for the step size. For the mixed derivatives we used the following formula,\cite{abramowitzstegun2008}
\begin{equation}
\partial_x\partial_y f(x,y)= \left\{ f(x+h,y+h) - f(x+h,y-h) -f(x-h,y+h) + f(x-h,y-h)  \right\}/4h^2
\end{equation}
We note that the convergence of the Newton-Raphson procedure to the correct lattice parameters for a specific $\alpha$ value describing the Burgers path is very sensitive to the starting values $\vec{x}_0$ and to the parameter $\lambda$.

The original variable seven parameter space $\vec{x}=(a,\alpha,\gamma_1,\gamma_2,\beta_1,\beta_2,\beta_3)^\top$ was reduced to five parameters because we found by calculations that the two conditions $\gamma_1=1$ and $\beta_3=\gamma_2$ are fulfilled along the (symmetry broken) Burgers path. This can be further reduced to the four parameter space $\vec{x}=(a,\alpha,\gamma_2,\beta_2)^\top$ because of the condition
\begin{equation}
\beta_1=f(\alpha)\beta_2 \quad \text{with} \quad f(\alpha)=\sqrt{\frac{3-\alpha}{1+\alpha}}
\end{equation}


\end{document}